\documentclass[final,1p,times]{elsarticle}
\usepackage{cases}
\usepackage{amssymb}
\usepackage{lineno}
\usepackage{color}

\newtheorem{proposition}{Proposition}

\newcommand{\ds}{\displaystyle}

\def\bfU{\mbox{\boldmath$U$}}
\def\bfA{\mbox{\boldmath$A$}}
\def\bfB{\mbox{\boldmath$B$}}
\def\bfS{\mbox{\boldmath$S$}}
\def\bfF{\mbox{\boldmath$F$}}
\def\bfG{\mbox{\boldmath$G$}}
\def\bfH{\mbox{\boldmath$H$}}
\def\bfZero{\mbox{\boldmath$0$}}
\def\bfP{\mbox{\boldmath$P$}}
\def\bfC{\mbox{\boldmath$C$}}
\def\bfCu{\mbox{\boldmath$C^u$}}
\def\bfV{\mbox{\boldmath$V$}}
\def\bfT{\mbox{\boldmath$T$}}
\def\bfGamma{\mbox{\boldmath$\Gamma$}}
\def\bfVs{\mbox{\boldmath$v_s$}}
\def\bfW{\mbox{\boldmath$w$}}
\def\bfSigma{\mbox{\boldmath$\sigma$}}
\def\bfEps{\mbox{\boldmath$\varepsilon$}}
\def\bfBeta{\mbox{\boldmath$\beta$}}
\def\bfPsi{\mbox{\boldmath$\psi$}}

\definecolor{mygreen}{rgb}{0.15,0.7,0.15}
\definecolor{myred}{rgb}{0.9,0.05,0.05}
\definecolor{myblue}{rgb}{0.0352,0.4981,0.6509}

\usepackage[normalem]{ulem}

\journal{Journal of Computational Physics}

\begin{document}

%\linenumbers

\graphicspath{{./figure/}}

\begin{frontmatter}

%\title{Simulation of transient poroelastic waves in transversely\\isotropic media with fractional attenuation} 
%\title{A Cartesian grid approach for wave simulation in heterogeneous transversely isotropic porous media with fractional attenuation}
\title{Wave simulation in 2D heterogeneous transversely isotropic porous media with fractional attenuation:\\ a Cartesian grid approach}

\author[IT]{Emilie Blanc}
\ead{emilie.blanc@it.uu.se}
\author[M2P2]{Guillaume Chiavassa  \corref{cor1}}
\ead{guillaume.chiavassa@centrale-marseille.fr}
\author[LMA]{Bruno Lombard}
\ead{lombard@lma.cnrs-mrs.fr}
\cortext[cor1]{Corresponding author. Tel.: +33 491 05 46 69.}
\address[IT]{Division of Scientific Computing, Department of Information Technology, Uppsala University, P.O. Box 337, SE-75105 Uppsala, Sweden}
\address[LMA]{Laboratoire de M\'{e}canique et d'Acoustique, UPR 7051 CNRS, 31 chemin Joseph Aiguier, 13402 Marseille, France}
\address[M2P2]{Centrale Marseille, M2P2, UMR 7340 - CNRS, Aix-Marseille Univ., 13451 Marseille, France}

\begin{abstract}
A time-domain numerical modeling of transversely isotropic Biot poroelastic waves is proposed in two dimensions. The viscous dissipation occurring in the pores is described using the dynamic permeability model developed by Johnson-Koplik-Dashen (JKD). Some of the coefficients in the Biot-JKD model are proportional to the square root of the frequency. In the time-domain, these coefficients introduce shifted fractional derivatives of order $1/2$, involving a convolution product. Based on a diffusive representation, the convolution kernel is replaced by a finite number of memory variables that satisfy local-in-time ordinary differential equations, resulting in the Biot-DA (diffusive approximation) model. The properties of both the Biot-JKD and the Biot-DA model are analyzed: hyperbolicity, decrease of energy, dispersion. To determine the coefficients of the diffusive approximation, two approaches are analyzed: Gaussian quadratures and optimization methods in the frequency range of interest. The nonlinear optimization is shown to be the better way of determination. A splitting strategy is then applied to approximate numerically the Biot-DA equations. The propagative part  is discretized using a fourth-order ADER scheme on a Cartesian grid, whereas the diffusive part is solved exactly. An immersed interface method is implemented to take into account heterogeneous media on a Cartesian grid and  to discretize the jump conditions at interfaces. Numerical experiments are presented. Comparisons with analytical solutions show the efficiency and the accuracy of the approach, and some numerical experiments are performed to investigate wave phenomena in complex media, such as multiple scattering across a set of random scatterers.
\end{abstract}


\end{frontmatter}

%------------------------------------------------------------------------------------------
%------------------------------------------------------------------------------------------

\section{Introduction}\label{SecIntro}

A porous medium consists of a solid matrix saturated with a fluid that circulates freely through the pores \cite{BIOT56A,BOURBIE,CARCIONE07}. Such media are involved in many applications, modeling for instance natural rocks, engineering composites \cite{GIBSON94} and biological materials \cite{COWIN89}. The most widely-used model describing the propagation of mechanical waves in porous media has been proposed by Biot in 1956 \cite{BIOT56A,BIOT56B}. It includes two classical waves (one "fast" compressional wave and one shear wave), in addition to a second "slow" compressional wave, which is highly dependent on the saturating fluid. This slow wave was observed experimentally in 1980 \cite{PLONA80}, thus confirming the validity of Biot's theory.

Two frequency regimes have to be distinguished when dealing with poroelastic waves. In the low-frequency range (LF), the flow inside the pores is of Poiseuille type \cite{BIOT56A}. The viscous effects are then proportional to the relative velocity of the motion between the fluid and the solid components. In the high-frequency range (HF), modeling the dissipation is a more delicate task. Biot first presented an expression for particular pore geometries \cite{BIOT56B}. In 1987, Johnson-Koplik-Dashen (JKD) published a general expression for the HF dissipation in the case of random pores \cite{JKD87}, where the viscous efforts depend on the square root of the frequency. No particular difficulties are raised by the HF regime if the solution is computed in the space-frequency domain \cite{Carcione96b,Santos05}. On the contrary, the computation of HF waves in the space-time domain is much more challenging. Time fractional derivatives are then introduced, involving convolution products \cite{LUBICH86}. The past of the solution must be stored, which dramatically increases the computational cost of the simulations.

The present work is proposes an efficient numerical model to simulate the transient poroelastic waves in the full frequency range of Biot's model. In the high-frequency range, only two numerical approaches have been proposed in the literature to integrate the Biot-JKD equations directly in the time-domain. The first approach consists in a straightforward discretization of the fractional derivatives defined by a convolution product in time \cite{MASSON10}. In the example given in \cite{MASSON10}, the solution is stored over $20$ time steps. The second approach is based on the diffusive representation of the fractional derivative \cite{LU05}. The convolution product is replaced by a continuum of memory variables satisfying local differential equations \cite{MATIGNON10}. This continuum is then discretized using Gaussian quadrature formulae \cite{YUAN02,DIETHELM08,BIRK10}, resulting in the Biot-DA (diffusive approximation) model. In the example proposed in \cite{LU05}, $25$ memory variables are used, which is equivalent, in terms of memory requirement, to storing $25$ time steps. The idea of using memory variables to avoid convolution products is close to the strategy commonly used in viscoelasticity \cite{EMMERICH87}.

The concern of realism leads us also to tackle with anisotropic porous media. Transverse isotropy is commonly used in practice. It is often induced by Backus averaging, which replaces isotropic layers much thinner than the wavelength by a homogeneous isotropic transverse medium \cite{GELINSKY97}. To our knowledge, the earliest numerical work combining low-frequency Biot's model and transverse isotropy is based on an operator splitting in conjonction with a Fourier pseudospectral method \cite{CARCIONE96}. Recently, a Cartesian-grid finite volume method has been developed \cite{LEVEQUE13}. One of the first work combining anistropic media and high-frequency range is proposed in \cite{HANYGA05}. However, the diffusive approximation proposed in the latter article has three  limitations. Firstly, the quadrature formulae make the convergence towards the original fractional operator very slow. Secondly, in the case of low frequencies, the Biot-DA model does not converge towards the Biot-LF model. Lastly, the number of memory variables required for a given accuracy is not specified.

The present work extends and improves our previous contributions about the modeling of poroelastic waves. In \cite{CHIAVASSA10}, we addressed 1D equations in the low-frequency range, introducing a splitting of the PDE. 2D generalizations for isotropic media required to implement space-time mesh refinement \cite{CHIAVASSA11,CHIAVASSA13}. Diffusive approximation of the fractional derivatives in the high-frequency range were introduced in \cite{BLANC_JCP13} and generalized in 2D in \cite{BLANC_JASA13}. Compared with \cite{BLANC_JASA13}, the originality of the present paper is threefold:
\begin{enumerate}
\item incorporation of anisotropy. The numerical scheme and the discretization of the interfaces need to be largely modified accordingly;
\item new procedure to determine the coefficients of the diffusive approximation. In \cite{BLANC_JCP13,BLANC_JASA13}, we used a classical least-squares optimization. It is much more accurate than the Gauss-Laguerre technique proposed in \cite{LU05}. But in counterpart, some coefficients are negative, which prevents to conclude about well-posedness of the diffusive model. Here, we fix this problem by using optimization with constraint of positivity, based on Shor's algorithm. Moreover, the accuracy of this new method  is largely improved compared with the linear optimization;
\item theoretical analysis. A new result about the eigenvalues of the diffusion matrix is introduced and the energy analysis is extended to anisotropy.
\end{enumerate}

This article is organized as follows. The original Biot-JKD model is outlined in $\S$ \ref{sec:phys} and the diffusive representation of fractional derivatives is described. The energy decrease is proven, and a dispersion analysis  is done. In $\S$ \ref{sec:DA}, an approximation of the diffusive model is presented, leading to the Biot-DA system. The properties of this system are also analyzed: energy, hyperbolicity and dispersion. Determination of the quadrature coefficients involved in the Biot-DA model are investigated in $\S$ \ref{sec:DA:coeff}. Gaussian quadrature formulae and optimization methods are successively proposed and compared, the latter being finally preferred. The numerical modeling of the Biot-DA system is addressed in $\S$ \ref{sec:num}, where the equations of evolution are split into two parts: the propagative part is discretized using a fourth-order finite-difference scheme, and the diffusive part is solved exactly. An immersed interface method is implemented to account for the jump conditions and for the geometry of the interfaces on a Cartesian grid when dealing with heterogeneous media. Numerous numerical experiments are presented in $\S$ \ref{sec:exp}, validating the method developed in this paper. In $\S$ \ref{sec:conclu}, a conclusion is drawn and some futures lines of research are suggested.  
 
%------------------------------------------------------------------------------------------
%------------------------------------------------------------------------------------------

\section{Physical modeling}\label{sec:phys}

\subsection{Biot model}\label{sec:phys:Biot}

\begin{figure}[htbp]
\begin{center}
\begin{tabular}{cc}
(a) & \hspace{0.8cm}(b)\\
\includegraphics[scale=0.4]{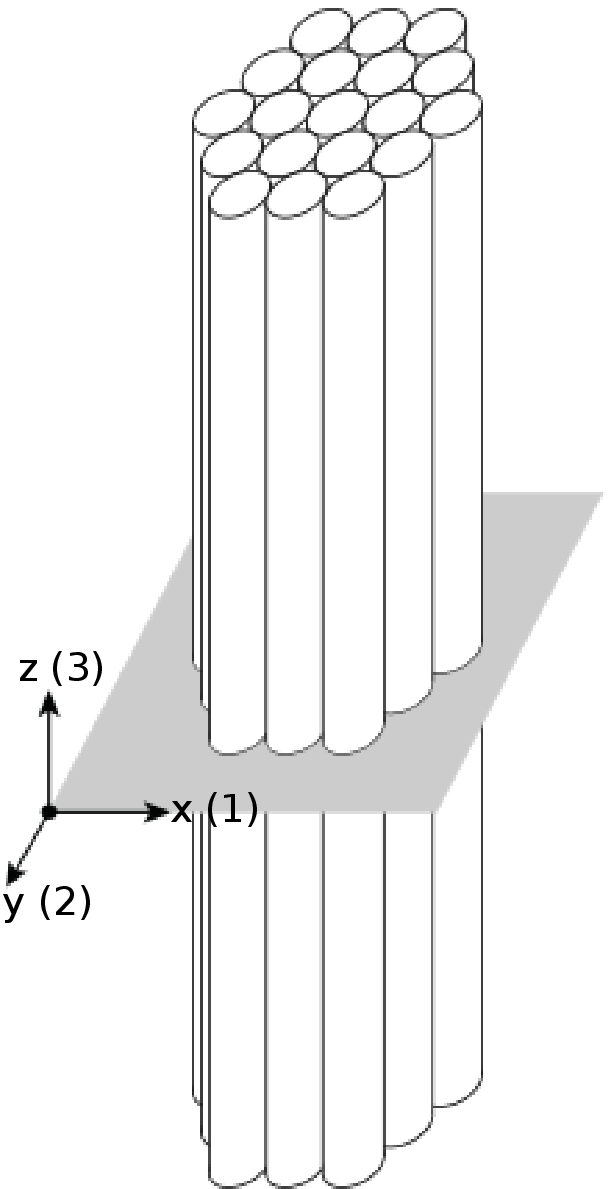} &
\hspace{0.8cm}
\includegraphics[scale=0.7]{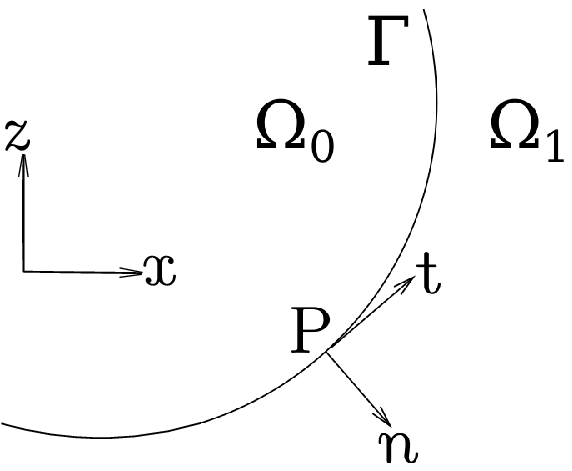} 
\end{tabular}
\end{center}
\caption{medium under study. (a): the physical properties are symmetric about the axis $z$ that is normal to the plane $(x,y)$ of isotropy. (b): interface $\Gamma$ separating two poroelastic media $\Omega_0$ and $\Omega_1$. The normal and tangential vectors at a point $P$ along $\Gamma$ are denoted by ${\bf n}$ and ${\bf t}$, respectively.}
\label{fig:interface}
\end{figure}

We consider a transversely isotropic porous medium, consisting of a solid matrix saturated with a fluid that circulates freely through the pores \cite{BIOT56A,BOURBIE,CARCIONE07}.  The subscripts $1$, $3$ represent the $x$, $z$ axes, where $z$ is the symmetry axis (figure \ref{fig:interface}). The perturbations propagate with a wavelength $\lambda$.

The  Biot model involves 15 positive physical parameters: the density $\rho_f$, the dynamic viscosity $\eta$ and the bulk modulus $K_f$ of the fluid, the density $\rho_s$ and the bulk modulus $K_s$ of the grains, the porosity $0\leqslant\phi\leqslant1$, the tortuosities ${\cal T}_1\geqslant 1$, ${\cal T}_3\geqslant 1$, the absolute permeabilities at null frequency $\kappa_1$, $\kappa_3$, and the symmetric definite positive drained elastic matrix $\bfC$
\begin{equation}
\bfC = \left( \begin{array}{cccc}
c_{11} & c_{13} & 0 & 0 \\
c_{13} & c_{33} & 0 & 0 \\
0 & 0 & c_{55} & 0 \\
0 & 0 & 0 & \displaystyle \frac{c_{11} - c_{12}}{2}
\end{array}
\right).
\label{eq:drained_elastic_matrix}
\end{equation}
The linear Biot model is valid if the following hypotheses are satisfied \cite{BIOT62}:
\begin{itemize}
\item[${\cal H}_1$]: the wavelength $\lambda$ is large in comparison with the characteristic radius of the pores $r$;
\item[${\cal H}_2$]: the amplitudes of the waves in the solid and in the fluid are small;
\item[${\cal H}_3$]: the single fluid phase is continuous;
\item[${\cal H}_4$]: the solid matrix is purely elastic;
\item[${\cal H}_5$]: the thermo-mechanical effects are neglected, which is justified when the saturating fluid is a liquid.
\end{itemize}
In the validity domain of homogenization theory (${\cal H}_1$), two frequency ranges have to be distinguished. The frontier between the low-frequency (LF) range and the high-frequency (HF) range is reached when the viscous efforts are similar to the inertial effects. The frequency transitions are given by \cite{BIOT56A}
\begin{equation}
\displaystyle f_{ci} = \frac{\eta\,\phi}{2\,\pi\,{\cal T}_i\,\kappa_i\,\rho_f}=\frac{\omega_{ci}}{2\,\pi},\quad i=1,3.
\label{eq:fc}
\end{equation}
Denoting $\mbox{\boldmath$u_s$}$ and $\mbox{\boldmath$u_f$}$ the solid and fluid displacements, the unknowns in a velocity-stress formulation are the solid velocity $\bfVs = \frac{\partial \,\mbox{\scriptsize\boldmath$u_s$}}{\partial\,t}$, the filtration velocity $\bfW = \frac{\partial\,\mbox{\scriptsize\boldmath${\cal W}$}}{\partial\,t} = \frac{\partial}{\partial\,t}\phi\,(\mbox{\boldmath$u_f$}-\mbox{\boldmath$u_s$})$, the elastic symmetric stress tensor $\underline{\bfSigma}$ and the acoustic pressure $p$. Under the hypothesis of small perturbations (${\cal H}_2$), the symmetric strain tensor $\underline{\bfEps}$ is
\begin{equation}
\underline{\bfEps} = \frac{1}{2}\,(\nabla{\mbox{\boldmath$u_s$}} + \nabla{\mbox{\boldmath$u_s$}}^T).
\label{eq:strain_tensor}
\end{equation}
Using the Voigt notation, the stress tensor and the strain tensor are arranged into vectors $\bfSigma$ and $\bfEps$
\begin{equation}
\displaystyle \bfSigma=(\sigma_{11}\,,\,\sigma_{33}\,,\,\sigma_{13})^T, \quad \displaystyle \bfEps=(\varepsilon_{11}\,,\,\varepsilon_{33}\,,\,2\,\varepsilon_{13})^T.
%\left\lbrace 
%\begin{array}{l}
%\displaystyle \bfSigma=(\sigma_{11}\,,\,\sigma_{33}\,,\,\sigma_{13})^T,\\
%[5pt]
%\displaystyle \bfEps=(\varepsilon_{11}\,,\,\varepsilon_{33}\,,\,2\,\varepsilon_{13})^T.
%\end{array}
%\right. 
\label{eq:notation_voigt}
\end{equation}
Setting
\begin{subnumcases}{\label{eq:def_Cu}}
\displaystyle \xi = -\nabla.\mbox{\boldmath${\cal W}$},\quad\bfCu = \bfC + m\,\bfBeta\,\bfBeta^T,\label{eq:def_Cu_a}\\
\displaystyle \bfBeta = (\beta_1\,,\,\beta_1\,,\,\beta_3)^T,\quad \beta_1 = 1 - \frac{c_{11} + c_{12} + c_{13}}{3\,K_s},\quad\beta_3 = 1 - \frac{2\,c_{13} + c_{33}}{3\,K_s},\label{eq:def_Cu_b}\\
\displaystyle K=K_s\,(1 + \phi\,(K_s/K_f - 1)),\quad m=\frac{K_s^2}{K-(2\,c_{11} + c_{33} + 2\,c_{12} + 4\,c_{13})/9},\label{eq:def_Cu_c}
\end{subnumcases}
where $\bfCu$ is the undrained elastic matrix and $\xi$ the rate of fluid flow, the poroelastic linear constitutive laws are \cite{CARCIONE07}
\begin{equation}
\displaystyle \bfSigma = \bfCu\,\bfEps - m\,\bfBeta\,\xi, \quad \displaystyle p = m\,\left( \xi - \bfBeta^T\,\bfEps \right).
\label{eq:biot_comportement}
\end{equation}
%\begin{subnumcases}{\label{eq:biot_comportement}}
%\displaystyle \bfSigma = \bfCu\,\bfEps - m\,\bfBeta\,\xi,\label{eq:biot_comportement_a}\\
%[5pt]
%\displaystyle p = m\,\left( \xi - \bfBeta^T\,\bfEps \right).\label{eq:biot_comportement_b}
%\end{subnumcases}
Using (\ref{eq:def_Cu_a}) and (\ref{eq:def_Cu_b}), we obtain equivalently
\begin{equation}
\bfSigma = \bfC\,\bfEps - \bfBeta\,p, \quad p = m\,\left( \xi - \bfBeta^T\,\bfEps \right).
%\left\lbrace 
%\begin{array}{l}
%\bfSigma = \bfC\,\bfEps - \bfBeta\,p,\\
%[5pt]
%\displaystyle p = m\,\left( \xi - \bfBeta^T\,\bfEps \right).
%\end{array}
%\right. 
\label{eq:biot_comportement_bis}
\end{equation}
The symmetry of $\underline{\bfSigma}$ implies compatibility conditions between spatial derivatives of the stresses and the pressure, leading to the Beltrami-Michell equation \cite{COUSSY95,RICE76}
\begin{equation}
\begin{array}{l}
\displaystyle \frac{\partial^2\sigma_{13}}{\partial\,x\,\partial\,z} = \Theta_0\,\frac{\partial^2\sigma_{11}}{\partial\,x^2} + \Theta_1\,\frac{\partial^2\sigma_{33}}{\partial\,x^2} + \Theta_2\,\frac{\partial^2p}{\partial\,x^2} + \Theta_3\,\frac{\partial^2\sigma_{11}}{\partial\,z^2} + \Theta_0\,\frac{\partial^2\sigma_{33}}{\partial\,z^2} + \Theta_4\,\frac{\partial^2p}{\partial\,z^2},\\
[15pt]
\displaystyle \Theta_0 = -c_{55}\,\frac{c_{13}}{c_{11}\,c_{33} - c_{13}^2},\quad\Theta_1 = -\frac{c_{11}}{c_{13}}\,\Theta_0,\quad\Theta_2 = \beta_1\,\Theta_0 + \beta_3\,\Theta_1,\\
[15pt]
\displaystyle \Theta_3 = -\frac{c_{33}}{c_{13}}\,\Theta_0,\quad\Theta_4 = \beta_3\,\Theta_0 + \beta_1\,\Theta_3.
\end{array}
\label{eq:beltrami_ani}
\end{equation}
If the medium is isotropic and in the elastic limit case ($\beta_1 = \beta_3 = 0$), we recover the usual equation of Barr\'e de Saint-Venant.

Introducing the densities
\begin{equation}
\rho = \phi\,\rho_f + (1-\phi)\,\rho_s,\quad \rho_{wi} = \frac{{\cal T}_i}{\phi}\,\rho_f,\quad i=1,3,
\end{equation}
the conservation of momentum yields
\begin{subnumcases}{\label{eq:biot_dynamique}}
\displaystyle \rho\,\frac{\partial\,\bfVs}{\partial\,t} + \rho_f\,\frac{\partial\,\bfW}{\partial\,t} = \nabla\,.\,\underline{\bfSigma},\label{eq:biot_dynamique_a}\\
[3pt]
\displaystyle \rho_f\,\frac{\partial\,\bfVs}{\partial\,t} + \mathrm{diag}\left(\rho_{wi}\right)\,\frac{\partial\,\bfW}{\partial\,t} + \mathrm{diag}\left(\frac{\eta}{\kappa_i}\,F_i(t)\right)*\bfW = -\nabla\,p,\label{eq:biot_dynamique_b}
\end{subnumcases}
where $\mathrm{diag}\left(d_i\right)$ denotes the $2\times 2$ diagonal matrix $({d_1 \atop 0} {0 \atop d_3})$, $*$ denotes the time convolution product and $F_i(t)$ are viscous operators. In LF, the flow in the pores is of Poiseuille type, and the dissipation efforts in (\ref{eq:biot_dynamique_b}) are given by
\begin{equation}
F_i(t) \equiv F_i^{LF}(t) = \delta(t) \Longleftrightarrow F_i^{LF}(t)*w_i(x,z,t) = w_i(x,z,t),\quad i=1,3,
\label{eq:F_lf}
\end{equation} 
where $\delta$ is the Dirac distribution, which amounts to the Darcy's law.

%------------------------------------------------------------------------------------------

\subsection{High frequency dissipation: the JKD model}\label{sec:phys:JKD}

In HF, a Prandtl boundary layer occurs at the surface of the pores, where the effects of viscosity are significant. Its width is inversely proportional to the square root of the frequency. Biot (1956) presented an expression of the dissipation process for particular pore geometries \cite{BIOT56B}. A general expression for the viscous operator for random networks of pores with constant radii has been proposed by Johnson, Koplik and Dashen (1987) \cite{JKD87}. This function is the most-simple one fitting the LF and HF limits and leading to a causal model. The only additional parameters are the viscous characteristic length $\Lambda_i$. We take \cite{MASSON10}
\begin{equation}
P_i = \frac{4\,{\cal T}_i\,\kappa_i}{\phi\,\Lambda_i^2},\quad\Omega_i = \frac{\omega_{ci}}{P_i} = \frac{\eta\,\phi^2\,\Lambda_i^2}{4\,{\cal T}_i^2\,\kappa_i^2\,\rho_f},\quad i=1,3,
\label{eq:coef_hf}
\end{equation}
where $P_i$ is the Pride number. The Pride number describes the geometry of the pores: $P_i=1/2$ corresponds to a set of non-intersecting canted tubes, whereas $P_i=1/3$ describes a set of canted slabs of fluids \cite{CARCIONE07}. Based on the Fourier transform in time, $\widehat{F}_i(\omega) = {\cal F}\left( F_i(t) \right) = \int _{\mathbb R} F_i(t)e^{-j\omega t}\,dt$, the viscous operators given by the JKD model are \cite{JKD87}
\begin{equation}
%\begin{array}{ll}
\widehat{F}_i^{JKD}(\omega)  \displaystyle = \left( 1+j\,\omega\,\frac{4\,{\cal T}_i^2\,\kappa_i^2\,\rho_f}{\eta\,\Lambda_i^2\,\phi^2}\right)^{1/2} = \left( 1+j\,P_i\,\frac{\omega}{\omega_{ci}}\right)^{1/2}=  \frac{1}{\sqrt{\Omega_i}}\,(\Omega_i +j\,\omega)^{1/2}.\\
%[15pt]
%& \displaystyle = \left( 1+j\,P_i\,\frac{\omega}{\omega_{ci}}\right)^{1/2},\\
%[12pt]
%& \displaystyle = \frac{1}{\sqrt{\Omega_i}}\,(\Omega_i +j\,\omega)^{1/2}.
%\end{array}
\label{eq:F_omega}
\end{equation}
Therefore, the terms $F_i(t)*w_i(x,z,t)$ involved in (\ref{eq:biot_dynamique_b}) are
\begin{equation}
\begin{array}{ll}
F_i^{JKD}(t)*w_i(x,z,t) & \displaystyle = {\cal F}^{-1}\left( \frac{1}{\sqrt{\Omega_i}}\,(\Omega_i + j\,\omega)^{1/2}\widehat{w}_i(x,z,\omega)\right),\\
[13pt]
& \displaystyle = \frac{1}{\sqrt{\Omega_i}}\,(D+\Omega_i)^{1/2}w_i(x,z,t).
\end{array}
\label{eq:F_t}
\end{equation}
In the last relation of (\ref{eq:F_t}), $(D+\Omega_i)^{1/2}$ is an operator. $D^{1/2}$ is a fractional derivative in time of order $1/2$, generalizing the usual derivative characterized by $\frac{\partial\,w_i}{\partial\,t} = {\cal F}^{-1}\left( j\,\omega\,\widehat{w}_i\right)$. The notation $\left(D+\Omega_i\right) ^{1/2}$ accounts for the shift $\Omega_i$ in (\ref{eq:F_t}).

%------------------------------------------------------------------------------------------

\subsection{The Biot-JKD equations of evolution}\label{sec:phys:EDP}

The system (\ref{eq:biot_dynamique}) is rearranged by separating $\frac{\partial\,\mbox{\scriptsize\boldmath$v_s$}}{\partial\,t}$ and $\frac{\partial\,\mbox{\scriptsize\boldmath$w$}}{\partial\,t}$ and using the definitions of $\bfEps$ and $\xi$. Taking
\begin{equation}
\gamma_i = \frac{\eta}{\kappa_i}\,\frac{\rho}{\chi_i}\,\frac{1}{\sqrt{\Omega_i}},\quad i=1,3,
\end{equation}
one obtains the following system of evolution equations
\begin{subnumcases}{\label{eq:2D_ani_syst_hyp_jkd_scalar}}
\displaystyle \frac{\partial\,v_{s1}}{\partial\,t} - \frac{\rho_{w1}}{\chi_1}\,\left( \frac{\partial\,\sigma_{11}}{\partial\,x}+\frac{\partial\,\sigma_{13}}{\partial\,z}\right)  - \frac{\rho_f}{\chi_1}\,\frac{\partial\,p}{\partial\,x} = \frac{\rho_f}{\rho}\,\gamma_1\,(D+\Omega_1)^{1/2}\,w_1 + G_{v_{s1}},\label{eq:2D_ani_syst_hyp_jkd_scalar_a}\\
[5pt]
\displaystyle \frac{\partial\,v_{s3}}{\partial\,t} - \frac{\rho_{w3}}{\chi_3}\,\left( \frac{\partial\,\sigma_{13}}{\partial\,x}+\frac{\partial\,\sigma_{33}}{\partial\,z}\right)  - \frac{\rho_f}{\chi_3}\,\frac{\partial\,p}{\partial\,z} = \frac{\rho_f}{\rho}\,\gamma_3\,(D+\Omega_3)^{1/2}\,w_3 + G_{v_{s3}},\label{eq:2D_ani_syst_hyp_jkd_scalar_b}\\
[5pt]
\displaystyle \frac{\partial\,w_1}{\partial\,t} + \frac{\rho_f}{\chi_1}\,\left( \frac{\partial\,\sigma_{11}}{\partial\,x}+\frac{\partial\,\sigma_{13}}{\partial\,z}\right) + \frac{\rho}{\chi_1}\,\frac{\partial\,p}{\partial\,x} = -\gamma_1\,(D+\Omega_1)^{1/2}\,w_1 + G_{w_1},\label{eq:2D_ani_syst_hyp_jkd_scalar_c}\\
[5pt]
\displaystyle \frac{\partial\,w_3}{\partial\,t} + \frac{\rho_f}{\chi_3}\,\left( \frac{\partial\,\sigma_{13}}{\partial\,x}+\frac{\partial\,\sigma_{33}}{\partial\,z}\right) + \frac{\rho}{\chi_3}\,\frac{\partial\,p}{\partial\,z} = -\gamma_3\,(D+\Omega_3)^{1/2}\,w_3 + G_{w_3},\label{eq:2D_ani_syst_hyp_jkd_scalar_d}\\
[5pt]
\displaystyle \frac{\partial\,\sigma_{11}}{\partial\,t} - c_{11}^u\,\frac{\partial\,v_{s1}}{\partial\,x} -c_{13}^u\,\frac{\partial\,v_{s3}}{\partial\,z} - m\,\beta_1\,\left( \frac{\partial\,w_1}{\partial\,x} + \frac{\partial\,w_3}{\partial\,z}\right) = G_{\sigma_{11}},\label{eq:2D_ani_syst_hyp_jkd_scalar_e}\\
[5pt]
\displaystyle \frac{\partial\,\sigma_{13}}{\partial\,t}-c_{55}^u\,\left( \frac{\partial\,v_{s3}}{\partial\,x}+\frac{\partial\,v_{s1}}{\partial\,z}\right) = G_{\sigma_{13}},\label{eq:2D_ani_syst_hyp_jkd_scalar_f}\\
[5pt]
\displaystyle \frac{\partial\,\sigma_{33}}{\partial\,t}-c_{13}^u\,\frac{\partial\,v_{s1}}{\partial\,x}- c_{33}^u\,\frac{\partial\,v_{s3}}{\partial\,z} - m\,\beta_3\,\left( \frac{\partial\,w_1}{\partial\,x} + \frac{\partial\,w_3}{\partial\,z}\right) = G_{\sigma_{33}},\label{eq:2D_ani_syst_hyp_jkd_scalar_g}\\
[5pt]
\displaystyle \frac{\partial\,p}{\partial\,t} + m\, \left( \beta_1\,\frac{\partial\,v_{s1}}{\partial\,x}+\beta_3\,\frac{\partial\,v_{s3}}{\partial\,z} +\frac{\partial\,w_1}{\partial\,x} +\frac{\partial\,w_3}{\partial\,z} \right) = G_p.\label{eq:2D_ani_syst_hyp_jkd_scalar_h}
\end{subnumcases}
The source terms $G_{v_{s1}}$, $G_{v_{s3}}$, $G_{w_1}$, $G_{w_3}$, $G_{\sigma_{11}}$, $G_{\sigma_{13}}$, $G_{\sigma_{33}}$ and $G_p$ have been introduced to model the forcing.

%------------------------------------------------------------------------------------------

\subsection{The diffusive representation}\label{sec:phys:DR}

The shifted fractional derivatives in (\ref{eq:F_t}) can be written \cite{HANYGA01}
\begin{equation}
(D+\Omega_i )^{1/2}w_i(x,z,t) = \int _0 ^t \frac{e^{-\Omega_i (t-\tau)}}{\sqrt{\pi\,(t-\tau)}}\,\left(\frac{\partial \,w_i}{\partial \,t}(x,z,\tau)+\Omega_i\,w_i(x,z,\tau)\,\right)d\tau,\hspace{0.5cm}i=1,3.
\label{eq:Dfrac}
\end{equation}
The operators $(D+\Omega_i)^{1/2}$ are not local in time and involve the entire time history of $\bfW$. Based on Euler's Gamma function, the diffusive representation of the totally monotone function $\frac{1}{\sqrt{\pi\,t}}$ is \cite{MATIGNON10}
\begin{equation}
\displaystyle \frac{1}{\sqrt{\pi\,t}} = \frac{1}{\pi}\,\int _0 ^\infty\,\frac{1}{\sqrt{\theta}}\,e^{-\theta t}d\theta .
\label{eq:fonction_diffu}
\end{equation}
Substituting (\ref{eq:fonction_diffu}) into (\ref{eq:Dfrac}) gives
\begin{equation}
(D+\Omega_i)^{1/2}w_i(x,z,t) = \frac{1}{\pi}\, \int _0 ^{\infty}\frac{1}{\sqrt{\theta}}\,\psi_{i}(x,z,\theta,t)\,d\theta,
\label{eq:derivee_frac}
\end{equation}
where the memory variables are defined as
\begin{equation}
\psi_{i}(x,z,\theta,t)=\int_0^t e^{-(\theta + \Omega_i)(t-\tau)}\,\left(\frac{\partial \,w_i}{\partial \,t}(x,z,\tau)+\Omega_i\,w_i(x,z,\tau)\,\right)\,d\tau.
\label{eq:variable_diffu}
\end{equation}
For the sake of clarity, the dependence on $\Omega_i$ and $w_i$ are omitted in $\psi_{i}$. From (\ref{eq:variable_diffu}), it follows that the two memory variables $\psi_{i}$ satisfy the ordinary differential equation
\begin{subnumcases}{\label{eq:EDO_psi}}
\displaystyle \frac{\partial\,\psi_{i}}{\partial\,t} = -(\theta + \Omega_i)\,\psi_{i} + \frac{\partial \,w_i}{\partial \,t}+\Omega_i\,w_i,\label{eq:EDO_psi_a} \\
[5pt]
\displaystyle \psi_{i}(x,z,\theta,0) = 0.\label{eq:EDO_psi_b}
\end{subnumcases}
The diffusive representation therefore transforms a non-local problem (\ref{eq:Dfrac}) into a continuum of local problems (\ref{eq:derivee_frac}). It should be emphasized at this point that no approximation have been made up to now. The computational advantages of the diffusive representation will be seen in $\S$ \ref{sec:DA} and \ref{sec:exp}, where the discretization of (\ref{eq:derivee_frac}) and (\ref{eq:EDO_psi_a}) will yield a numerically tractable formulation.

%------------------------------------------------------------------------------------------

\subsection{Energy of Biot-JKD}\label{sec:phys:NRJ}

Now, we express the energy of the Biot-JKD model (\ref{eq:2D_ani_syst_hyp_jkd_scalar}).
\begin{proposition}[\bf Decrease of the energy]
Let us consider the Biot-JKD model (\ref{eq:2D_ani_syst_hyp_jkd_scalar}) without forcing, and let us denote 
\begin{equation}
E=E_1+E_2+E_3,
\label{eq:E1E2E3_JKD}
\end{equation}
with
\begin{equation}
\begin{array}{l}
\displaystyle E_1 = \frac{1}{2}\,\int_{\mathbb{R}^2}\left(\rho\,\bfVs^T\,\bfVs + 2\,\rho_f\,\bfVs^T\,\bfW + \bfW^T\,\mathrm{diag}\left(\rho_{wi}\right)\,\bfW\right)\,dx\,dz,\\
[15pt]
\displaystyle E_2 = \frac{1}{2}\,\int_{\mathbb{R}^2}\left(\left( \bfSigma + p\,\bfBeta \right)^T\,\bfC^{-1}\,\left( \bfSigma + p\,\bfBeta \right) + \frac{1}{m}\,p^2\right)\,dx\,dz,\\
[15pt]
\displaystyle E_3 = \frac{1}{2}\,\int_{\mathbb{R}^2}\frac{\eta}{\pi}\,\int_0^{\infty}(\bfW-\bfPsi)^T\,\mathrm{diag}\left(\frac{1}{\kappa_i\,\sqrt{\Omega_i\,\theta}\,(\theta+2\,\Omega_i)}\right)\,(\bfW-\bfPsi)\,d\theta\,dx\,dz.
\end{array}
\label{eq:energieJKD}
\end{equation}
Then, $E$ is an energy which satisfies
\begin{equation}
\begin{array}{l}
\displaystyle \frac{\textstyle d\,E}{\textstyle d\,t} = -\int_{\mathbb{R}^2}\frac{\eta}{\pi}\,\int_0^{\infty} \displaystyle \left\lbrace \bfPsi^T\,\mathrm{diag}\left(\frac{\theta+\Omega_i}{\kappa_i\,\sqrt{\Omega_i\,\theta}\,(\theta+2\,\Omega_i)}\right)\,\bfPsi\right.\\
[15pt]
\hspace{1.1cm}
\left. \displaystyle + \bfW^T\,\mathrm{diag}\left(\frac{\Omega_i}{\kappa_i\,\sqrt{\Omega_i\,\theta}\,(\theta+2\,\Omega_i)}\right)\,\bfW \right\rbrace \,d\theta\,dx\,dz \;\leqslant\; 0 .
\end{array}
\label{eq:dEdtJKD}
\end{equation}
\label{prop:nrjJKD}
\end{proposition}
Proposition \ref{prop:nrjJKD} is proven in \ref{annexe:proof_nrj}. It calls for the following comments:
\begin{itemize}
\item the Biot-JKD model is stable;
\item when the viscosity of the saturating fluid is neglected ($\eta=0$), the energy of the system is conserved;
\item the terms $E_1$ and $E_2$ in (\ref{eq:energieJKD}) have a clear physical significance: $E_1$ is the kinetic energy, and $E_2$ is the strain energy;
\item the energy analysis is valid for continuously variable parameters.
\end{itemize}

%------------------------------------------------------------------------------------------

\subsection{Dispersion analysis}\label{sec:phys:dispersion}

In this section, we derive the dispersion relation of the waves which propagate in a poroelastic medium. This relation describes the frequency dependence of phase velocities and attenuations of waves. For this purpose, we search for a general plane wave solution of (\ref{eq:2D_ani_syst_hyp_jkd_scalar})
\begin{equation}
\left\lbrace 
\begin{array}{l}
\bfV=(v_1\,,\,v_3\,,\,w_1\,,\,w_3)^T = \mbox{\boldmath$V_0$}\,e^{j(\omega t-\mbox{\scriptsize\boldmath$k$}.\mbox{\scriptsize\boldmath$r$})},\\
[5pt]
\bfT=(\sigma_{11}\,,\,\sigma_{13}\,,\,\sigma_{33}\,,\,-p)^T = \mbox{\boldmath$T_0$}\,e^{j(\omega t-\mbox{\scriptsize\boldmath$k$}.\mbox{\scriptsize\boldmath$r$})},
\end{array}
\right. 
\label{eq:plane_wave}
\end{equation}
where $\mbox{\boldmath$k$} = k\,(\cos(\varphi),\;\sin(\varphi))^T$ is the wavevector, $k$ is the wavenumber, $\mbox{\boldmath$V_0$}$ and $\mbox{\boldmath$T_0$}$ are the polarizations, $\mbox{\boldmath$r$} = (x,\;z)^T$ is the position, $\omega = 2\,\pi\,f$ is the angular frequency and $f$ is the frequency. By substituting equation (\ref{eq:plane_wave}) into equations (\ref{eq:2D_ani_syst_hyp_jkd_scalar_e})-(\ref{eq:2D_ani_syst_hyp_jkd_scalar_h}), we obtain the $4\times 4$ linear system:
\begin{equation}
\begin{array}{ccc}
\omega\,\bfT = -k & \underbrace{\left( \begin{array}{cccc}
c_{11}^u\,c_{\varphi} & c_{13}^u\,s_{\varphi} & \beta_1\,m\,c_{\varphi} & \beta_1\,m\,s_{\varphi}\\
[5pt]
c_{55}^u\,s_{\varphi} & c_{55}^u\,c_{\varphi} & 0 & 0\\
[5pt]
c_{13}^u\,c_{\varphi} & c_{33}^u\,s_{\varphi} & \beta_3\,m\,c_{\varphi} & \beta_3\,m\,s_{\varphi}\\
[5pt]
\beta_1\,m\,c_{\varphi} & \beta_3\,m\,s_{\varphi} & m\,c_{\varphi} & m\,s_{\varphi}
\end{array}\right)}  & \bfV,\\
 & \mbox{\boldmath${\cal C}$} & 
\end{array}
\label{eq:dispersion_matC}
\end{equation}
where $c_{\varphi} = \cos(\varphi)$ and $s_{\varphi} = \sin(\varphi)$. Then, substituting (\ref{eq:plane_wave}) into (\ref{eq:2D_ani_syst_hyp_jkd_scalar_a})-(\ref{eq:2D_ani_syst_hyp_jkd_scalar_d}) gives another $4\times 4$ linear system:
\begin{equation}
\begin{array}{ccccc}
-k & \underbrace{\left( \begin{array}{cccc}
c_{\varphi} & s_{\varphi} & 0 & 0\\
[5pt]
0 & c_{\varphi} & s_{\varphi} & 0\\
[5pt]
0 & 0 & 0 & c_{\varphi}\\
[5pt]
0 & 0 & 0 & s_{\varphi}
\end{array}\right) } & \bfT = \omega & \underbrace{\left( \begin{array}{cccc}
\rho & 0 & \rho_f & 0\\
[5pt]
0 & \rho & 0 & \rho_f\\
[5pt]
\rho_f & 0 & \displaystyle \frac{\widehat{Y}_1^{JKD}(\omega)}{j\,\omega} & 0\\
[5pt]
0 & \rho_f & 0 & \displaystyle \frac{\widehat{Y}_3^{JKD}(\omega)}{j\,\omega}
\end{array}\right) } & \bfV,\\
& \mbox{\boldmath${\cal L}$} & & \bfGamma & 
\end{array}
\label{eq:dispersion_matGammaL}
\end{equation}
where $\widehat{Y}_1^{JKD}$ and $\widehat{Y}_3^{JKD}$ are the viscodynamic operators \cite{NORRIS86}:
\begin{equation}
\widehat{Y}_i^{JKD} = j\,\omega\,\rho_{wi} + \frac{\eta}{\kappa_i}\,\widehat{F}_i^{JKD}(\omega),\quad i=1,3.
\label{eq:operateur_visco_jkd}
\end{equation}
Since the matrix $\bfGamma$ is invertible, the equations (\ref{eq:dispersion_matC}) and (\ref{eq:dispersion_matGammaL}) lead to the eigenproblem
\begin{equation}
\bfGamma^{-1}\,\mbox{\boldmath${\cal L}$}\,\mbox{\boldmath${\cal C}$}\,\bfV =  \left( \frac{\omega}{k} \right)^2\,\bfV.
\label{eq:syst_dispersion_gen}
\end{equation}
The equation (\ref{eq:syst_dispersion_gen}) is solved numerically, leading to two quasi-compressional waves denoted $qP_f$ (fast) and $qP_s$ (slow), and to one quasi-shear wave denoted $qS$ \cite{CARCIONE96}. The wavenumbers thus obtained depend on the frequency and on the angle $\varphi$. One of the eigenvalues is zero with multiplicity two, and the other non-zero eigenvalues correspond to the wave modes $\pm k_{pf}(\omega,\varphi)$, $\pm k_{ps}(\omega,\varphi)$ and $\pm k_{s}(\omega,\varphi)$. Therefore three waves propagates symmetrically along the directions $\cos(\varphi)\,x + \sin(\varphi)\,z$ and $-\cos(\varphi)\,x - \sin(\varphi)\,z$.

The wavenumbers give the phase velocities $c_{pf}(\omega,\varphi) = \omega/\Re\mbox{e}(k_{pf})$, $c_{ps}(\omega,\varphi) = \omega/\Re\mbox{e}(k_{ps})$, and $c_{s}(\omega,\varphi) = \omega/\Re\mbox{e}(k_{s})$, with $0<c_{ps}<c_{pf}$ and $0<c_{s}$. The attenuations $\alpha_{pf}(\omega,\varphi) = -\Im\mbox{m}(k_{pf})$, $\alpha_{ps}(\omega,\varphi) = -\Im\mbox{m}(k_{ps})$ and $\alpha_{s}(\omega,\varphi) = -\Im\mbox{m}(k_{s})$ are also deduced. Both the phase velocities and the attenuations of Biot-LF and Biot-JKD are strictly increasing functions of the frequency. The high-frequency limits ($\omega \rightarrow \infty$ in (\ref{eq:syst_dispersion_gen})) of phase velocities $c_{pf}^{\infty}(\varphi)$, $c_{ps}^{\infty}(\varphi)$ and $c_s^{\infty}(\varphi)$ are obtained by diagonalizing the left-hand side of (\ref{eq:2D_ani_syst_hyp_jkd_scalar}).

\begin{figure}[htbp]
\begin{center}
\begin{tabular}{cc}
phase velocity of the $P_f$ wave & attenuation  of the $P_f$ wave\\
\includegraphics[scale=0.34]{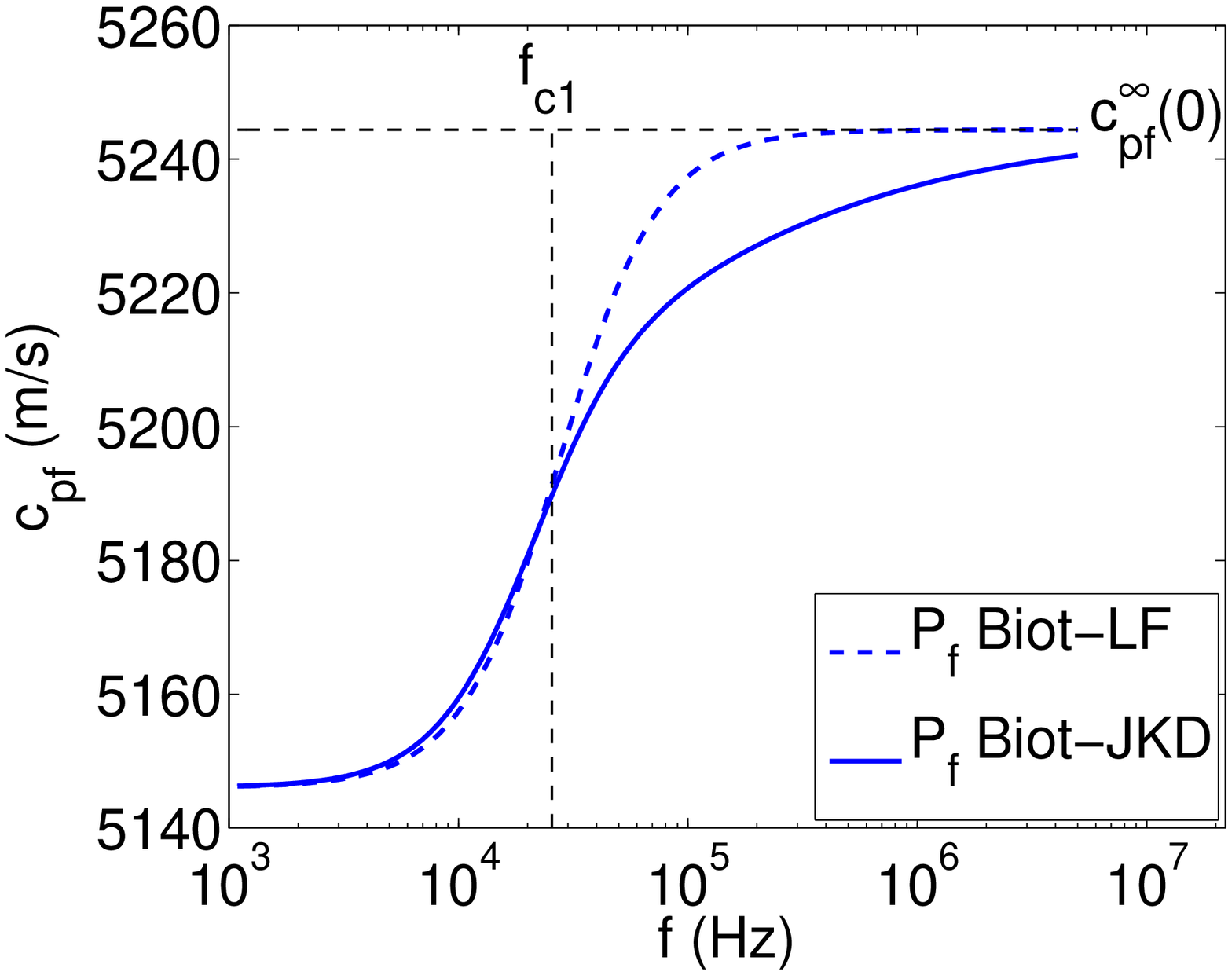} &
\includegraphics[scale=0.34]{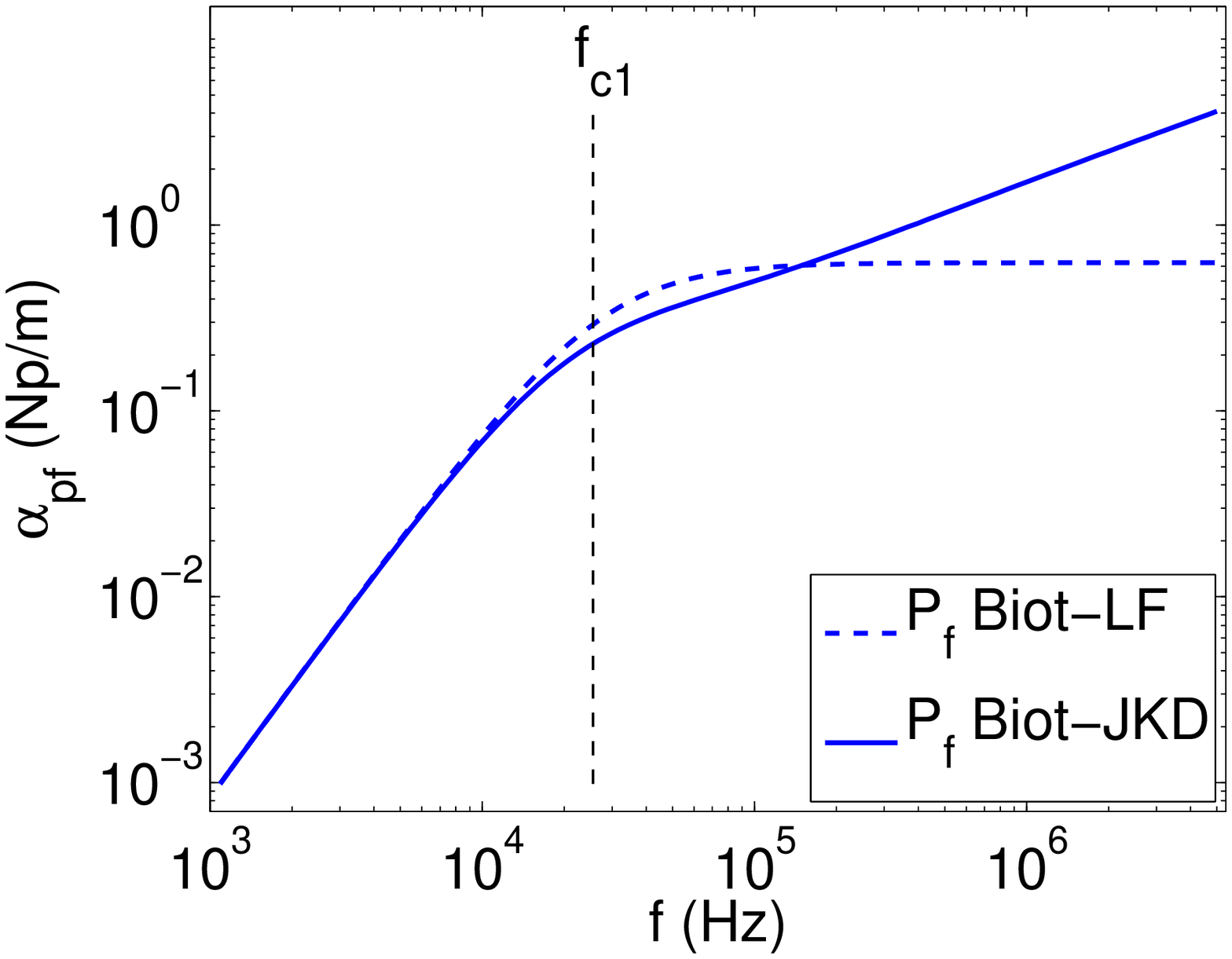}\\
[10pt]
phase velocity of the $S$ wave & attenuation  of the $S$ wave\\
\includegraphics[scale=0.34]{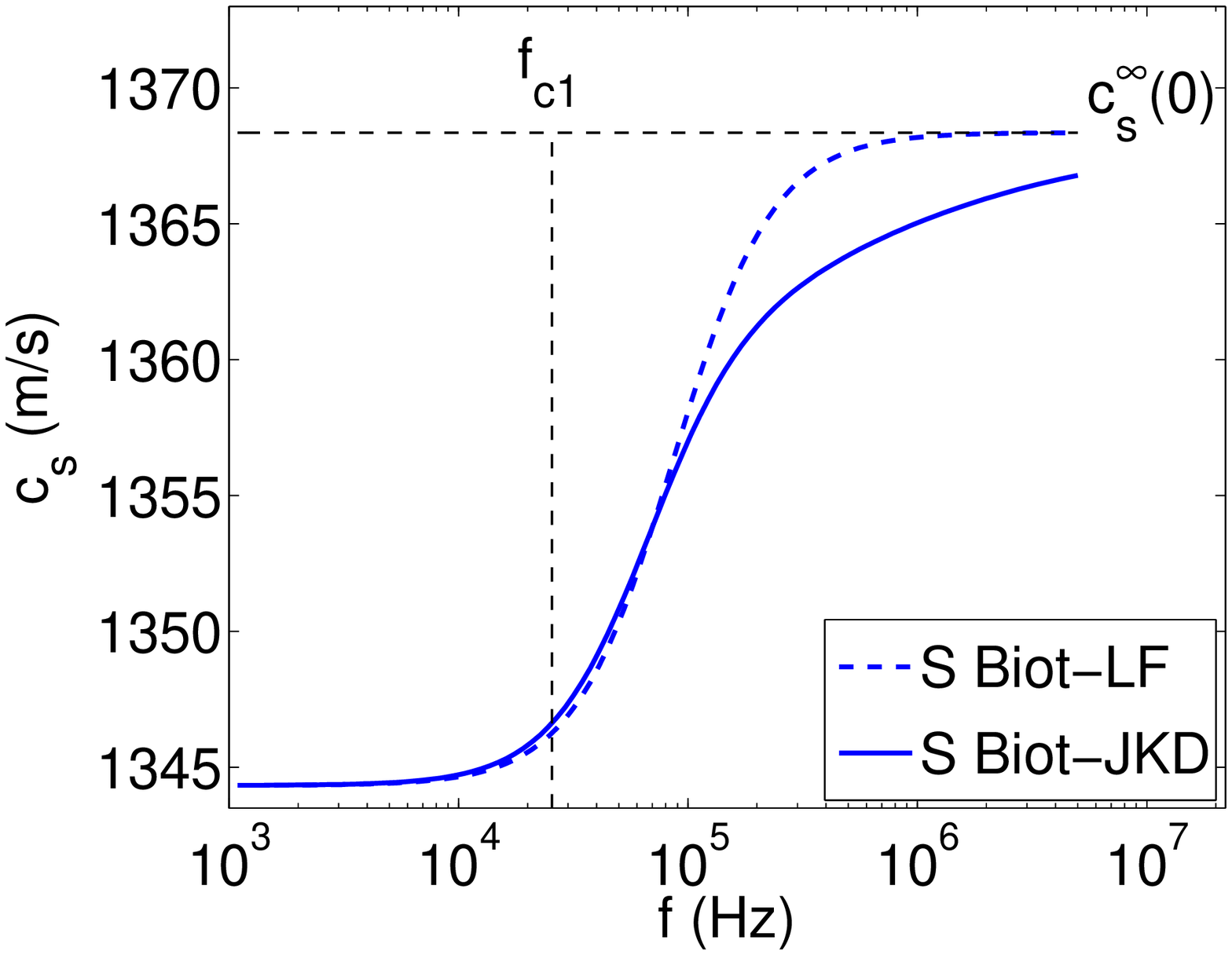} &
\includegraphics[scale=0.34]{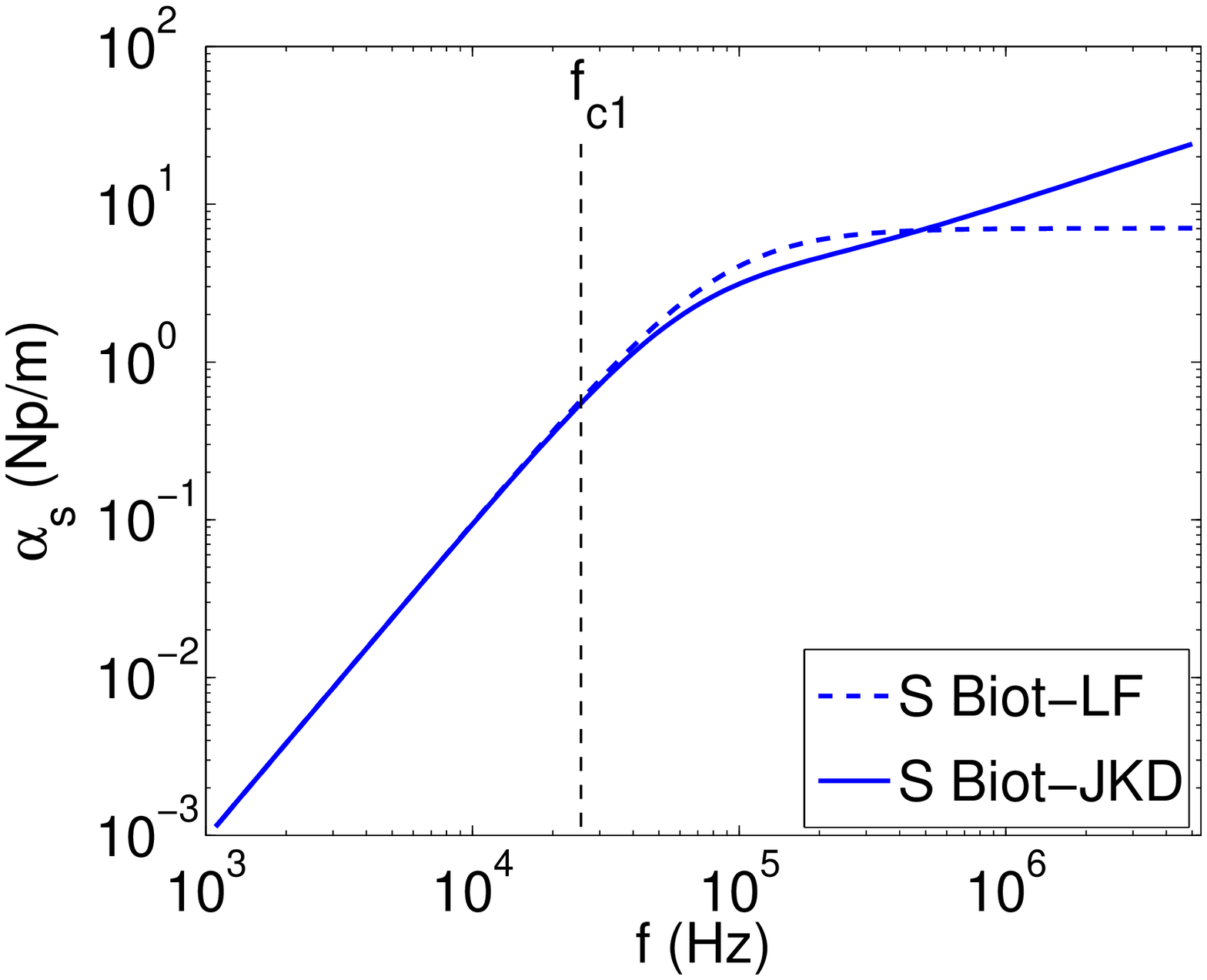}\\
[10pt]
phase velocity of the $P_s$ wave & attenuation  of the $P_s$ wave\\
\includegraphics[scale=0.34]{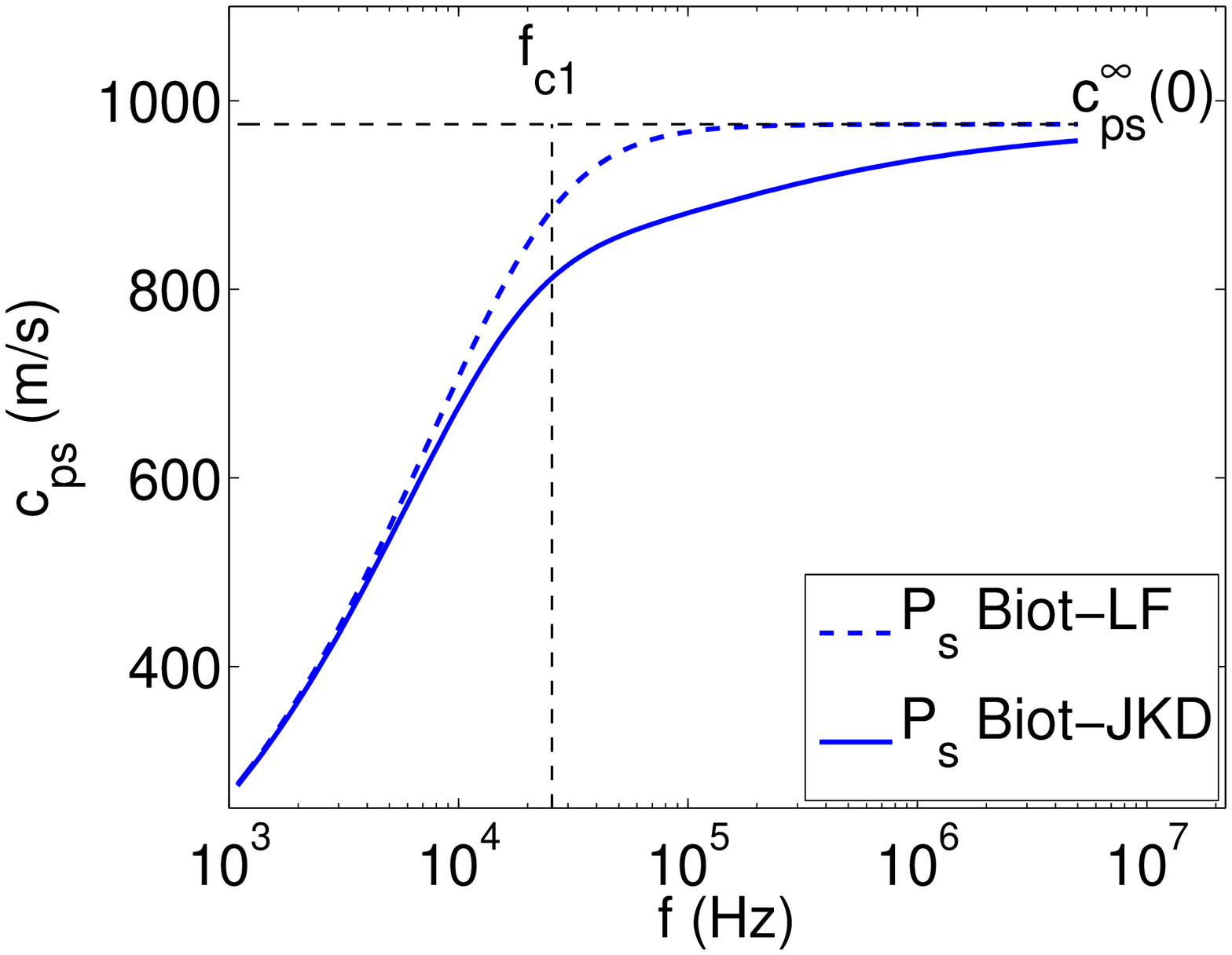} &
\includegraphics[scale=0.34]{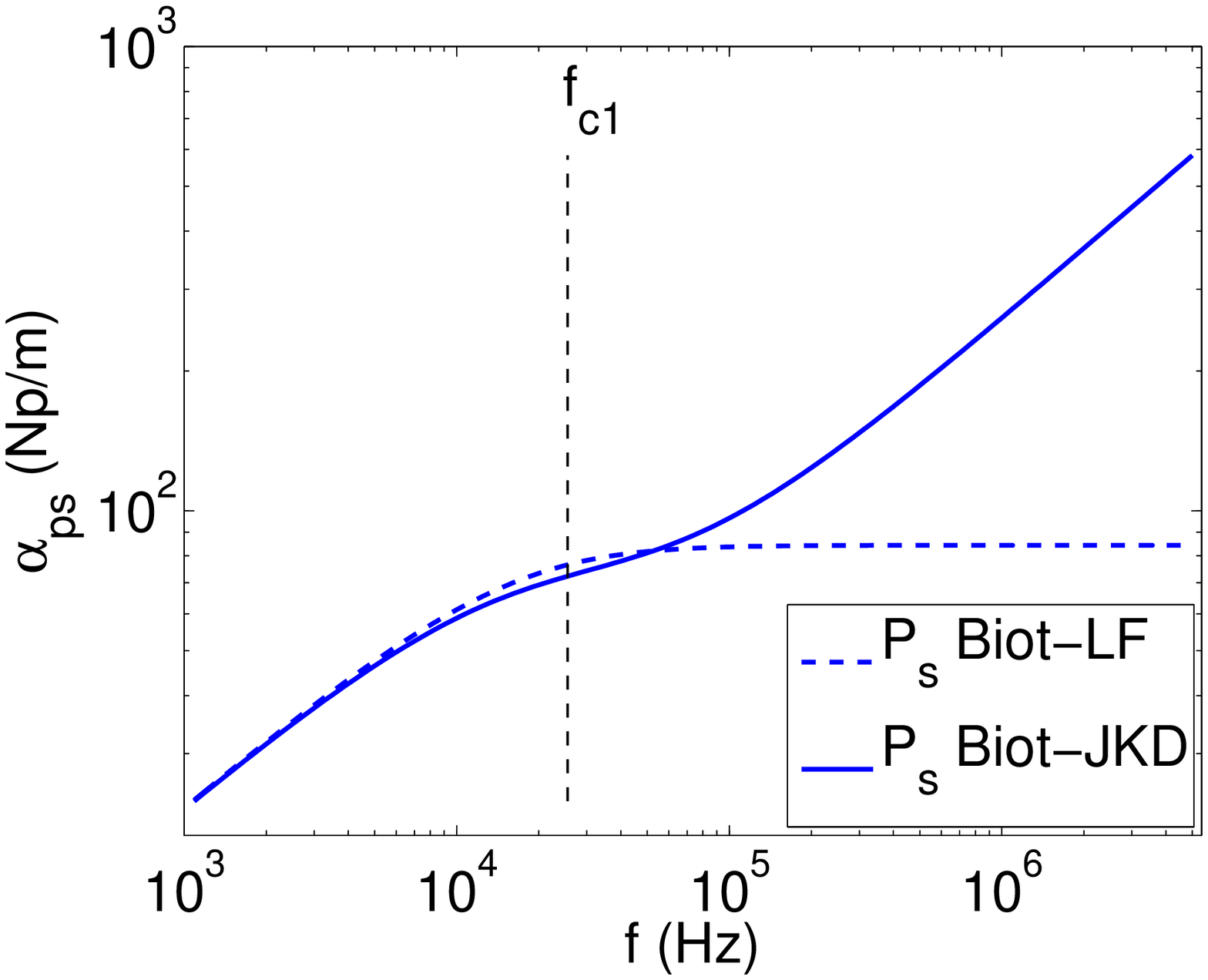}
\end{tabular}
\end{center}
\caption{dispersion curves in terms of the frequency. Comparison between Biot-LF and Biot-JKD models at $\varphi = 0$ rad. The vertical dotted line denotes the critical frequency separating low-frequency and high frequency regimes. The horizontal dotted lines in the left row denote the maximal phase velocity at infinite frequency.}
\label{fig:dispersion_bf_freq}
\end{figure}

Various authors have illustrated the effect of the JKD correction on the phase velocity and on the attenuation \cite{MASSON06}. In figure \ref{fig:dispersion_bf_freq}, the physical parameters are those of medium $\Omega_0$ (cf table \ref{table:para_phy_ani}), where the frequencies of transition are $f_{c1} = 25.5$ kHz, $f_{c3} = 85$ kHz. The dispersion curves are shown in terms of the frequency at $\varphi = 0$ rad. The high-frequency limit of the phase velocities of the quasi-compressional waves are $c_{pf}^{\infty}(0) = 5244$ m/s and $c_{ps}^{\infty}(0) = 975$ m/s, which justifies the denomination "fast" and "slow".
%\begin{figure}[htbp]
%\begin{center}
%\begin{tabular}{cc}
%phase velocity& attenuation\\
%\includegraphics[scale=0.32]{figure/vitesse_phase.eps} &
%\includegraphics[scale=0.32]{figure/attenuation.eps}\\
%\end{tabular}
%\end{center}
%\caption{dispersion curves in terms of the frequency. Comparison between Biot-LF and Biot-JKD models at $\varphi = 0$ rad.}
%\label{fig:dispersion_bf_freq}
%\end{figure}
Figure \ref{fig:dispersion_bf_freq} calls for the following comments \cite{BOURBIE}:
\begin{itemize}
\item when $f<f_{ci}$, the Biot-JKD and Biot-LF dispersion curves are very similar as might be expected, since $\widehat{F}^{JKD}_i(0) = \widehat{F}^{LF}_i(0) = 1$;
\item the frequency evolution of the phase velocity and of the attenuation is radically different for the three waves, whatever the chosen model (LF or JKD): the effect of viscous losses is negligible on the fast wave, small on the shear wave, whereas it is very important on the slow wave;
\item when $f\ll f_{ci}$, the slow compressional wave is almost static \cite{CHANDLER81,RICE76}. When $f>f_{ci}$, the slow wave propagates but is greatly attenuated.
\end{itemize}

Taking
\begin{equation}
\bfU_1 = \left( \begin{array}{cccc}
1 & 0 & 0 & 0\\
[5pt]
0 & 0 & 1 & 0\\
[5pt]
0 & 0 & 0 & 1\\
[5pt]
0 & 0 & 0 & 0
\end{array}\right) ,\quad \bfU_3 = \left( \begin{array}{cccc}
0 & 0 & 1 & 0\\
[5pt]
0 & 1 & 0 & 0\\
[5pt]
0 & 0 & 0 & 0\\
[5pt]
0 & 0 & 0 & 1
\end{array}\right) ,
\label{eq:matrix_energy_velocity}
\end{equation} 
the energy velocity vector $\bfV_e$ is \cite{CARCIONE96,CARCIONE93}:
\begin{equation}
\left\lbrace 
\begin{array}{l}
\displaystyle \bfV_e = \frac{\left\langle \bfP \right\rangle}{\left\langle E_s + E_k \right\rangle} = \frac{\left\langle \bfP \right\rangle}{\left\langle E \right\rangle},\\
[10pt]
\displaystyle \left\langle \bfP \right\rangle = -\frac{1}{2}\,\Re\mbox{e}\left( 
\left( 
\overrightarrow{e_x}\,(\bfU_1.\bfT)^T + \overrightarrow{e_z}\,(\bfU_3.\bfT)^T
\right).\overline{\bfV}
\right),\\
[10pt]
\displaystyle \left\langle E \right\rangle = \frac{1}{4}\,\Re\mbox{e}\left( \left( 1 + \frac{(\omega/k)^2}{\left| \omega/k \right| ^2} \right) \bfV^T\,\bfGamma\,\overline{\bfV} \right) ,
\end{array}
\right. 
\label{eq:energy_velocity}
\end{equation}
where $\overline{\bfV}$ is the complex conjugate of $\bfV$, $\left\langle \bfP \right\rangle$ is the Umov-Poynting vector, $\left\langle E_k \right\rangle$ and $\left\langle E_s \right\rangle$ are the average kinetic and strain energy densities, and $\left\langle E \right\rangle$ is the mean energy density. The theoretical wavefronts are the locus of the end of energy velocity vector $\bfV_e$ multiplied by the time of propagation.  We will use this property in $\S$ \ref{sec:exp} to validate the simulations.

%------------------------------------------------------------------------------------------
%------------------------------------------------------------------------------------------

\section{The Biot-DA (diffusive approximation) model}\label{sec:DA}

The aim of this section is to approximate the Biot-JKD model, using a numerically tractable approach. 

\subsection{Diffusive approximation}\label{sec:DA:DA}

The diffusive representation of fractional derivatives (\ref{eq:derivee_frac}) is approximated by using a quadrature formula on $N$ points, with weights $a_{\ell}^i$ and abcissae $\theta_{\ell}^i$ ($i=1,3$):
\begin{equation}
\begin{array}{ll}
(D+\Omega_i)^{1/2}w_i(x,z,t) & \displaystyle = \frac{1}{\pi}\,\int_0^{\infty}\frac{1}{\sqrt{\theta}}\psi^i(x,z,\theta,t)\,d\theta
\simeq \sum\limits_{\ell=1}^N a_{\ell}^i\,\psi^i(x,z,\theta_{\ell}^i,t),\\
[10pt]
%& \displaystyle \simeq \sum\limits_{\ell=1}^N a_{\ell}^i\,\psi^i(x,z,\theta_{\ell}^i,t),\\
%[15pt]
& \displaystyle \equiv \sum_{\ell=1}^N a_{\ell}^i\,\psi_{\ell}^i(x,z,t).
\end{array}
\label{eq:DA}
\end{equation}
From (\ref{eq:EDO_psi_a}), the $2\,N$ memory variables $\psi_{\ell}^i$ satisfy the ordinary differential equations
\begin{equation}
\left\lbrace 
\begin{array}{l}
\displaystyle \frac{\partial\,\psi_{\ell}^i}{\partial\,t} = -(\theta_{\ell}^i + \Omega_i)\,\psi_{\ell}^i + \frac{\partial\,w_i}{\partial\,t} + \Omega_i\,w_i,\\
[8pt]
\displaystyle \psi_{\ell}^i(x,z,0) = 0.
\end{array}
\right. 
\label{eq:EDO_psi_DA_ani}
\end{equation}

%------------------------------------------------------------------------------------------

\subsection{The Biot-DA first-order system}\label{sec:DA:EDP}

The fractional derivatives involved in the Biot-JKD system (\ref{eq:2D_ani_syst_hyp_jkd_scalar}) are replaced by their diffusive approximation (\ref{eq:DA}), with evolution equations (\ref{eq:EDO_psi_DA_ani}). After some algebraic operations, the Biot-DA system is written as a first-order system in time and in space, used in the numerical simulations of $\S$ \ref{sec:exp} ($j=1,\cdots N$)
\begin{equation}
\left\lbrace 
\begin{array}{l}
\displaystyle \frac{\partial\,v_{s1}}{\partial\,t} - \frac{\rho_{w1}}{\chi_1}\,\left( \frac{\partial\,\sigma_{11}}{\partial\,x}+\frac{\partial\,\sigma_{13}}{\partial\,z}\right)  - \frac{\rho_f}{\chi_1}\,\frac{\partial\,p}{\partial\,x} = \frac{\rho_f}{\rho}\,\gamma_1\,\sum\limits_{\ell=1}^N a_{\ell}^1\,\psi_{\ell}^1 + G_{v_{s1}},\\
[10pt]
\displaystyle \frac{\partial\,v_{s3}}{\partial\,t} - \frac{\rho_{w3}}{\chi_3}\,\left( \frac{\partial\,\sigma_{13}}{\partial\,x}+\frac{\partial\,\sigma_{33}}{\partial\,z}\right)  - \frac{\rho_f}{\chi_3}\,\frac{\partial\,p}{\partial\,z} = \frac{\rho_f}{\rho}\,\gamma_3\,\sum\limits_{\ell=1}^N a_{\ell}^3\,\psi_{\ell}^3 + G_{v_{s3}},\\
[10pt]
\displaystyle \frac{\partial\,w_1}{\partial\,t} + \frac{\rho_f}{\chi_1}\,\left( \frac{\partial\,\sigma_{11}}{\partial\,x}+\frac{\partial\,\sigma_{13}}{\partial\,z}\right)  + \frac{\rho}{\chi_1}\,\frac{\partial\,p}{\partial\,x} = -\,\gamma_1\,\sum\limits_{\ell=1}^N a_{\ell}^1\,\psi_{\ell}^1 + G_{w_1},\\
[10pt]
\displaystyle \frac{\partial\,w_3}{\partial\,t} + \frac{\rho_f}{\chi_3}\,\left( \frac{\partial\,\sigma_{13}}{\partial\,x}+\frac{\partial\,\sigma_{33}}{\partial\,z}\right)  + \frac{\rho}{\chi_3}\,\frac{\partial\,p}{\partial\,z} = -\,\gamma_3\,\sum\limits_{\ell=1}^N a_{\ell}^3\,\psi_{\ell}^3 + G_{w_3},\\
[10pt]
\displaystyle \frac{\partial\,\sigma_{11}}{\partial\,t} - c_{11}^u\,\frac{\partial\,v_{s1}}{\partial\,x} -c_{13}^u\,\frac{\partial\,v_{s3}}{\partial\,z} - m\,\beta_1\,\left( \frac{\partial\,w_1}{\partial\,x} + \frac{\partial\,w_3}{\partial\,z}\right) = G_{\sigma_{11}},\\
[10pt]
\displaystyle \frac{\partial\,\sigma_{13}}{\partial\,t}-c_{55}^u\,\left( \frac{\partial\,v_{s3}}{\partial\,x}+\frac{\partial\,v_{s1}}{\partial\,z}\right) = G_{\sigma_{13}},\\
[10pt]
\displaystyle \frac{\partial\,\sigma_{33}}{\partial\,t}-c_{13}^u\,\frac{\partial\,v_{s1}}{\partial\,x}- c_{33}^u\,\frac{\partial\,v_{s3}}{\partial\,z} - m\,\beta_3\,\left( \frac{\partial\,w_1}{\partial\,x} + \frac{\partial\,w_3}{\partial\,z}\right) = G_{\sigma_{33}},\\
[10pt]
\displaystyle \frac{\partial\,p}{\partial\,t} + m\, \left( \beta_1\,\frac{\partial\,v_{s1}}{\partial\,x}+\beta_3\,\frac{\partial\,v_{s3}}{\partial\,z} +\frac{\partial\,w_1}{\partial\,x} +\frac{\partial\,w_3}{\partial\,z} \right) = G_p,\\
[10pt]
\displaystyle \frac{\partial\,\psi_j^1}{\partial\,t} + \frac{\rho_f}{\chi_1}\left( \frac{\partial\,\sigma_{11}}{\partial\,x} + \frac{\partial\,\sigma_{13}}{\partial\,z}\right) + \frac{\rho}{\chi_1}\,\frac{\partial\,p}{\partial\,x} = \Omega_1\,w_1 - \gamma_1\,\sum\limits_{\ell=1}^N a_{\ell}^1\,\psi_{\ell}^1 - (\theta_j^1 + \Omega_1)\,\psi_j^1 + G_{w_1},\\
[10pt]
\displaystyle \frac{\partial\,\psi_j^3}{\partial\,t} + \frac{\rho_f}{\chi_3}\left( \frac{\partial\,\sigma_{13}}{\partial\,x} + \frac{\partial\,\sigma_{33}}{\partial\,z}\right) + \frac{\rho}{\chi_3}\,\frac{\partial\,p}{\partial\,z} = \Omega_3\,w_3 - \gamma_3\,\sum\limits_{\ell=1}^N a_{\ell}^3\,\psi_{\ell}^3 - (\theta_j^3 + \Omega_3)\,\psi_j^3 + G_{w_3}.
\end{array}
\right. 
\label{eq:2D_ani_syst_hyp_ad}
\end{equation}
Taking the vector of unknowns
\begin{equation}
\bfU = (v_{s1}\,,\,v_{s3}\,,\,w_1\,,\,w_3\,,\,\sigma_{11}\,,\,\sigma_{13}\,,\,\sigma_{33}\,,\,p\,,\,\psi_1^1\,,\,\psi_1^3\,,\,\cdots\,,\,\psi_N^1\,,\,\psi_N^3)^T,
\label{eq:vect_U_ad}
\end{equation}
and the forcing
\begin{equation}
\bfG = \left( G_{v_{s1}}\,,\,G_{v_{s3}}\,,\,G_{w_1}\,,\,G_{w_3}\,,\,G_{\sigma_{11}}\,,\,G_{\sigma_{13}}\,,\,G_{\sigma_{33}}\,,\,G_p\,,\,G_{w_1}\,,\,G_{w_3}\,,\,G_{w_1}\,,\,G_{w_3} \right)^T,
\label{eq:forcing_DA}
\end{equation}
the system (\ref{eq:2D_ani_syst_hyp_ad}) is written in the form:
\begin{equation}
\frac{\partial\,\bfU}{\partial\,t} + \bfA\,\frac{\partial\,\bfU}{\partial\,x} + \bfB\,\frac{\partial\,\bfU}{\partial\,z} = -\bfS\,\bfU + \bfG,
\label{eq:2D_ani_syst_hyp_tens_ad}
\end{equation}
where $\bfA$ and $\bfB$ are the ($2\,N+8)\times(2\,N+8)$ propagation matrices and $\bfS$ is the diffusive matrix (given in  \ref{annexe:matABS}). The number of unknowns increases linearly with the number of memory variables. Only the matrix $\bfS$ depends on the coefficients of the diffusive approximation.

%------------------------------------------------------------------------------------------

\subsection{Properties}\label{sec:DA:prop}

Some properties are stated to characterize the first-order differential system (\ref{eq:2D_ani_syst_hyp_ad}). First, one notes that the only difference between the Biot-LF model, the Biot-JKD model and the Biot-DA model occurs in the viscous operators
\begin{equation}
\widehat{F}_i(\omega) = \left\lbrace \begin{array}{ll}
\displaystyle \widehat{F}_i^{LF}(\omega) = 1 \qquad\; & \mbox{Biot-LF},\\
[5pt]
\displaystyle \widehat{F}_i^{JKD}(\omega) = \frac{1}{\sqrt{\Omega_i}}\,(\Omega_i + j\,\omega)^{1/2} & \mbox{Biot-JKD},\\
[10pt]
\displaystyle \widehat{F}_i^{DA}(\omega) = \frac{\Omega_i + j\,\omega}{\sqrt{\Omega_i}}\,\sum\limits_{\ell=1}^N \frac{a_{\ell}^i}{\theta_{\ell}^i + \Omega_i + j\,\omega} & \mbox{Biot-DA}.
\end{array}\right.
\label{eq:fonction_correction}
\end{equation}
The dispersion analysis of the Biot-DA model is obtained by replacing the viscous operators $\widehat{F}_i^{JKD}(\omega)$ by $\widehat{F}_i^{DA}(\omega)$ in (\ref{eq:operateur_visco_jkd}). One of the eigenvalues of $\bfGamma^{-1}\,\mbox{\boldmath${\cal L}$}\,\mbox{\boldmath${\cal C}$}$ (\ref{eq:syst_dispersion_gen}) is still zero with multiplicity two, and the other non-zero eigenvalues correspond to the wave modes $\pm k_{pf}(\omega,\varphi)$, $\pm k_{ps}(\omega,\varphi)$ and $\pm k_{s}(\omega,\varphi)$. Consequently, the diffusive approximation does not introduce spurious wave.

\begin{proposition}
The eigenvalues of the matrix $\mbox{\boldmath$M$} = \cos(\varphi)\,\bfA + \sin(\varphi)\,\bfB$ are
\begin{equation}
sp(\mbox{\boldmath$M$}) = \left\lbrace 0\,,\,\pm c_{pf}^{\infty}(\varphi)\,,\,\pm c_{ps}^{\infty}(\varphi)\,,\,\pm c_{s}^{\infty}(\varphi)\right\rbrace,
\end{equation}
with $0$ being of multiplicity $2\,N+2$.
\end{proposition}

\noindent
The non-zero eigenvalues do not depend on the viscous operators $\widehat{F}_i(\omega)$. Consequently, the high-frequency limits of the phase velocities $c_{pf}^{\infty}(\varphi)$, $c_{ps}^{\infty}(\varphi)$ and $c_s^{\infty}(\varphi)$, defined in $\S$ \ref{sec:phys:dispersion}, are the same for both Biot-LF, Biot-JKD and Biot-DA models. An argumentation similar to \cite{LEVEQUE13} shows that the matrix $\mbox{\boldmath$M$}$ is diagonalizable for all $\varphi$ in $[0,2\,\pi[$, with real eigenvalues. The three models are therefore hyperbolic.

\begin{proposition}[\bf Decrease of the energy]
An energy analysis of (\ref{eq:2D_ani_syst_hyp_ad}) is performed. Let us consider the Biot-DA model (\ref{eq:2D_ani_syst_hyp_ad}) without forcing, and let us denote
\begin{equation}
E=E_1+E_2+E_3,
\label{eq:E1E2E3_AD}
\end{equation}
where $E_1$, $E_2$ are defined in equations (\ref{eq:energieJKD}) and 
\begin{equation}
\begin{array}{l}
%\displaystyle E_1 = \frac{1}{2}\,\int_{\mathbb{R}^2}\left(\rho\,\bfVs^T\,\bfVs + 2\,\rho_f\,\bfVs^T\,\bfW + \bfW^T\,\mathrm{diag}\left(\rho_{wi}\right)\,\bfW\right)\,dx\,dz,\\
%[15pt]
%\displaystyle E_2 = \frac{1}{2}\,\int_{\mathbb{R}^2}\left(\left( \bfSigma + p\,\bfBeta \right)^T\,\bfC^{-1}\,\left( \bfSigma + p\,\bfBeta \right) + \frac{1}{m}\,p^2\right)\,dx\,dz,\\
%[15pt]
\displaystyle E_3 = \frac{1}{2}\,\int_{\mathbb{R}^2}\frac{\eta}{\pi}\,\sum\limits_{\ell=1}^N(\bfW-\mbox{\boldmath$\psi_{\ell}$})^T\,\mathrm{diag}\left(\frac{a_{\ell}^i}{\kappa_i\,\sqrt{\Omega_i\,\theta_{\ell}^i}\,(\theta_{\ell}^i+2\,\Omega_i)}\right)\,(\bfW-\mbox{\boldmath$\psi_{\ell}$})\,dx\,dz.
\end{array}
\label{eq:energieAD}
\end{equation}
Then, $E$ satisfies
\begin{equation}
\begin{array}{l}
\displaystyle \frac{\textstyle d\,E}{\textstyle d\,t} = -\int_{\mathbb{R}^2}\frac{\eta}{\pi}\,\sum\limits_{\ell=1}^N \displaystyle \left\lbrace \mbox{\boldmath$\psi_{\ell}$}^T\,\mathrm{diag}\left(\frac{a_{\ell}^i\,(\theta_{\ell}^i+\Omega_i)}{\kappa_i\,\sqrt{\Omega_i\,\theta_{\ell}^i}\,(\theta_{\ell}^i+2\,\Omega_i)}\right)\,\mbox{\boldmath$\psi_{\ell}$}\right. \\
[20pt]
\hspace{1.1cm}\displaystyle \left. + \bfW^T\,\mathrm{diag}\left(\frac{a_{\ell}^i\,\Omega_i}{\kappa_i\,\sqrt{\Omega_i\,\theta_{\ell}^i}\,(\theta_{\ell}^i+2\,\Omega_i)}\right)\,\bfW \right\rbrace \,dx\,dz.
\end{array}
\label{eq:dEdtAD}
\end{equation}
\label{prop:nrjAD}
\end{proposition}
\noindent
The proof of the proposition \ref{prop:nrjAD} is similar to the proof of the proposition \ref{prop:nrjJKD} and will not be repeated here. Proposition \ref{prop:nrjAD} calls the following comments:
\begin{itemize}
%\item the terms $E_1$ and $E_2$ are the same in both the Biot-DA and Biot-JKD models;
\item only $E_3$ and the time evolution of $E$ are modified by the diffusive approximation;
\item the abscissae $\theta_{\ell}^i$ are always positive, as explained in $\S$ \ref{sec:DA:coeff}, but not necessarily the weights $a_{\ell}^i$. Consequently, in the general case, we cannot say that the Biot-DA model is stable. However, in the particular case where the coefficients $\theta_{\ell}^i$, $a_{\ell}^i$ are all positive, $E$ is an energy, and $\frac{d\,E}{d\,t} < 0$: the Biot-DA model is therefore stable in this case.
\end{itemize}

\begin{proposition}
Let us assume that the abscissae $\theta_{\ell}^i$ have been sorted in increasing order
\begin{equation}
\theta_1^i < \theta_2^i < \cdots < \theta_N^i,\quad i=1,3,\\
\label{eq:poids_croissant}
\end{equation}
and that the coefficients $\theta_{\ell}^i$, $a_{\ell}^i$ of the diffusive approximation (\ref{eq:DA}) are positive. Then zero is an eigenvalue with multiplicity $6$ of $\bfS$. Moreover, the $2\,N + 2$ non-zero eigenvalues of $\bfS$ (denoted $s_{\ell}^i$, $\ell = 1,\cdots,N+1$) are real positive, and satisfy
\begin{equation}
\begin{array}{l}
0 < s_1^i < \theta_1^i + \Omega_i < \cdots < s_N^i < \theta_N^i + \Omega_i <  s_{N+1}^i,\quad i=1,3.
\end{array}
\label{eq:vpS}
\end{equation}
\label{prop:diffusive_part_vpS}
\end{proposition}
\noindent
Proposition \ref{prop:diffusive_part_vpS} is proven in  \ref{annexe:proof_vpS}. As we will see in $\S$ \ref{sec:num}, the proposition \ref{prop:diffusive_part_vpS} ensures the stability of the numerical method. Positivity of quadrature abscissae and weights is again the fundamental hypothesis.

%------------------------------------------------------------------------------------------

\subsection{Determining the Biot-DA parameters}\label{sec:DA:coeff} 

For the sake of clarity, the space coordinates and the subscripts due to the anisotropy are omitted. The quadrature coefficients aim to approximate improper integrals of the form
\begin{equation}
(D+\Omega)^{1/2}w(t) = \frac{1}{\pi}\,\int_0^{\infty}\frac{1}{\sqrt{\theta}}\psi(t,\theta)\,d\theta \simeq \sum\limits_{\ell=1}^N a_{\ell}\,\psi(t,\theta_{\ell}).
\label{eq:quadrature_AD}
\end{equation}
Moreover, the positivity of the quadrature coefficients is crucial for the stability of the Biot-DA model and its numerical implementation, as shown in propositions \ref{prop:nrjAD} and \ref{prop:diffusive_part_vpS}. Two approaches can be employed for this purpose. While the most usual one is based on orthogonal polynomials, the second approach is associated with an optimization procedure applied to the viscous operators (\ref{eq:fonction_correction}). 

%------------------------------------------------------------------------------------------

\subsubsection{Gaussian quadratures\label{section:gauss_jacobi}}

Various orthogonal polynomials exist to evaluate the improper integral (\ref{eq:quadrature_AD}). The first method, proposed in \cite{YUAN02}, is to use the Gauss-Laguerre quadrature formula, which approximates improper integrals over $\mathbb{R}^+$. Slow convergence of this method is explained and corrected in \cite{DIETHELM08}. It consists in replacing the Gauss-Laguerre quadrature by a Gauss-Jacobi quadrature, more suitable for functions which decrease algebraically. A last improvement, proposed in \cite{BIRK10}, consists in using a modified Gauss-Jacobi quadrature formula, recasting the improper integral (\ref{eq:quadrature_AD}) as
\begin{equation}
\displaystyle \frac{1}{\pi}\,\int_0^{\infty} \frac{1}{\sqrt{\theta}}\,\psi(\theta)\,d\theta=\frac{1}{\pi}\,\int_{-1}^{+1}(1-\tilde{\theta})^\gamma(1+\tilde{\theta})^\delta\tilde{\psi}(\tilde{\theta})\,d\tilde{\theta}\simeq \frac{1}{\pi}\,\sum\limits_{\ell = 1}^N\tilde{a}_\ell\,\tilde{\psi}(\tilde{\theta}_\ell),
\label{eq:quadrature_derivee_frac_birk}
\end{equation}
with the modified memory variable $\tilde{\psi}$ defined as
\begin{equation}
\tilde{\psi}(\tilde{\theta})=\frac{4}{(1-\tilde{\theta})^{\gamma-1}(1+\tilde{\theta})^{\delta+3}}\,\left(\frac{1+\tilde{\theta}}{1-\tilde{\theta}}\right)\psi\left(\left(\frac{1-\tilde{\theta}}{1+\tilde{\theta}}\right)^2\right).
\label{eq:fonction_gauss_jacobi}
\end{equation}
The abscissae $\tilde{\theta}_{\ell}$, which are the zeros of the Gauss-Jacobi polynomials, and the weights $\tilde{a}_{\ell}$ can be computed by standard routines \cite{NUM_RECIPES}. In \cite{BIRK10}, the author proves that for fractional derivatives of order $1/2$, the optimal coefficients to use are $\gamma = 1$ and $\delta = 1$. The coefficients of the diffusive approximation $\theta_{\ell}$ and $a_{\ell}$ (\ref{eq:quadrature_AD}) are therefore related to the coefficients $\tilde{\theta}_{\ell}$ and $\tilde{a}_{\ell}$ (\ref{eq:quadrature_derivee_frac_birk}) by
\begin{equation}
\theta_{\ell} = \left(\frac{1-\tilde{\theta}_{\ell}}{1+\tilde{\theta}_{\ell}}\right)^2,\quad a_{\ell} = \frac{1}{\pi}\,\frac{4\,\tilde{a}_{\ell}}{(1-\tilde{\theta}_{\ell})\,(1+\tilde{\theta}_{\ell})^3}.
\label{eq:coef_quad_birk}
\end{equation}
By construction, they are strictly positive.

%------------------------------------------------------------------------------------------

\subsubsection{Optimization procedures\label{section:solvopt}}

In \cite{BLANC_JCP13,BLANC_JASA13}, we proposed a different method to determine the  coefficients $\theta_{\ell}$ and $a_{\ell}$ of the diffusive approximation (\ref{eq:quadrature_AD}).  This method is based on the frequency expressions of the viscous operators and takes into account the frequency content of the source.  Our requirement is therefore to approximate the viscous operator $\widehat{F}^{JKD}(\omega)$ by $\widehat{F}^{DA}(\omega)$ (\ref{eq:fonction_correction}) in the frequency range of interest $I = [\omega_{\min},\omega_{\max}]$, centered on the central angular frequency of the source. This leads to the  minimization of  the quantity $\chi^2$ with respect to the abcissae $\theta_{\ell}$ and to the weights $a_{\ell}$
\begin{equation}
\chi^2 = \sum\limits_{k=1}^K\left| \frac{\widehat{F}^{DA}(\omega_k)}{\widehat{F}^{JKD}(\omega_k)} - 1 \right| ^2
= \sum\limits_{k=1}^K\left| \sum\limits_{\ell=1}^N a_{\ell}\,\frac{(\Omega + j\,\omega_k)^{1/2}}{\theta_{\ell}+\Omega + j\,\omega_k} - 1 \right| ^2,
\label{eq:chi2}
\end{equation}
where the angular frequencies $\omega_k$ are distributed linearly in $I$ on a logarithmic scale of $K$ points
\begin{equation}
\omega_k = \omega_{\min}\,\left(\frac{\omega_{\max}}{\omega_{\min}}\right) ^{\frac{k-1}{K-1}},\qquad k=1\cdots K.
\label{eq:tilde_omega_k}
\end{equation}
In \cite{BLANC_JCP13,BLANC_JASA13}, the abcissae $\theta_{\ell}$ were arbitrarily put linearly on a logarithmic scale, as (\ref{eq:tilde_omega_k}). Only the weights $a_{\ell}$ were optimized with a linear least-squares minimization procedure of (\ref{eq:chi2}). Some negative weights were obtained, which represents a major drawback, at least theoretically, since the stability of the Biot-DA model can not be guaranteed. 

To remove this drawback and improve the minimization of $\chi^2$, a nonlinear constrained optimization is developed, where both the abcissae and the weights are optimized. The coefficients $\theta_{\ell}$ and $a_{\ell}$ are now constrained to be positive. An additional constraint $\theta_{\ell} \leqslant \theta_{max}$ is also introduced to ensure the computational accuracy in the forthcoming numerical method ($\S$ \ref{sec:num}). Setting
\begin{equation}
\theta_{\ell} = (\theta'_{\ell})^2,\quad a_{\ell} = (a'_{\ell})^2,
\label{eq:coef_min_shor}
\end{equation}
the number of constraints decreases from $3\,N$ to $N$ leading to the following minimization problem:
\begin{equation}
\displaystyle \min\limits_{(\theta_{\ell}',a_{\ell}')} \chi^2,\quad\theta_{\ell}'\leqslant \sqrt{\theta_{\max}}.
\label{eq:pb_opti_2}
\end{equation}
The constrained minimization problem (\ref{eq:pb_opti_2}) is nonlinear and non-quadratic with respect to abscissae $\theta_{\ell}'$. To solve it, we implement the program SolvOpt \cite{KAPPEL00,SHOR85}, used in viscoelasticity \cite{REKIK11}. Since this Shor's algorithm is iterative, it requires an initial estimate $\theta_{\ell}'^0$, $a_{\ell}'^0$ of the coefficients which satisfies the constraints of the minimization problem (\ref{eq:pb_opti_2}). For this purpose, $\theta_{\ell}^0$ and $a_{\ell}^0$ are initialized with the method based on the modified Gauss-Jacobi quadrature formula (\ref{eq:coef_quad_birk}).
%\begin{equation}
%\theta_{\ell}'^0 = \frac{1-\tilde{\theta}_{\ell}}{1+\tilde{\theta}_{\ell}},\quad a_{\ell}'^0 = \sqrt{\frac{1}{\pi}\,\frac{4\,\tilde{a}_{\ell}}{(1-\tilde{\theta}_{\ell})\,(1+\tilde{\theta}_{\ell})^3}}.
%\label{eq:initial_estimate}
%\end{equation}
Different initial guess have been used, derived from Gaus-Legendre and Gauss-Jacobi methods, leading to the same final coefficients $\theta_{\ell}$ and $a_{\ell}$.

In what follows, we always use the parameters
%\begin{equation}
$\omega_{\min} = \omega_0/10,\quad\omega_{\max}=10\,\omega_0,\quad\theta_{\max}=100\,\omega_0,\quad K=2\,N,$
%\end{equation}
where $\omega_0=2\,\pi\,f_0$ is the central angular frequency of the source.

%------------------------------------------------------------------------------------------

\subsubsection{Discussion\label{section:choix_opti}}

\begin{figure}[htbp]
\begin{center}
\begin{tabular}{cc}
(a) & (b)\\
\includegraphics[scale=0.33]{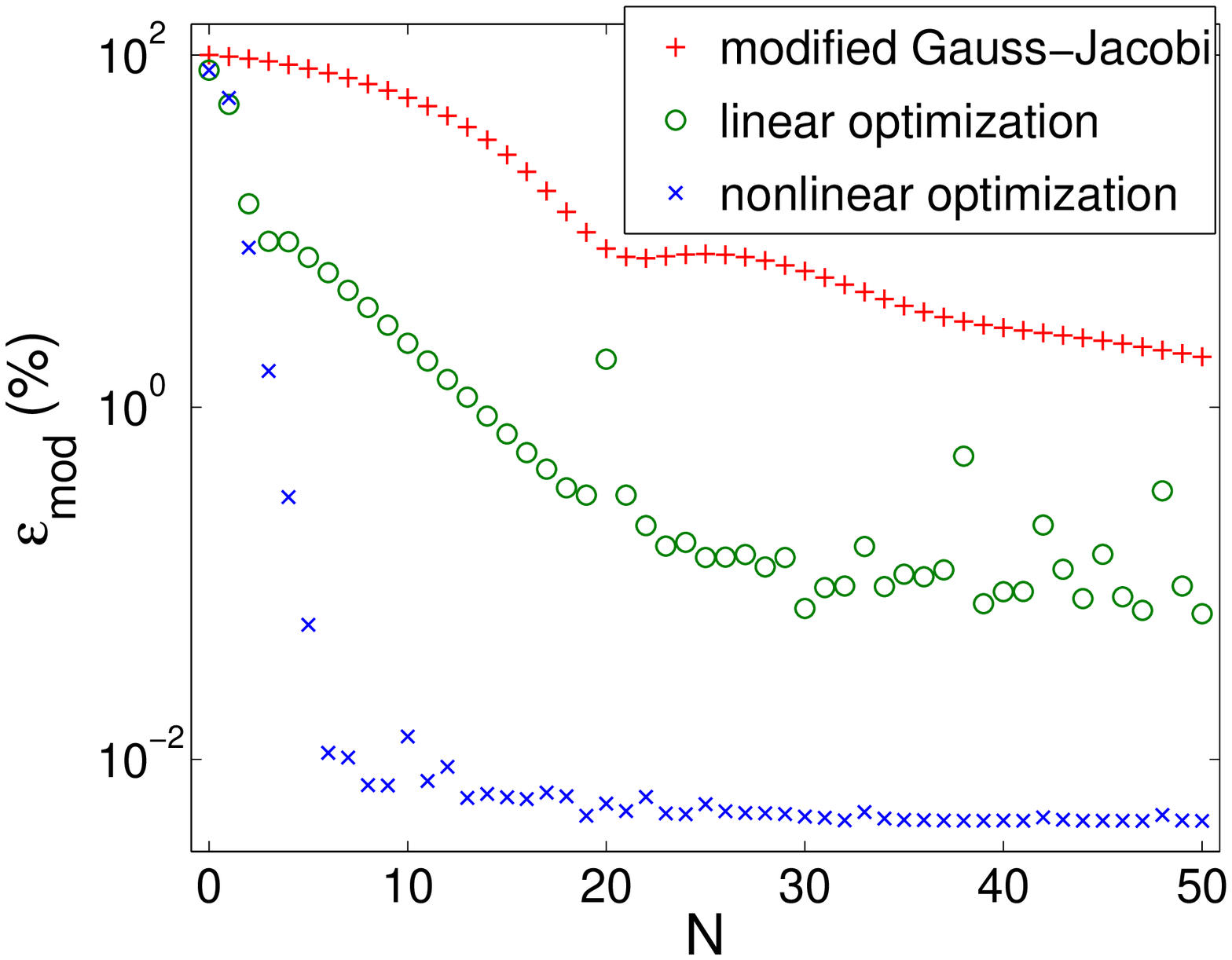} & 
\includegraphics[scale=0.33]{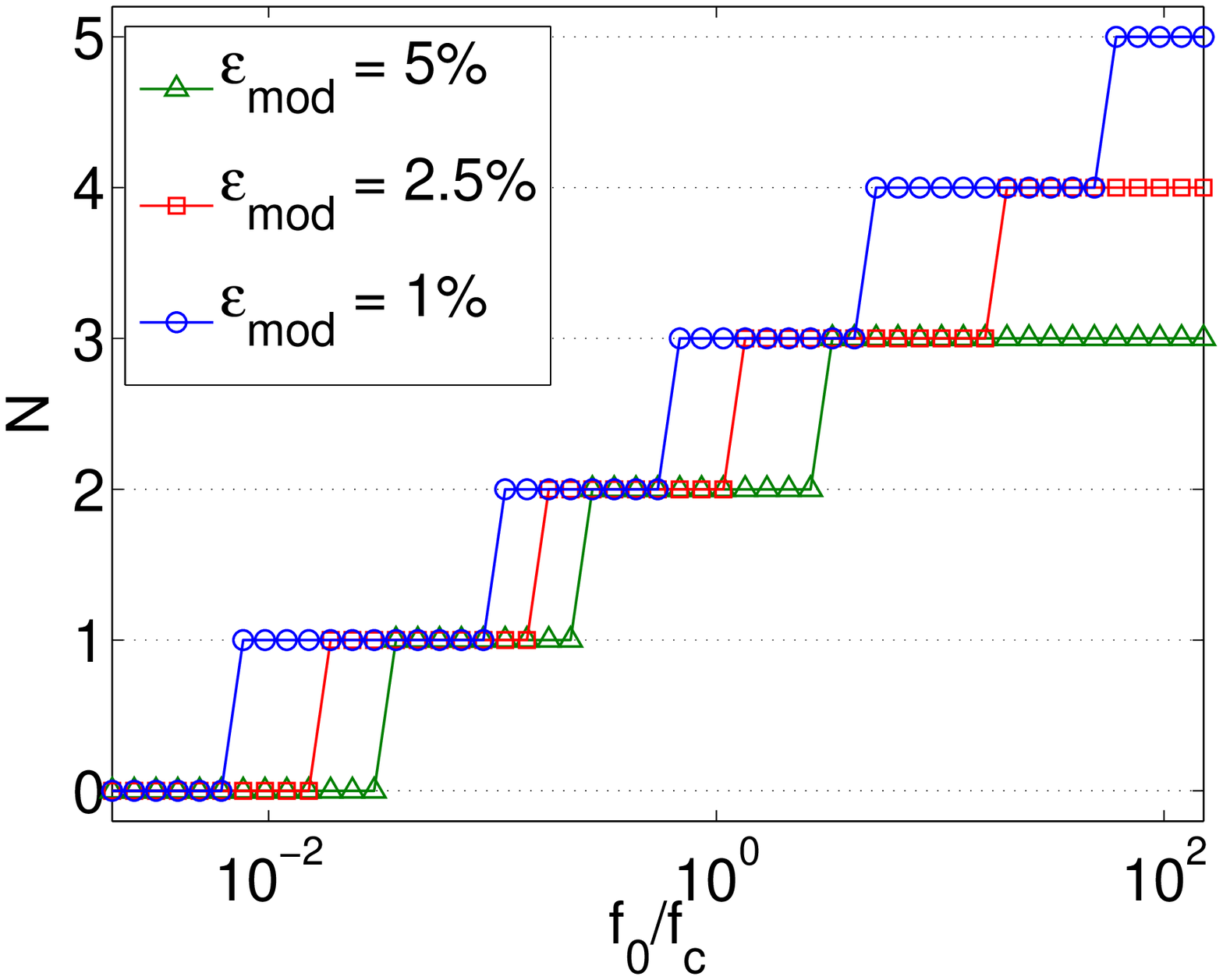}
\end{tabular}
\end{center}
\caption{(a): relative error $\varepsilon_{mod}$ in terms of $N$ for both the modified Gauss-Jacobi quadrature and the nonlinear constrained optimization. (b): required values of $N$ in terms of $f_0/f_{c1}$ and the required accuracy $\varepsilon_{mod}$ for the nonlinear optimization.}
\label{fig:erreur_opti}
\end{figure}

To compare the quadrature methods presented in $\S$ \ref{section:gauss_jacobi} and \ref{section:solvopt}, we first define the error of model $\varepsilon_{mod}$ as
\begin{equation}
\varepsilon_{mod} = \,\left|\left| \frac{\widehat{F}^{DA}(\omega)}{\widehat{F}^{JKD}(\omega)}-1\right|\right|_{L_2} \, = \left( \int_{\omega_{\min}}^{\omega_{\max}}\left| \frac{\widehat{F}^{DA}(\omega)}{\widehat{F}^{JKD}(\omega)}-1\right|^2\,d\omega \right)^{1/2}.
\label{eq:norme2_relative_error}
\end{equation}
The variation of $\varepsilon_{mod}$ in terms of the number $N$ of memory variables, for $f_0 = 200$ kHz and $f_c = 3.84$ kHz, is represented on figure \ref{fig:erreur_opti}-a. The Gauss-Jacobi method converges very slowly, and the error is always larger than $1$ \% even for $N=50$. Moreover, for values of $N\leqslant10$, the error is always larger than $60$ \%.  For both the linear and the nonlinear optimizations, the errors decrease rapidly with $N$. Nevertheless, the nonlinear procedure outperforms the results obtained in the linear case. For $N=8$ for instance, the relative error of the nonlinear optimization ($\varepsilon_{mod} \simeq 7.16\,10^{-3}$ \%) is $514$ times smaller than the error of the linear optimization ($\varepsilon_{mod} \simeq 3.68$ \%).  For larger values of $N$, the system is poorly conditioned and the order of convergence deteriorates; in practice, this is not penalizing since large values of $N$ are not used. An example of a priori parametric determination of $N$ in terms of both the frequency range and the desired accuracy is also given in figure \ref{fig:erreur_opti}-b for the nonlinear procedure. The case $N=0$ corresponds to the Biot-LF model.

\begin{figure}[htbp]
\begin{center}
\begin{tabular}{cc}
(a) & (b)\\
\includegraphics[scale=0.30]{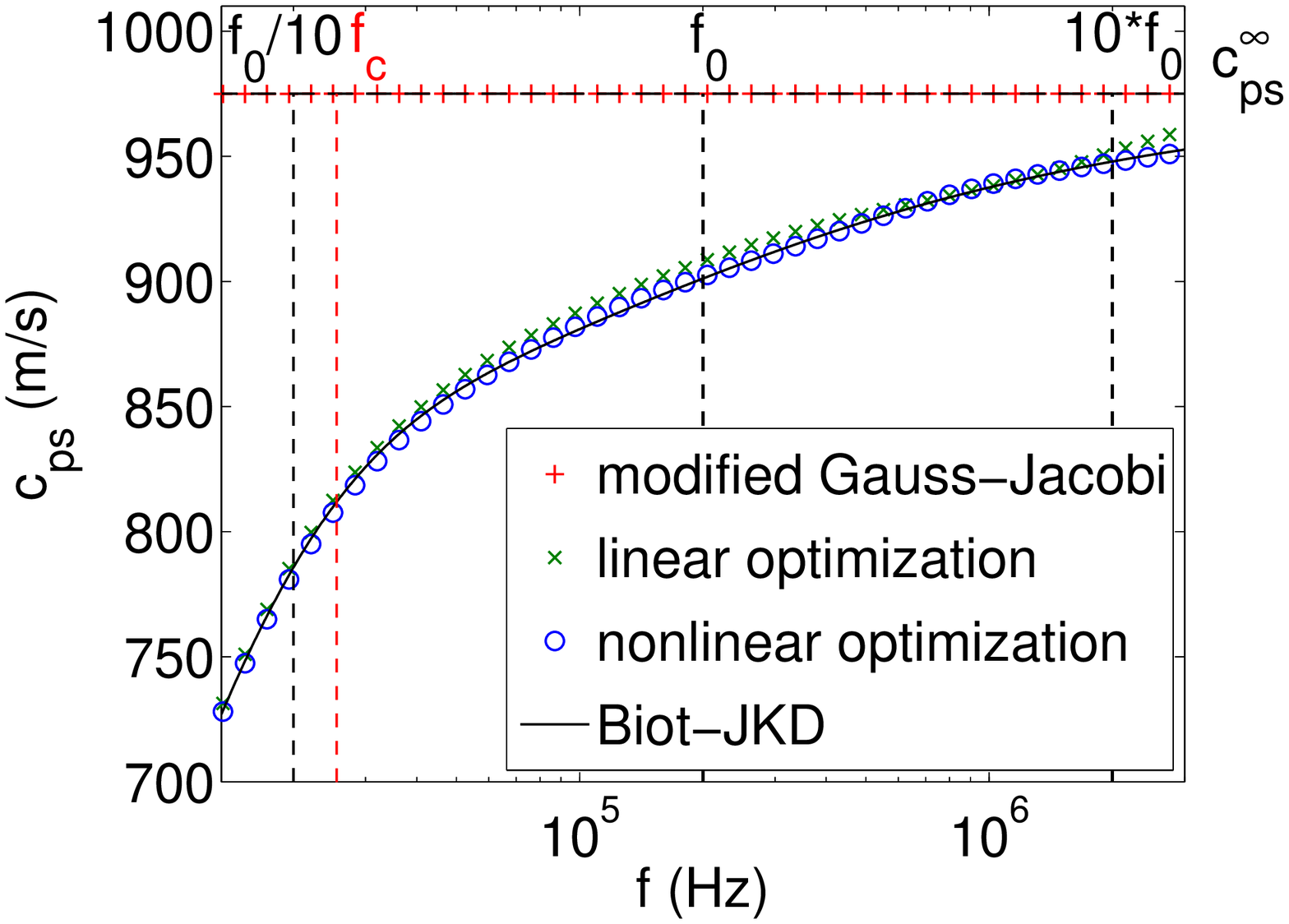} & 
\includegraphics[scale=0.30]{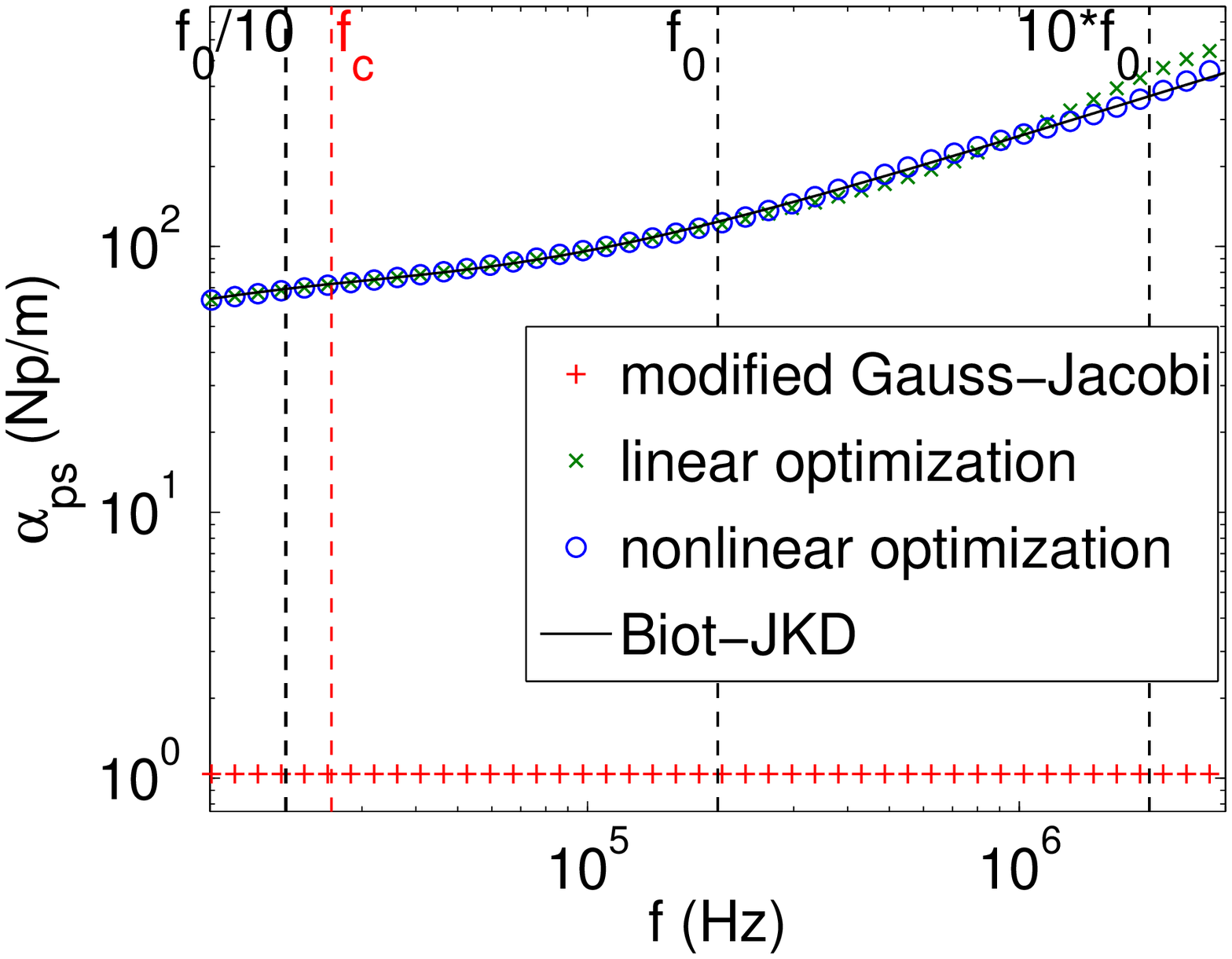}\\
%(c) & (d)\\
%\includegraphics[scale=0.30]{figure/compare_attps_3_zoom.eps} & 
%\includegraphics[scale=0.30]{figure/compare_attps_6_zoom.eps}\\
\end{tabular}
\end{center}
\caption{phase velocity (a)  and attenuation (b) of the slow quasi-compressional wave. Comparison between the Biot-DA model and the Biot-JKD model for $N=3$.}
\label{fig:fig_opti_dispersion}
\end{figure}

It is also important to compare the influence of the quadrature coefficient on the physical observables. For that purpose, we represent on figure \ref{fig:fig_opti_dispersion} the  phase velocity and the attenuation of the slow wave of the Biot-DA model, obtained with the different quadrature methods. As expected, the results given by the Gauss-Jacobi method are extremely poor. On the contrary, the linear and non-linear procedures are able to represent very accurately the variations of these quantities on the considered range of frequencies, even for the small values $N=3$. Based on these results and the positivity requirement, the nonlinear constrained optimization is therefore considered as the better way to determine the coefficients of the diffusive approximation. This method is used in all what follows.

%------------------------------------------------------------------------------------------
%------------------------------------------------------------------------------------------

\section{Numerical modeling}\label{sec:num}

\subsection{Splitting}\label{sec:num:splitting}

In order to integrate the Biot-DA system (\ref{eq:2D_ani_syst_hyp_tens_ad}), a uniform grid is introduced, with mesh size $\Delta\,x,_,\Delta\,z$ and time step $\Delta\,t$. The approximation of the exact solution $\bfU(x_i = i\,\Delta\,x,z_j = j\,\Delta\,z,t_n = n\,\Delta\,t)$ is denoted by $\bfU_{ij}^n$, with $0 \leqslant i \leqslant N_x$, $0 \leqslant j \leqslant N_z$. If $\Delta\,x = \Delta\,z$, a straightforward discretization of (\ref{eq:2D_ani_syst_hyp_tens_ad}) by an explicit time scheme typically leads  to the following condition of stability
\begin{equation}
\Delta t \leqslant \min\left( \Upsilon\,\frac{\Delta x}{\max\limits_{\varphi\in[0,\pi/2]}c_{pf}^{\infty}(\varphi)},\frac{2}{R(\bfS)}\right),
\label{eq:CFL_direct}
\end{equation}
where $R(\bfS)$ is the spectral radius of $\bfS$, and $\Upsilon > 0$ is obtained by a Von-Neumann analysis when $\bfS=\bfZero$. The first term of (\ref{eq:CFL_direct}), which depends of the propagation matrices $\bfA$ and $\bfB$, is the classical CFL condition. The second term of (\ref{eq:CFL_direct}) depends only on the diffusive matrix $\bfS$. From proposition \ref{prop:diffusive_part_vpS}, we deduce that the spectral radius of $\bfS$ satisfies
\begin{equation}
R(\bfS) > \max\limits_{\ell=1,\cdots,N}(\theta_{\ell}^1 + \Omega_1,\theta_{\ell}^3 + \Omega_3)
\label{chapter3:eq:spectral_radius_S}
\end{equation}
if the coefficients $\theta_{\ell}^i$ and $a_{\ell}^i$ of the diffusive approximation are positive. With highly dissipative fluids, the second term of (\ref{eq:CFL_direct}) can be so small that numerical computations are intractable.

A more efficient strategy is adopted here, based on the second-order Strang splitting \cite{LEVEQUE02}. It consists in splitting the original system (\ref{eq:2D_ani_syst_hyp_tens_ad}) into a propagative part
\begin{equation}
\frac{\partial\,\bfU}{\partial\,t} + \bfA\,\frac{\partial\,\bfU}{\partial\,x} + \bfB\,\frac{\partial\,\bfU}{\partial\,z} = \bfZero,\qquad(\bfH_p)
\label{eq:propagative_part}
\end{equation}
and a diffusive part with forcing
\begin{equation}
\frac{\partial\,\bfU}{\partial\,t} = -\bfS\,\bfU + \bfG, \qquad\qquad\qquad(\bfH_d)
\label{eq:diffusive_part}
\end{equation}
where $\bfH_p$ and $\bfH_d$ are the operators associated with each part. One solves alternatively the propagative part and the diffusive part:
\begin{equation}
\bfU^{n+1} = \bfH_d\left(t_{n+1},\frac{\Delta t}{2}\right)\circ\bfH_p(\Delta t)\circ\bfH_d\left(t_n,\frac{\Delta t}{2}
\right)\,\bfU^{n}.
\label{eq:strang_splitting}
\end{equation}

The discrete operator $\bfH_p$ associated with the propagative part (\ref{eq:propagative_part}) is an ADER 4 (Arbitrary DERivatives) scheme \cite{DUMBSER04}. This scheme is fourth-order accurate in space and time, is dispersive of order 4 and dissipative of order 6 \cite{STRIKWERDA99}, and has a stability limit $\Upsilon=1$. On Cartesian grids, ADER 4 amounts to a fourth-order Lax-Wendroff scheme. A general expression of the ADER scheme, together with its numerical analysis, can be found in the section 4-3 of the thesis \cite{THESE_BLANC}.

The solution of (\ref{eq:diffusive_part}) is given by
\begin{equation}
\begin{array}{ll}
\displaystyle \bfH_d\left(t_k,\frac{\Delta t}{2}\right)\,\bfU(t_0) & \displaystyle = e^{-\mbox{\scriptsize\boldmath$S$}\,\Delta t/2}\,\bfU\left(t_0\right) + \int_{t_0}^{t_0 + \Delta t/2} e^{-\mbox{\scriptsize\boldmath$S$}\,(t_0 + \Delta t/2-\tau)}\,\bfG(\tau)\,d\tau,\\
[12pt]
& \displaystyle \simeq e^{-\mbox{\scriptsize\boldmath$S$}\,\frac{\Delta t}{2}}\,\bfU(t_0) - (\mbox{\boldmath$I$} - e^{-\mbox{\scriptsize\boldmath$S$}\,\frac{\Delta t}{2}})\,\bfS^{-1}\,\bfG(t_k),
\end{array}
\label{eq:int_diffu_part_F}
\end{equation}
with $k=n$ or $n+1$. The exponential matrix $e^{-\mbox{\scriptsize\boldmath$S$}\,\Delta t/2}$ is computed numerically using the $(6,6)$ Pad\'e approximation in the "scaling and squaring method" \cite{MOLER03}. Proposition \ref{prop:diffusive_part_vpS} ensures that the numerical integration of the diffusive step (\ref{eq:diffusive_part}) is unconditionally stable \cite{BLANC_JCP13}.  Without forcing, i.e. $\bfG = \bfZero$, the integration of the diffusive part (\ref{eq:diffusive_part}) is exact.

The full algorithm is therefore stable under the optimum CFL condition of stability
\begin{equation}
\Delta t = \Upsilon\,\frac{\Delta x}{\max\limits_{\varphi\in[0,\pi/2]}c_{pf}^{\infty}(\varphi)},\quad\Upsilon\leqslant 1,
\label{eq:CFL_dsplit}
\end{equation}
which is always independent of the Biot-DA model coefficients. Since the matrices $\bfA$, $\bfB$ and $\bfS$ do not commute, the order of convergence decreases from $4$ to $2$. Using a fourth-order ADER scheme is nevertheless advantageous, compared with the second-order Lax-Wendroff scheme: the stability limit is improved, and numerical artifacts (dispersion, attenuation, anisotropy) are greatly reduced.

%------------------------------------------------------------------------------------------

\subsection{Immersed interface method}\label{sec:num:IIM}

Let us consider two transversely isotropic homogeneous poroelastic media $\Omega_0$ and $\Omega_1$ separated by a stationary interface $\Gamma$, as shown in figure \ref{fig:interface}. The governing equations (\ref{eq:2D_ani_syst_hyp_tens_ad}) in each medium have to be completed by a set of jump conditions. The simple case of perfect bonding and perfect hydraulic contact along $\Gamma$ is considered here, modeled by the jump conditions \cite{GUREVICH99}:
\begin{equation}
\left[ \bfVs \right] = {\bf 0},\quad \left[ \bfW.\mbox{\boldmath$n$} \right] = 0,\quad \left[ \underline{\bfSigma}.\mbox{\boldmath$n$} \right] = \bfZero,\quad \left[ p \right] = 0.
\label{eq:jump_condition}
\end{equation}
The discretization of the interface conditions requires special care. A straightforward stair-step representation of interfaces introduces first-order geometrical errors and yields spurious numerical diffractions. In addition, the jump conditions (\ref{eq:jump_condition}) are not enforced numerically if no special treatment is applied. Lastly, the smoothness requirements to solve (\ref{eq:propagative_part}) are not satisfied, decreasing the convergence rate of the ADER scheme.

To remove these drawbacks while maintaining the efficiency of Cartesian grid methods, immersed interface methods constitute a possible strategy \cite{LEVEQUE97,LI94,CHIAVASSA11}. The latter studies can be consulted for a detailed description of this method. The basic principle is as follows: at the irregular nodes where the ADER scheme crosses an interface, modified values of the solution are used on the other side of the interface instead of the usual numerical values.

Calculating these modified values is a complex task involving high-order derivation of jump conditions (\ref{eq:jump_condition}), high-order derivation of the Beltrami-Michell equation (\ref{eq:beltrami_ani}) and algebraic manipulation, such as singular value decompositions. All these time consuming procedures can be carried out during a preprocessing stage and only small matrix-vector multiplications need to be performed during the simulation. After optimizing the code, the extra CPU cost can be practically negligible, i.e. lower than 1\% of that required by the time-marching procedure.

Compared with $\S$ 3-3 of \cite{CHIAVASSA11}, the modifications induced by anisotropy concern 
\begin{itemize}
\item step 1: the derivation of the jump conditions, 
\item step 2: the derivation of the Beltrami-Michell equation.
\end{itemize}
These modifications are tedious and hence will not be repeated here. They are  deduced from the new expressions (\ref{eq:beltrami_ani}) and (\ref{eq:2D_ani_syst_hyp_ad}).

%------------------------------------------------------------------------------------------

\subsection{Summary of the algorithm}\label{sec:num:algo}

The numerical method can be summed up as follows:
\begin{enumerate}
\item pre-processing step
\begin{itemize}
\item diffusive coefficients: initialisation (\ref{eq:coef_quad_birk}), nonlinear optimisation (\ref{eq:coef_min_shor})-(\ref{eq:pb_opti_2})
\item numerical scheme: ADER matrices for (\ref{eq:propagative_part}), exponential of the diffusive matrix (\ref{eq:int_diffu_part_F})
\item immersed interface method: detection of irregular points, computation of extrapolation matrices
\end{itemize}
\item time iterations
\begin{itemize}
\item immersed interface method: computation of modified values near interfaces
\item diffusive half-step $\bfH_d$ (\ref{eq:int_diffu_part_F})
\item propagative step $\bfH_p$ (\ref{eq:propagative_part}), using modified values near interfaces
\item diffusive half-step $\bfH_d$ (\ref{eq:int_diffu_part_F})
\end{itemize}
\end{enumerate}

%------------------------------------------------------------------------------------------
%------------------------------------------------------------------------------------------

\section{Numerical experiments}\label{sec:exp}

\noindent
{\it Configuration}\label{sec:exp:config}

In order to demonstrate the ability of the present method to be applied to a wide range of applications, the numerical tests will be run on two different transversely isotropic porous media. The medium $\Omega_0$ is composed of thin layers of epoxy and glass, strongly anisotropic if the wavelengths are large compared to the thickness of the layers \cite{CARCIONE96}. The medium $\Omega_1$ is water saturated Berea sandstone, which is sedimentary rock commonly encountered in petroleum engineering. The grains are predominantly sand sized and composed of quartz bonded by silica \cite{CARCIONE96,DAI95}.

The values of the physical parameters are given in table \ref{table:para_phy_ani}. The viscous characteristic lengths $\Lambda_1$ and $\Lambda_3$ are obtained by setting the Pride numbers $P_1 = P_3 = 0.5$. We also report in these tables some useful values, such as phase velocities, critical frequencies, and quadrature parameters computed for each media. The central frequency of the source is $f_0=200$ kHz, and the quadrature coefficients $\theta_{\ell}^i$, $a_{\ell}^i$, $i=1,3,$ are determined by nonlinear constrained optimization with $N=3$ memory variables. The error of model $\varepsilon_{mod}$ (\ref{eq:norme2_relative_error}) is also given. We note that the transition frequencies $f_{c1}$ and $f_{c3}$ are the same for both $\Omega_0$ and $\Omega_1$. In this particular case, the coefficients of the diffusive approximation are therefore also the same.
\begin{table}[htbp]
\begin{center}\footnotesize
\begin{tabular}{llll}
 & Parameters & $\Omega_0$ & $\Omega_1$\\
\hline
\rule[-1mm]{0mm}{3mm} Saturating fluid & $\rho_f$ (kg/m$^3$) & $1040$ & $1040$\\
\rule[-1mm]{0mm}{3mm} & $\eta$ (Pa.s)  & $10^{-3}$ & $10^{-3}$\\
\rule[-1mm]{0mm}{3mm} & $K_f$ (GPa)  & $2.5$ & $2.5$\\
\rule[-1mm]{0mm}{3mm} Grain & $\rho_s$ (kg/m$^3$) & $1815$ & $2500$\\
\rule[-1mm]{0mm}{3mm} & $K_s$ (GPa) & $40$ & $80$\\
\rule[-1mm]{0mm}{3mm} Matrix & $\phi$ & $0.2$ & $0.2$\\
\rule[-1mm]{0mm}{3mm} & ${\cal T}_1$ & $2$ & $2$\\
\rule[-1mm]{0mm}{3mm} & ${\cal T}_3$ & $3.6$ & $3.6$\\
\rule[-1mm]{0mm}{3mm} & $\kappa_1$ (m$^2$) & $6.\,10^{-13}$ & $6.\,10^{-13}$\\
\rule[-1mm]{0mm}{3mm} & $\kappa_3$ (m$^2$) & $10^{-13}$ & $10^{-13}$\\
\rule[-1mm]{0mm}{3mm} & $c_{11}$ (GPa) & $39.4$ & $71.8$\\
\rule[-1mm]{0mm}{3mm} & $c_{12}$ (GPa) & $1$ & $3.2$\\
\rule[-1mm]{0mm}{3mm} & $c_{13}$ (GPa) & $5.8$ & $1.2$\\
\rule[-1mm]{0mm}{3mm} & $c_{33}$ (GPa) & $13.1$ & $53.4$\\
\rule[-1mm]{0mm}{3mm} & $c_{55}$ (GPa) & $3$ & $26.1$\\
\rule[-1mm]{0mm}{3mm} & $\Lambda_1$ (m) & $6.93\,10^{-6}$ & $2.19\,10^{-7}$\\
\rule[-1mm]{0mm}{3mm} & $\Lambda_3$ (m) & $3.79\,10^{-6}$ & $1.20\,10^{-7}$\\\hline
\rule[-1mm]{0mm}{4mm} Dispersion & $c_{pf}^{\infty}(0)$ (m/s) & $5244.40$ & $6004.31$\\
\rule[-1mm]{0mm}{4mm} & $c_{pf}(f_0,0)$ kHz (m/s) & $5227.10$ & $5988.50$\\
\rule[-1mm]{0mm}{4mm} & $c_{pf}^{\infty}(\pi/2)$ (m/s) & $3583.24$ & $5256.03$\\
\rule[-1mm]{0mm}{4mm} & $c_{pf}(f_0,\pi/2)$ (m/s) & $3581.42$ & $5245.84$\\
\rule[-1mm]{0mm}{4mm} & $c_{ps}^{\infty}(0)$ (m/s) & $975.02$ & $1026.45$\\
\rule[-1mm]{0mm}{4mm} & $c_{ps}(f_0,0)$ (m/s) & $901.15$ & $949.33$\\
\rule[-1mm]{0mm}{4mm} & $c_{ps}^{\infty}(\pi/2)$ (m/s) & $604.41$ & $745.59$\\
\rule[-1mm]{0mm}{4mm} & $c_{ps}(f_0,\pi/2)$ (m/s) & $534.88$ & $661.32$\\
\rule[-1mm]{0mm}{4mm} & $c_{s}^{\infty}(0)$ (m/s) & $1368.36$ & $3484.00$\\
\rule[-1mm]{0mm}{4mm} & $c_{s}(f_0,0)$ (m/s) & $1361.22$ & $3470.45$\\
\rule[-1mm]{0mm}{4mm} & $c_{s}^{\infty}(\pi/2)$ (m/s) & $1388.53$ & $3522.07$\\
\rule[-1mm]{0mm}{4mm} & $c_{s}(f_0,\pi/2)$ (m/s) & $1381.07$ & $3508.05$\\
\rule[-1mm]{0mm}{4mm} & $f_{c1}$ (Hz) & $2.55\,10^{4}$ & $2.55\,10^{4}$\\
\rule[-1mm]{0mm}{4mm} & $f_{c3}$ (Hz) & $8.50\,10^{4}$ & $8.50\,10^{4}$\\ \hline
\rule[-1mm]{0mm}{4mm} Optimization & $\theta_1^1$ (rad/s) & $1.64\,10^5$ & $1.64\,10^5$\\
\rule[-1mm]{0mm}{4mm} & $\theta_2^1$ (rad/s) & $2.80\,10^6$ & $2.80\,10^6$\\
\rule[-1mm]{0mm}{4mm} & $\theta_3^1$ (rad/s) & $3.58\,10^7$ & $3.58\,10^7$\\
\rule[-1mm]{0mm}{4mm} & $a_1^1$ (rad$^{1/2}$/s$^{1/2}$) & $5.58\,10^2$ & $5.58\,10^2$\\
\rule[-1mm]{0mm}{4mm} & $a_2^1$ (rad$^{1/2}$/s$^{1/2}$) & $1.21\,10^3$ & $1.21\,10^3$\\
\rule[-1mm]{0mm}{4mm} & $a_3^1$ (rad$^{1/2}$/s$^{1/2}$) & $7.32\,10^3$ & $7.32\,10^3$\\
\rule[-1mm]{0mm}{4mm} & $\varepsilon_{mod}^1$ (\%) & $1.61$ & $1.61$\\
\rule[-1mm]{0mm}{4mm} & $\theta_1^3$ (rad/s) & $3.14\,10^5$ & $3.14\,10^5$\\
\rule[-1mm]{0mm}{4mm} & $\theta_2^3$ (rad/s) & $5.06\,10^7$ & $5.06\,10^7$\\
\rule[-1mm]{0mm}{4mm} & $\theta_3^3$ (rad/s) & $4.50\,10^6$ & $4.50\,10^6$\\
\rule[-1mm]{0mm}{4mm} & $a_1^3$ (rad$^{1/2}$/s$^{1/2}$) & $7.57\,10^2$ & $7.57\,10^2$\\
\rule[-1mm]{0mm}{4mm} & $a_2^3$ (rad$^{1/2}$/s$^{1/2}$) & $8.79\,10^3$ & $8.79\,10^3$\\
\rule[-1mm]{0mm}{4mm} & $a_3^3$ (rad$^{1/2}$/s$^{1/2}$) & $1.38\,10^3$ & $1.38\,10^3$\\
\rule[-2mm]{0mm}{5mm} & $\varepsilon_{mod}^3$ (\%) & $0.53$ & $0.53$\\ \hline
\end{tabular}
\end{center}
\caption{Physical parameters of the transversely isotropic media used in the numerical experiments. The phase velocities $c_{pf}(f_0,\varphi)$, $c_{ps}(f_0,\varphi)$ and $c_{s}(f_0,\varphi)$ are computed at $f = f_0 = 200$ kHz when the wavevector $\mbox{\boldmath$k$}$ makes an angle $\varphi$ with the horizontal $x$-axis, and $c_{pf}^{\infty}(\varphi)$, $c_{ps}^{\infty}(\varphi)$, $c_{s}^{\infty}(\varphi)$ denote the high-frequency limit of the phases velocities.}
\label{table:para_phy_ani}
\end{table}
In all the numerical simulations, the time step is computed from the physical parameters of the media through relations (\ref{eq:CFL_dsplit}), setting the CFL number $\Upsilon=0.95$. The numerical experiments are performed on an Intel Core i7 processor at $2.80$ GHz.

In the first test and the third test, the computational domain $[-0.15,0.15]^2$ m is discretized with $N_x = N_z = 2250$ grid nodes in each direction, which amounts to 20 points per slow compressional wavelength in $\Omega_0$. In the other tests, the computational domain $[-0.1,0.1]^2$ m is discretized with $N_x= N_z = 1500$, which amounts also to 20 points per slow compressional wavelength in $\Omega_0$ and in $\Omega_1$.\\

%------------------------------------------------------------------------------------------

\noindent
{\it Test 1: homogeneous medium}\label{sec:exp:test1}

\begin{figure}[htbp]
\begin{center}
\begin{tabular}{cc}
$p$ & zoom\\
\hspace{-0.3cm}
\includegraphics[scale=0.36]{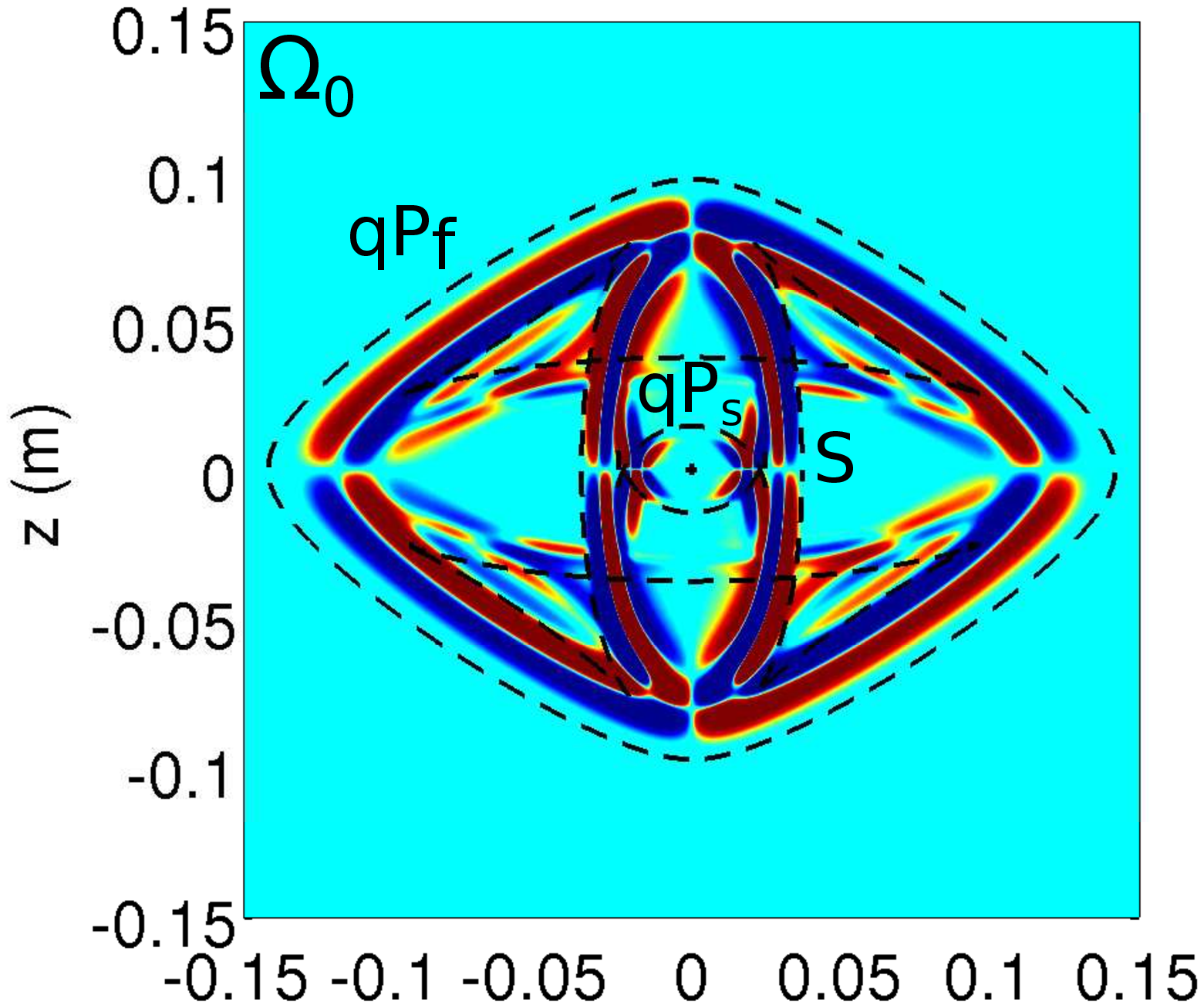} & 
\hspace{-0.3cm}
\includegraphics[scale=0.40]{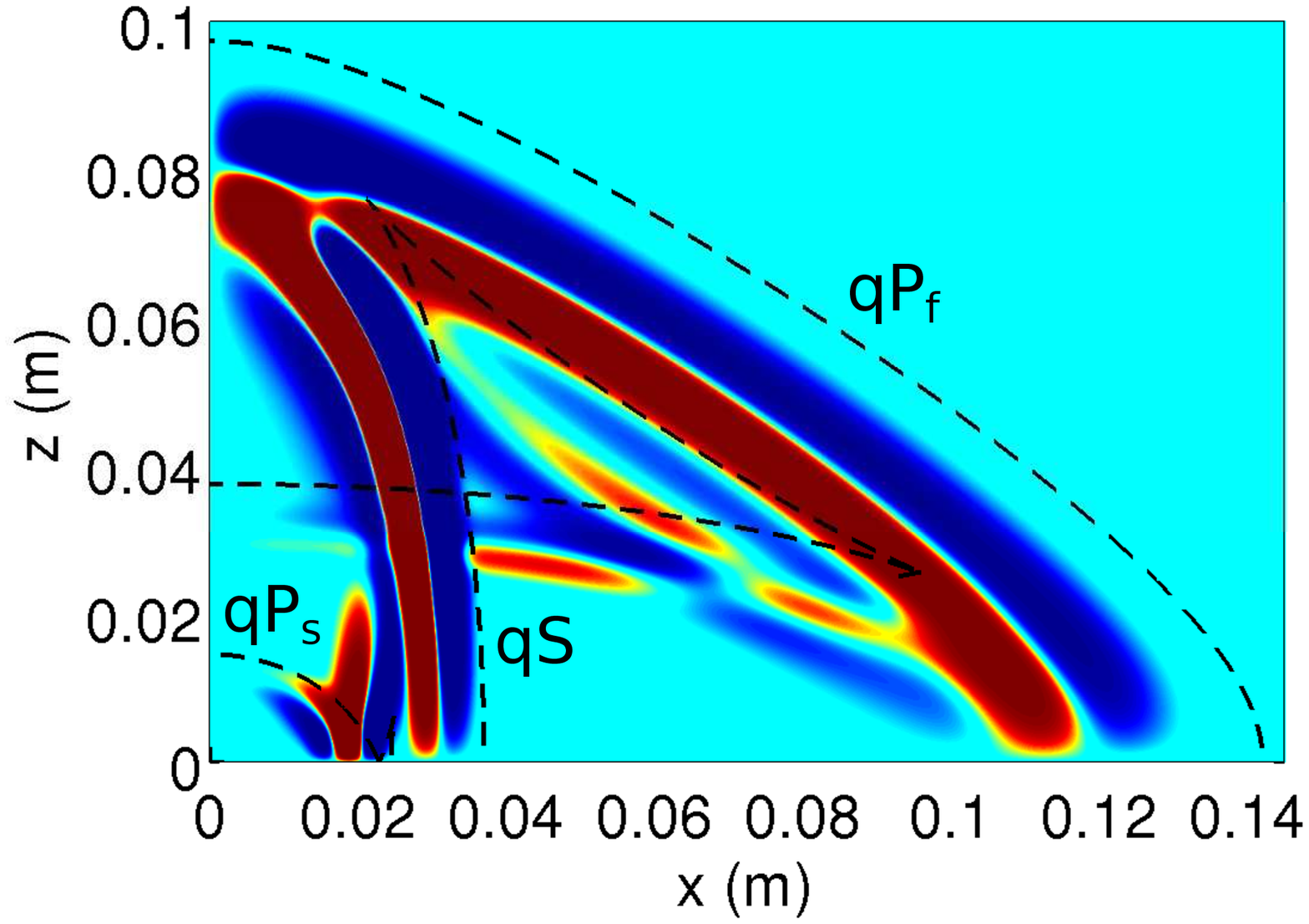}
\end{tabular}
\end{center}
\caption{test 1. Fast and slow quasi-compressional waves, respectively $qP_f$ and $qP_s$, and quasi-shear wave $qS$ emitted by a source point at $(0 \mbox{ m},0 \mbox{ m})$. Pressure at $t_1 \simeq 2.72\,10^{-5}$ s.}
\label{fig:2D_homogene_ani}
\end{figure}

In the first test, the homogeneous medium $\Omega_0$ (table \ref{table:para_phy_ani}) is excited by a source point located at $(0 \mbox{ m},0 \mbox{ m})$. The only non-null component of the forcing $\bfF$ (\ref{eq:forcing_DA}) is $G_{\sigma_{13}} = g(t)\,h(x,z)$, where $g(t)$ is a Ricker signal of central frequency $f_0$ and of time-shift $t_0=1/f_0=10^{-5}$ s:
\begin{equation}
g(t) =
\left\lbrace 
\begin{array}{ll}
\displaystyle \left( 2\,\pi^2\,f_0^2\,\left( t-t_0\right) ^2-1\right) \,\exp\left( -\pi^2\,f_0^2\,(t-t_0)^2\right) \; & \displaystyle \mbox{if}\;0\leqslant t\leqslant 2\,t_0,\\
[10pt]
\displaystyle 0 & \displaystyle \mbox{otherwise},
\end{array}
\right. 
\label{eq:ricker}
\end{equation}
and $h(x,z)$ is a truncated Gaussian centered at point $(0,0)$, of radius $R_0=6.56\,10^{-3}$ m and $\Sigma=3.28\,10^{-3}$ m:
\begin{equation}
h(x,z) =
\left\lbrace 
\begin{array}{ll}
\displaystyle \frac{1}{\pi\,\Sigma^2}\,\exp\left(-\frac{x^2+z^2}{\Sigma^2}\right) \; & \mbox{if}\;0\leqslant x^2+z^2\leqslant R_0^2,\\
[10pt]
\displaystyle 0 & \mbox{otherwise}.
\end{array}
\right. 
\label{eq:gaussienne}
\end{equation}
The time step is $\Delta t=2.41\,10^{-8}$ s. We use a truncated Gaussian for $h(x,z)$ rather than a Dirac distribution to avoid spurious numerical artifacts localized around the source point. This source generates cylindrical waves of all types: fast and slow quasi-compressional waves and quasi-shear waves, which are denoted by $qP_f$, $qP_s$ and $qS$, respectively, in figure \ref{fig:2D_homogene_ani}. The three waves are observed in the pressure field. Contrary to the isotropic case, where the pressure of the shear wave is null, pressure is visible in the qS wave.

A comparison is proposed with the theoretical wavefronts deduced from the dispersion analysis (section \ref{sec:phys:dispersion}) and the resolution of (\ref{eq:syst_dispersion_gen}). They are denoted by a black dotted line in figure \ref{fig:2D_homogene_ani}. It is observed that the computed waves are well positionned at the final instant $t_1 \simeq 2.72\,10^{-5}$ s (corresponding to 1125 time steps). No special care is applied to simulate outgoing waves (with PML, for instance), since the simulation is stopped before the waves have reached the edges of the computational domain. The cusp of the shear wave is seen in the numerical solution.\\

%------------------------------------------------------------------------------------------

\noindent
{\it Test 2: diffraction of a plane wave by a plane interface}\label{sec:exp:test2}

\begin{figure}[htbp]
\begin{center}
\begin{tabular}{cc}
(a) & (b)\\
\includegraphics[scale=0.40]{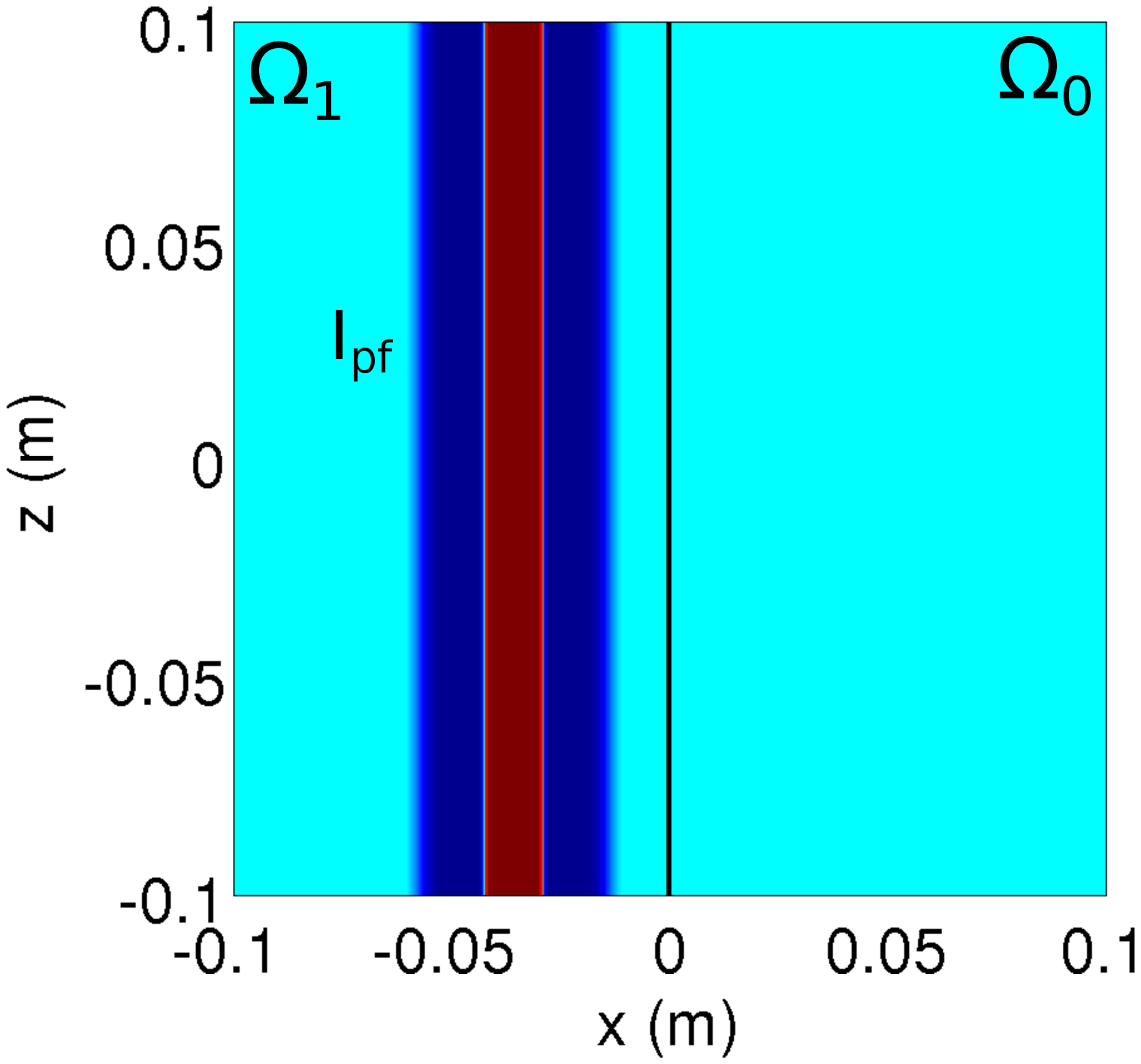} & 
\includegraphics[scale=0.40]{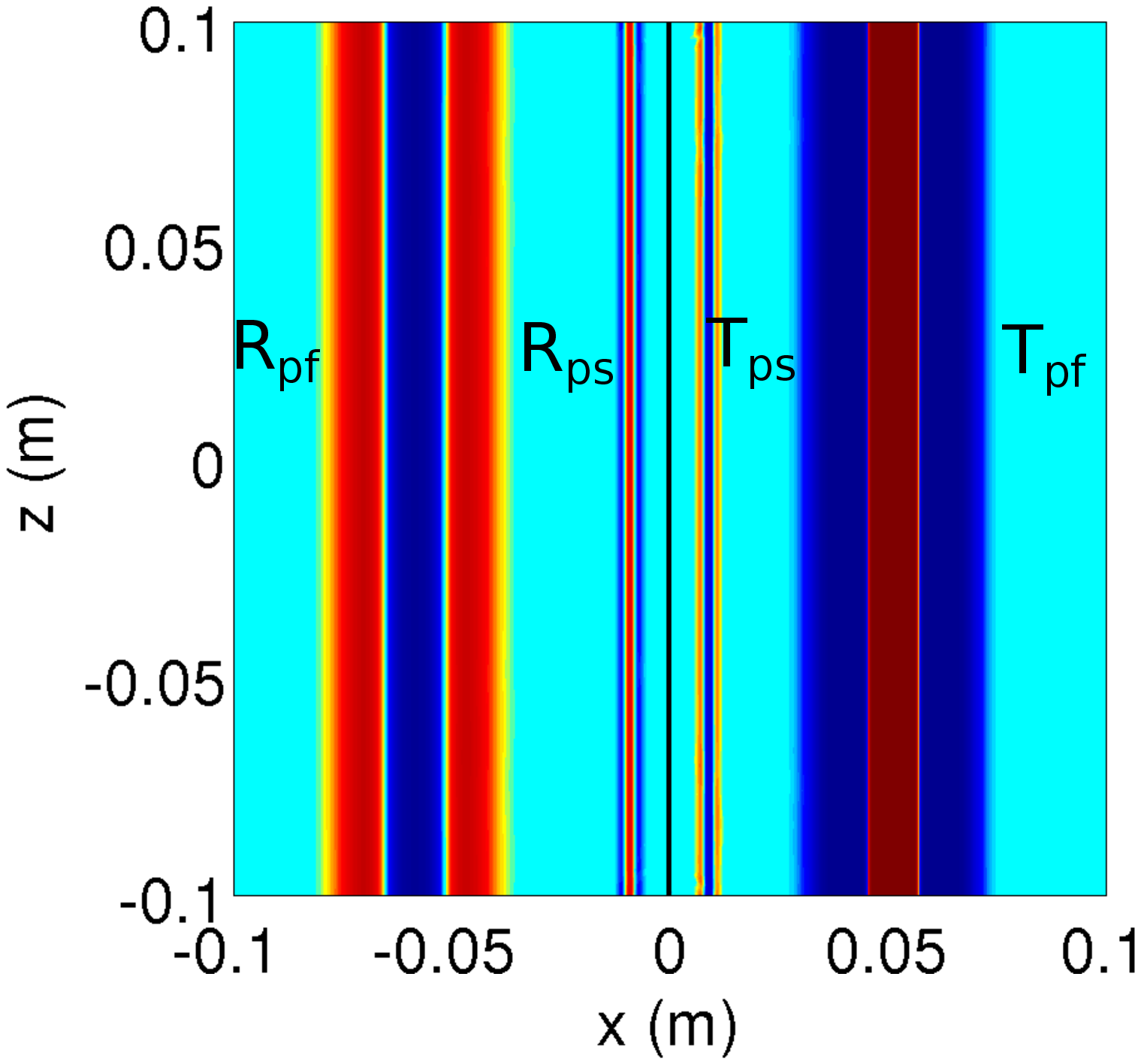}\\
\end{tabular}
\end{center}
\caption{test 2. Snapshot of pressure at initial time (a) and at $t_1 \simeq 1.48\,10^{-5}$ s (b). The plane interface is denoted by a straight black line, separating $\Omega_1$ (on the left) and $\Omega_0$ (on the right).}
\label{fig:validation_interface_ani_carte}
\end{figure}

\begin{figure}[htbp]
\begin{center}
\begin{tabular}{cc}
Pressure & Zoom on the slow compressional waves\\
\includegraphics[scale=0.33]{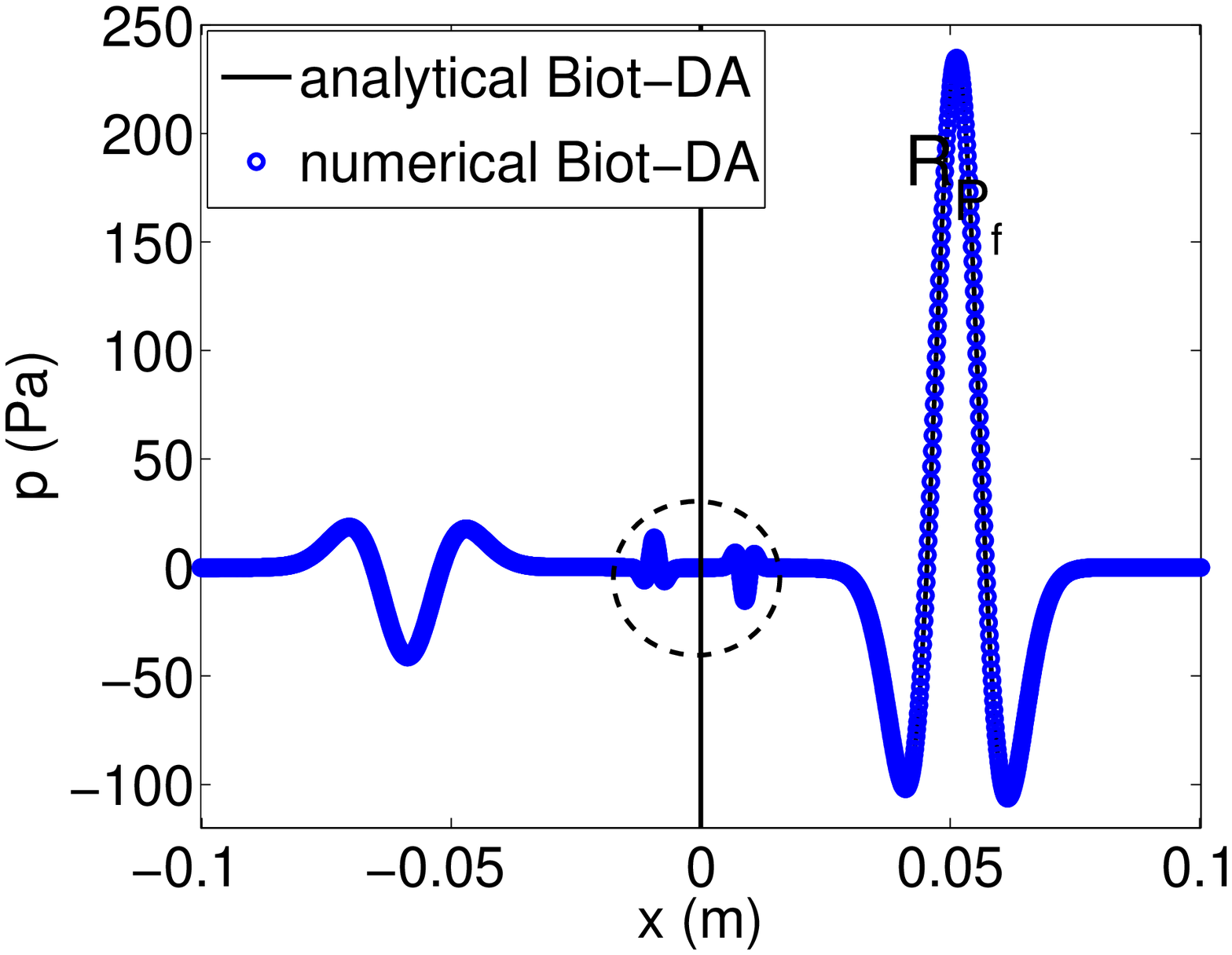} & 
\includegraphics[scale=0.33]{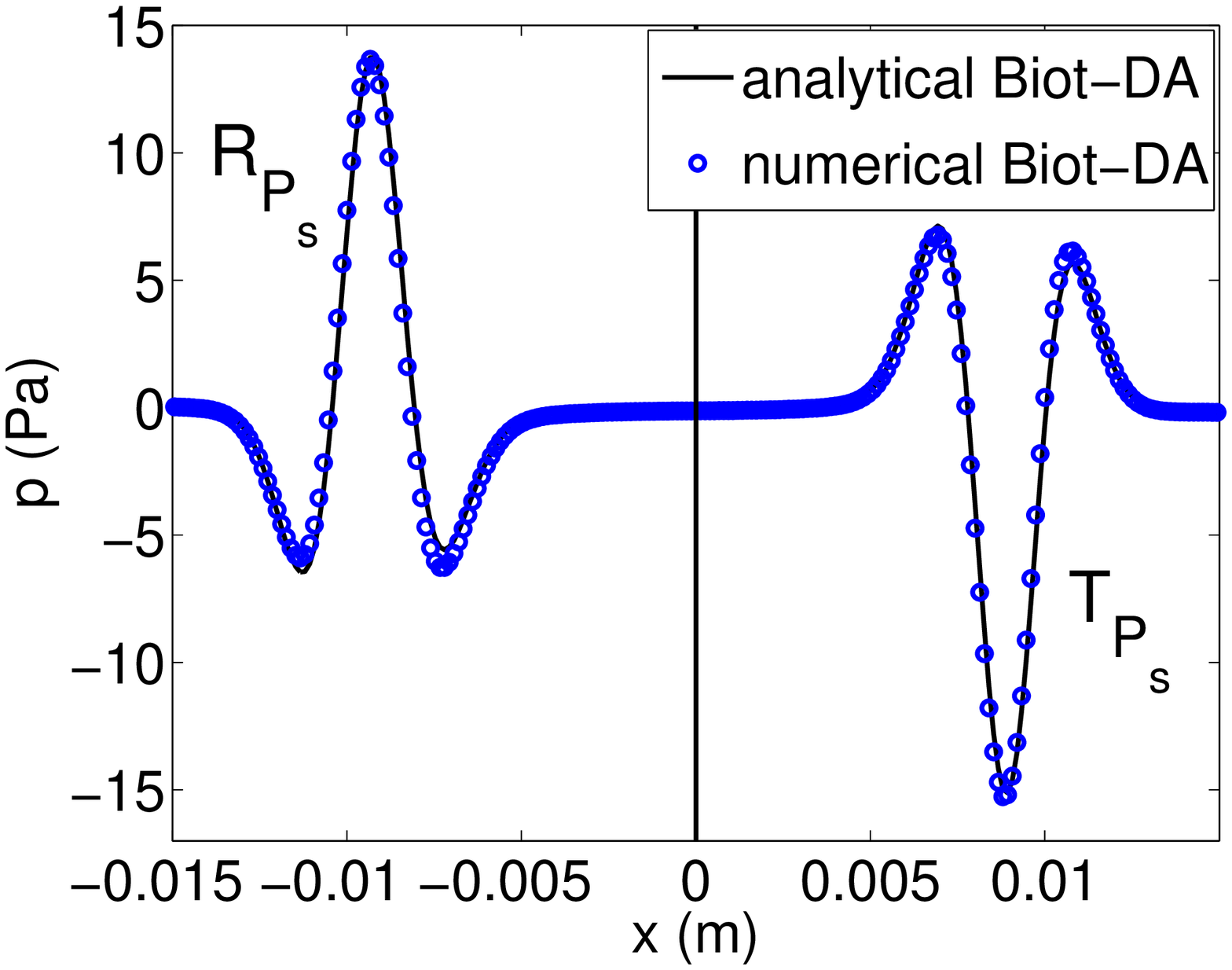}\\
\end{tabular}
\end{center}
\caption{test 2. Pressure along the line $z=0$ m; vertical line denotes the interface. Comparison between the numerical values (circle) and the analytical values (solid line) of $p$ at $t_1 \simeq 1.48\,10^{-5}$ s.}
\label{fig:validation_interface_ani_coupe}
\end{figure}

In the second test, the source is a plane fast compressional wave traveling in the positive direction of the $x$-axis, whose wavevector $\mbox{\boldmath$k$}$ is parallel to the direction of propagation. Its time evolution is the same Ricker signal as in the first test (\ref{eq:ricker}). We use periodic boundary conditions at the top and at the bottom of the domain. The validity of the method is checked in the particular case of heterogeneous transversely isotropic media, where a semi-analytical solution can be obtained easily. The media $\Omega_0$ and $\Omega_1$ are separated by a vertical wave plane interface at $x = 0$ m. The incident $P_f\,$-wave ($Ip_f$) propagates in the medium $\Omega_1$. The time step is $\Delta t=2.11\,10^{-8}$ s. The figure \ref{fig:validation_interface_ani_carte} shows a snapshot of the pressure at $t_1 \simeq 1.48\,10^{-5}$ s (corresponding to 750 time steps), on the whole computational domain. The reflected fast and slow quasi-compressional waves, denoted respectively $Rp_f$ and $Rp_s$, propagate in the medium $\Omega_1$; and the transmitted fast and slow quasi-compressional waves, denoted respectively $Tp_f$ and $Tp_s$, propagate in the medium $\Omega_0$.
In this case, we compute the exact solution of Biot-DA thanks to Fourier tools and poroelastic equations; a general overview of the analytical solution is given in the \ref{SecAppExact}. The figure \ref{fig:validation_interface_ani_coupe} shows the excellent agreement between the analytical and the numerical values of the pressure along the line $z=0$ m. Despite the relative simplicity of this configuration (1D evolution of the waves and lack of shear waves), it can be viewed as a validation of the numerical method which is fully 2D whatever the geometrical setting.\\

%------------------------------------------------------------------------------------------

\noindent
{\it Test 3: source point and plane interface or sinusoidal interface}\label{sec:exp:test3}

\begin{figure}[htbp]
\begin{center}
\begin{tabular}{cc}
(a) & (b)\\
\includegraphics[scale=0.37]{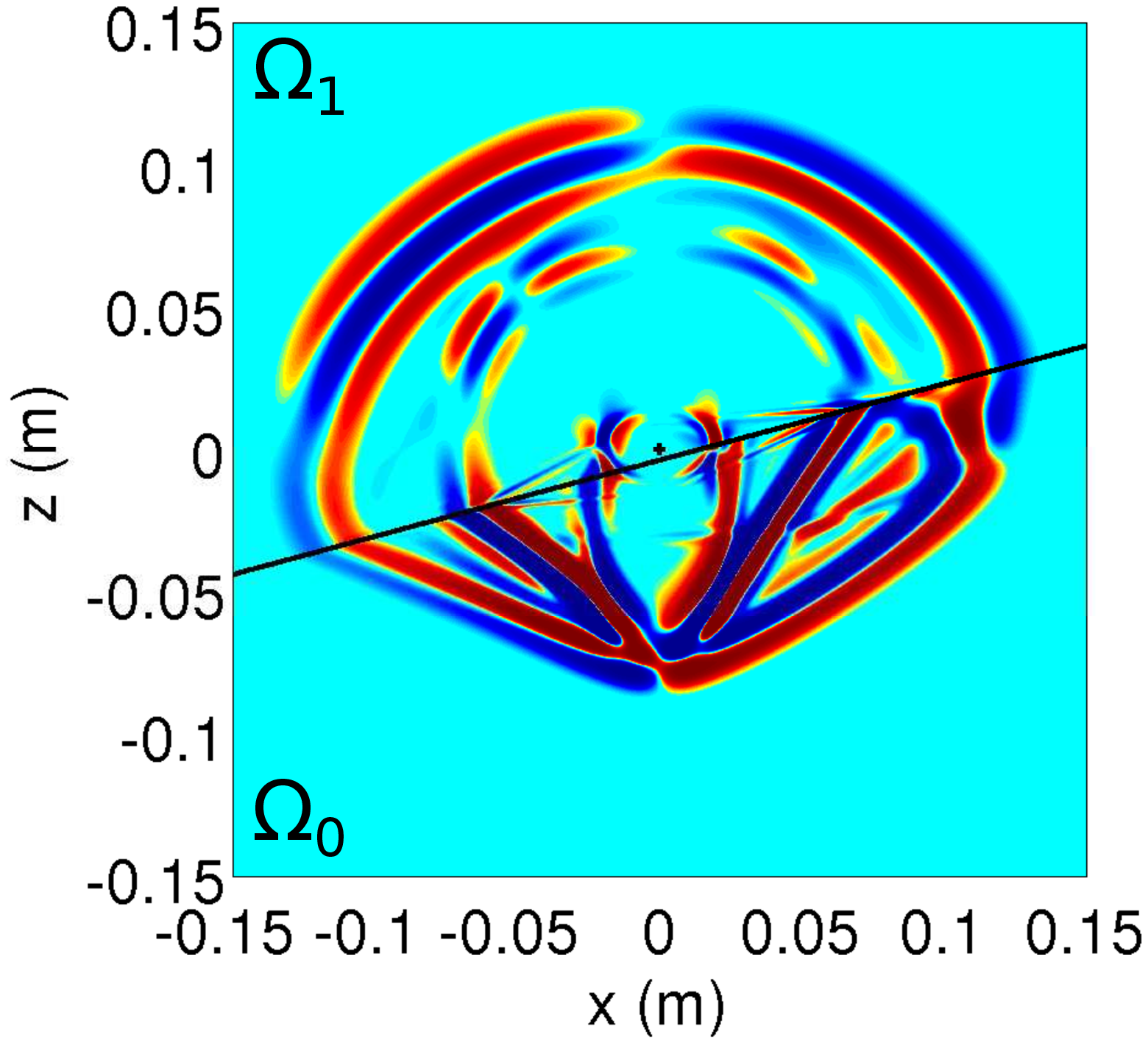} & 
\includegraphics[scale=0.37]{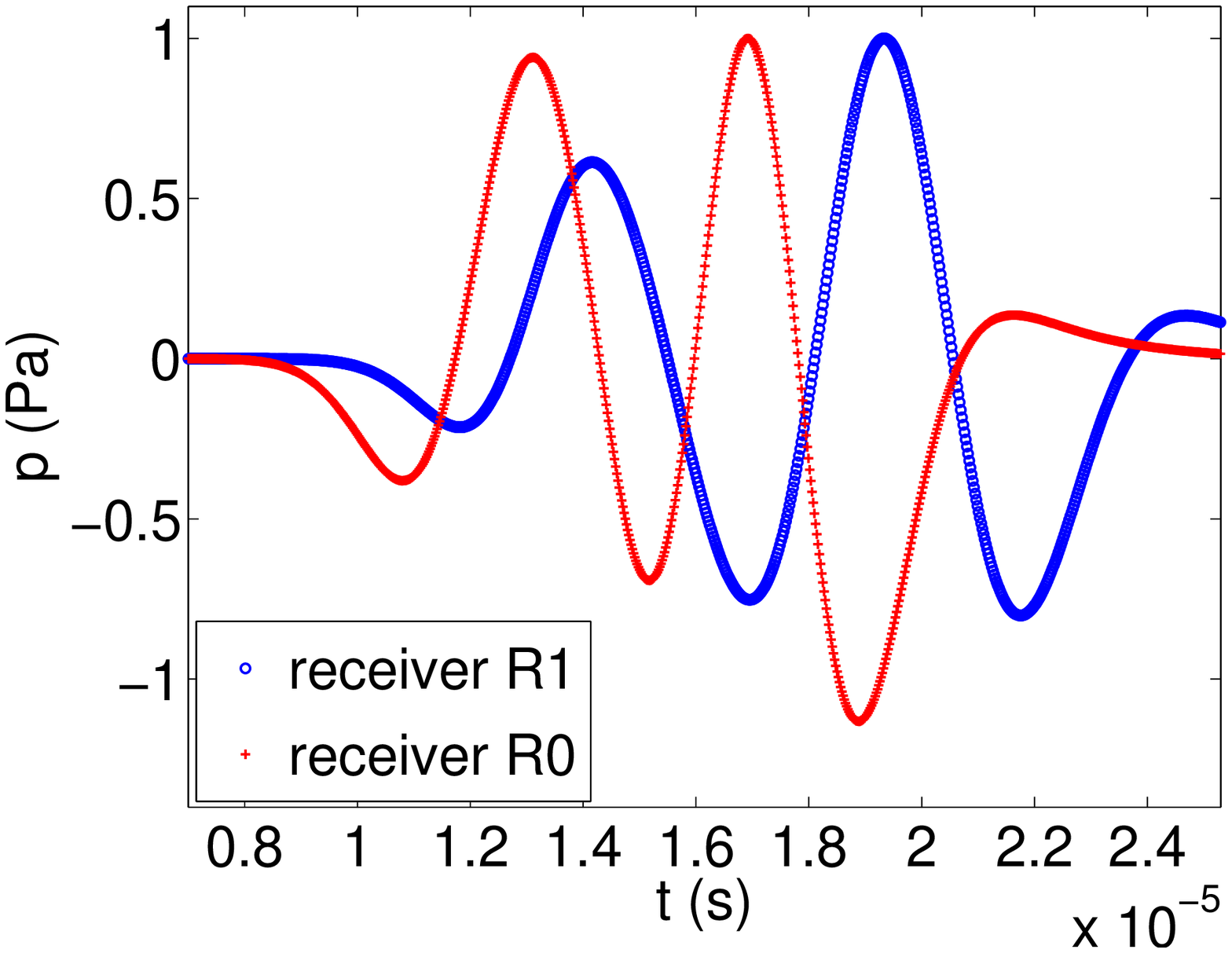}
\end{tabular}
\end{center}
\caption{test 3. Snapshot of pressure and at $t_1 \simeq 2.53\,10^{-5}$ s (a). The plane interface separating the media $\Omega_0$ and $\Omega_1$ is denoted by a straight black line. Time evolution of the pressure (b) at receiver $R_0$ in $\Omega_0$ (red) and at receiver $R_1$ in $\Omega_1$ (blue).}
\label{fig:droite_incline}
\end{figure}

In the previous test, the configuration was fully 1D, but more complex geometries can be handled on a Cartesian grid thanks to the immersed interface method. As an example, the media $\Omega_0$ and $\Omega_1$ are separated by a plane interface with slope $15$ degree with the horizontal $x$-axis, passing through the point $(0 \mbox{ m},-0.004 \mbox{ m})$. The homogeneous medium $\Omega_1$ is excited by the source point described in test 1. This source emits cylindrical waves which interact with the medium $\Omega_0$. The time step is $\Delta t = 2.11\,10^{-8}$ s. Snapshot of the pressure at time $t \simeq 2.53\,10^{-5}$ s (corresponding to 1200 time steps) and the time evolution of the pressure at receivers $R_0$ $(0.042 \mbox{ m},0.0068 \mbox{ m})$ in $\Omega_0$ and $R_1$ $(0.048 \mbox{ m},0.0071 \mbox{ m})$ in $\Omega_1$ are represented on figure \ref{fig:droite_incline}.

\begin{figure}[htbp]
\begin{center}
\begin{tabular}{cc}
(a) & (b)\\
\includegraphics[scale=0.37]{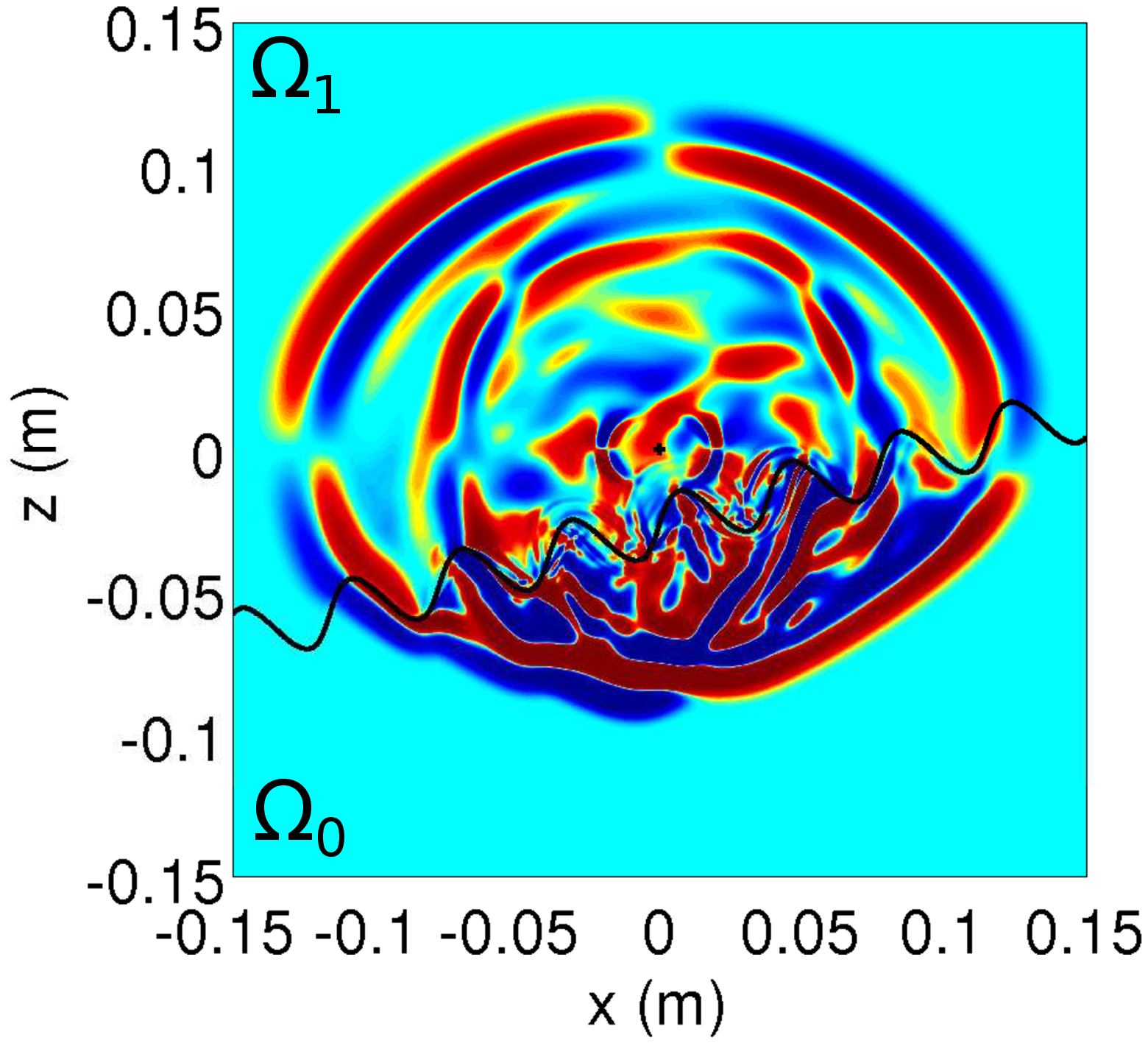} &
\includegraphics[scale=0.37]{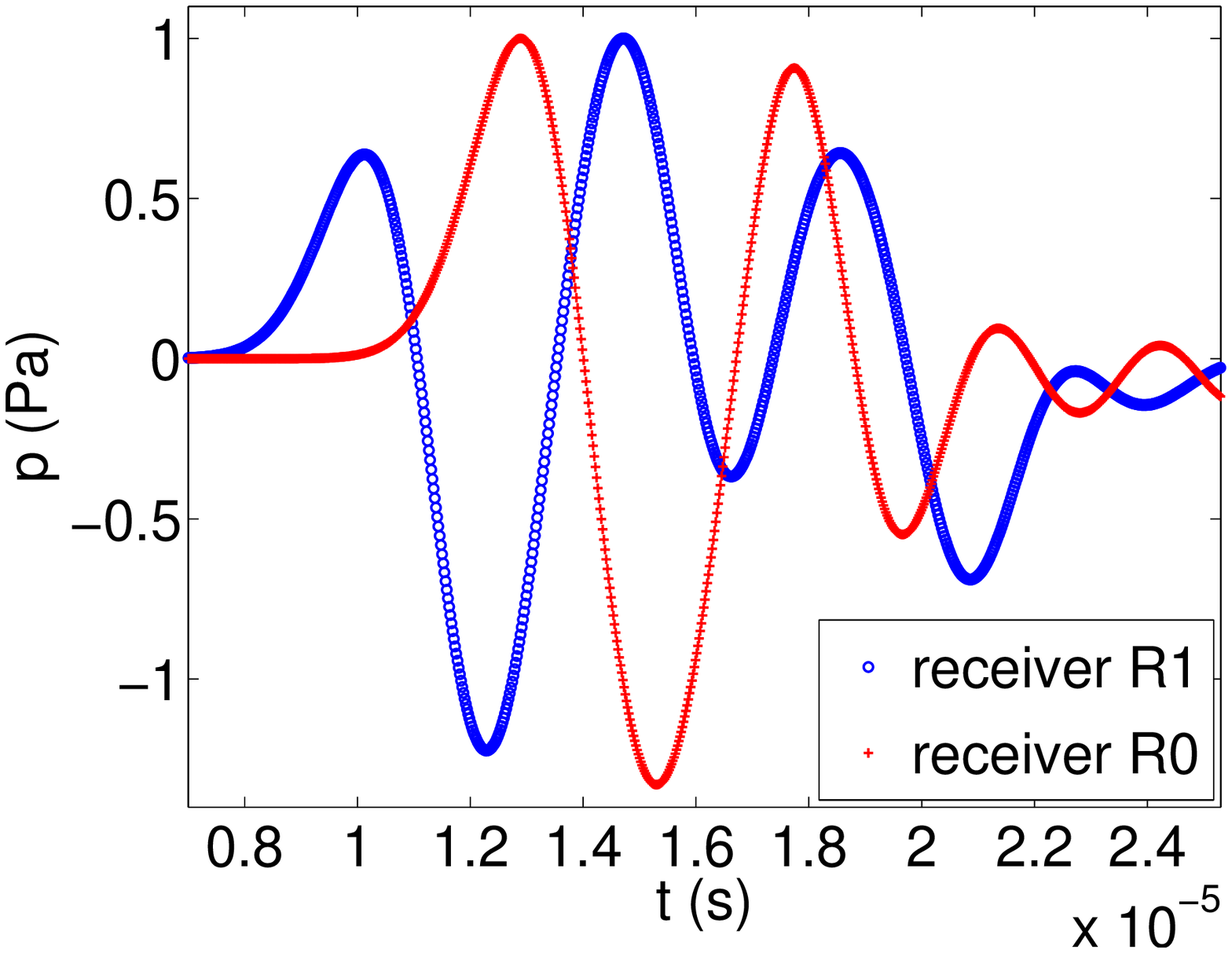}
\end{tabular}
\end{center}
\caption{test 3. Snapshot of pressure and at $t_1 \simeq 2.53\,10^{-5}$ s (a). The sinusoidal interface separating the media $\Omega_0$ and $\Omega_1$ is denoted by a straight black line. Time evolution of the pressure (b) at receiver $R_0$ in $\Omega_0$ (red) and at receiver $R_1$ in $\Omega_1$ (blue).}
\label{fig:sinus_incline}
\end{figure}

The plane interface can be easily replaced, for instance by a sinusoidal interface of equation
\begin{equation}
-\sin\theta\,(x-x_s) + \cos\theta\,(z-z_s) - A_s\,\sin\left(\omega_s\,\left(\cos\theta\,(x-x_s) + \sin\theta\,(z-z_s)\right)\right) = 0,
\label{eq:sinus}
\end{equation}
with $x_s = 0$ m, $z_s = -0.027$ m, $A_s = 0.01$ m, $\omega_s = 50\,\pi$ rad/s, $\theta = \pi/12$ rad. Snapshot of the pressure and the time evolution of the pressure at receivers $R_0$ $(0.036 \mbox{ m},-0.031 \mbox{ m})$ in $\Omega_0$ and $R_1$ $(0.036 \mbox{ m},-0.014 \mbox{ m})$ in $\Omega_1$ are represented on figure \ref{fig:sinus_incline}.
In both cases, no spurious diffraction is induced by the stair-step representation of the interface, thanks to the immersed interface method. Moreover, classical waves conversions and scattering phenomena are observed: reflected, transmitted and Stoneley waves. The shape of the transmitted waves, not circular, illustrates the strong anisotropy of the medium $\Omega_0$.\\

%------------------------------------------------------------------------------------------

\noindent
{\it Test 4: multiple ellipsoidal scatterers}\label{sec:exp:test4}

To illustrate the ability of the proposed numerical strategy to handle  complex geometries, $200$ ellipsoidal scatterers of medium $\Omega_1$, with major and minor radii of $0.025$ m and $0.02$ m, are randomly distributed in a matrix of medium $\Omega_0$. The computational domain is $[-0.8,0.8]^2$ m, hence the surfacic concentration of scatterers is $25$ \%. A uniform distribution of scatterers is used. The source is the same plane fast compressional wave than is the second test, and we use periodic boundary conditions at the top and at the bottom of the domain.

\begin{figure}[htbp]
\begin{center}
\begin{tabular}{cc}
(a) & (b)\\
\includegraphics[scale=0.42]{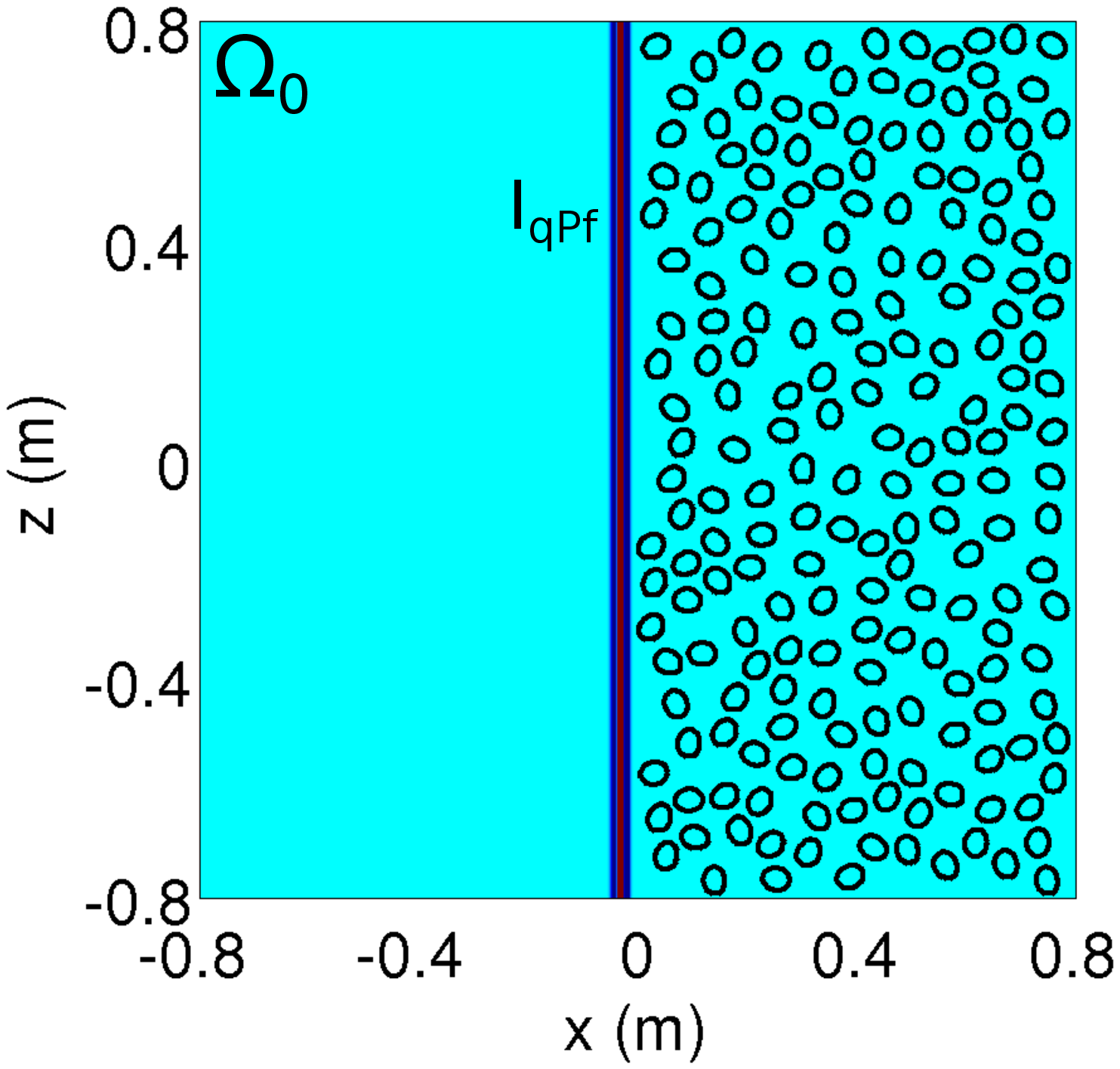}&
\includegraphics[scale=0.42]{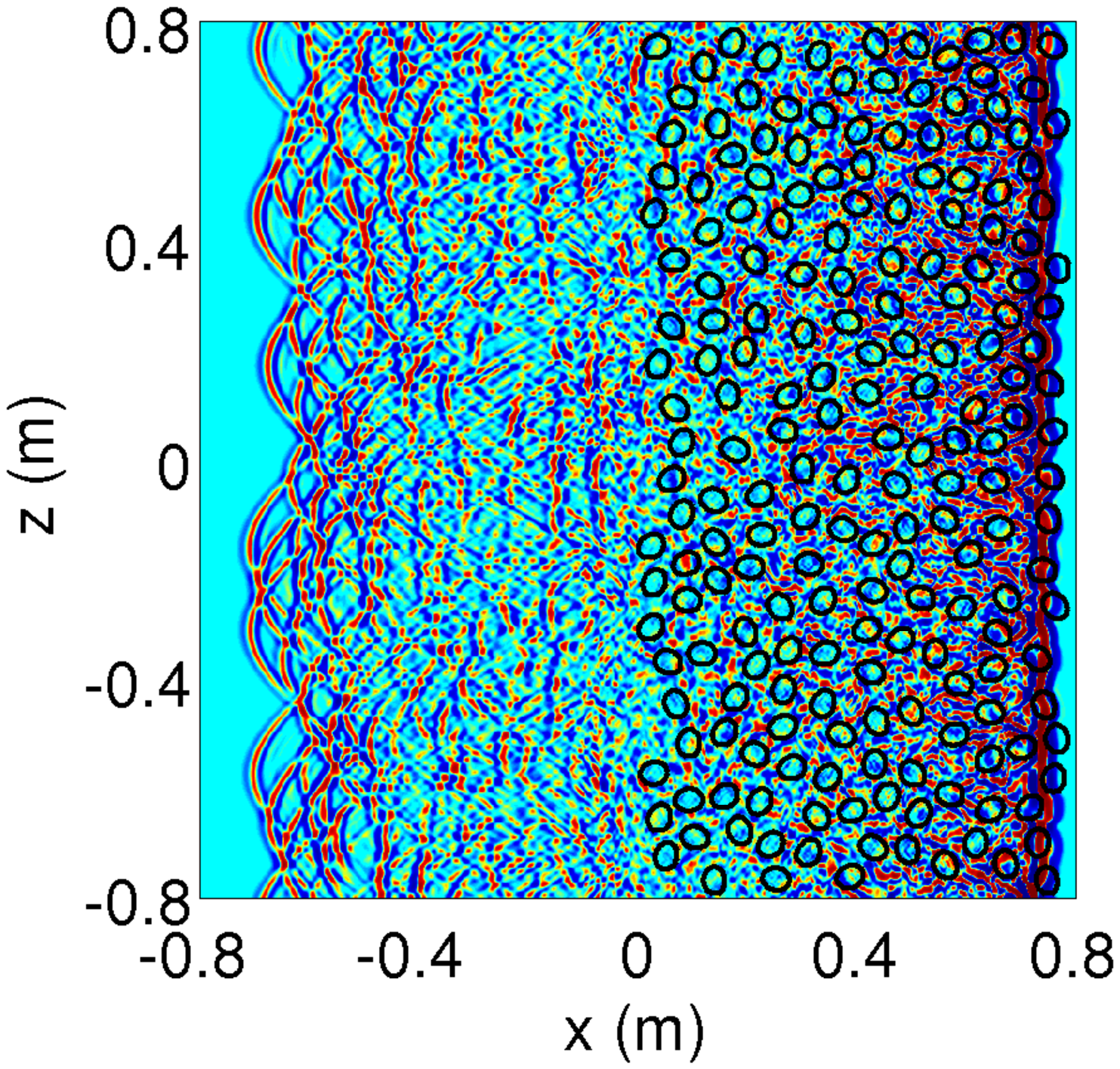}
\end{tabular}
\end{center}
\caption{test 4. Multiple scattering in random media. Snapshot of $p$ at the initial time (a) and at time $t_1 \simeq 1.43\,10^{-4}$ s. The matrix is $\Omega_0$, whereas the $200$ scatterers are $\Omega_1$.}
\label{fig:ellipses_ani}
\end{figure}

The time step is $\Delta t=3.37\,10^{-7}$ s. The pressure is represented at the initial time and at time $t_1 \simeq 1.43\,10^{-4}$ s (corresponding to 425 time steps) on figure \ref{fig:ellipses_ani}. This simulation has taken approximately $11.5$ h of preprocessing and $8.5$ h of time-stepping. Similar numerical experiments are also performed for a surfacic concentration of scatterers of $10$ \% and $15$ \%.

\begin{figure}[htbp]
\begin{center}
\begin{tabular}{cc}
(a) & (b)\\
\includegraphics[scale=0.35]{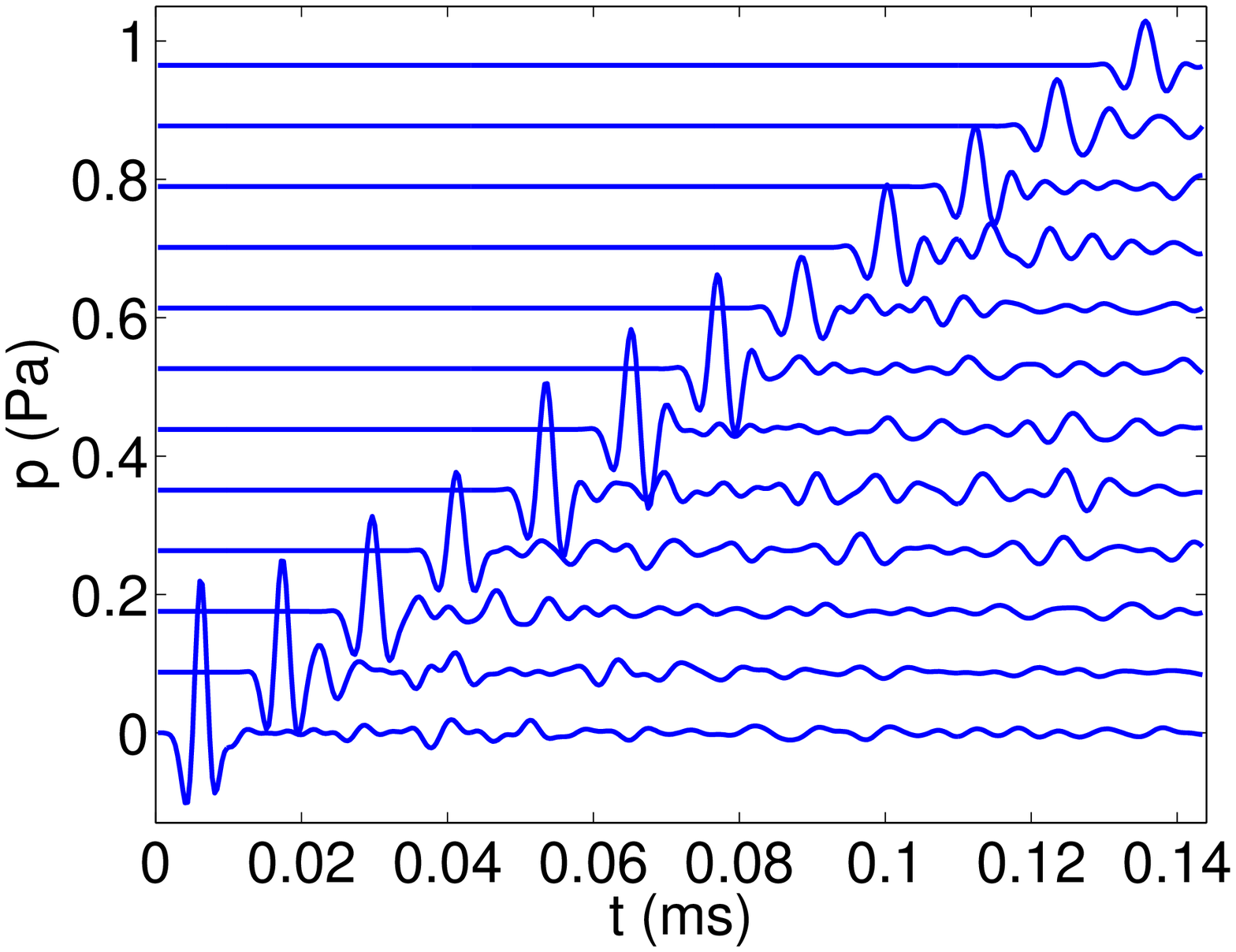} & 
\includegraphics[scale=0.35]{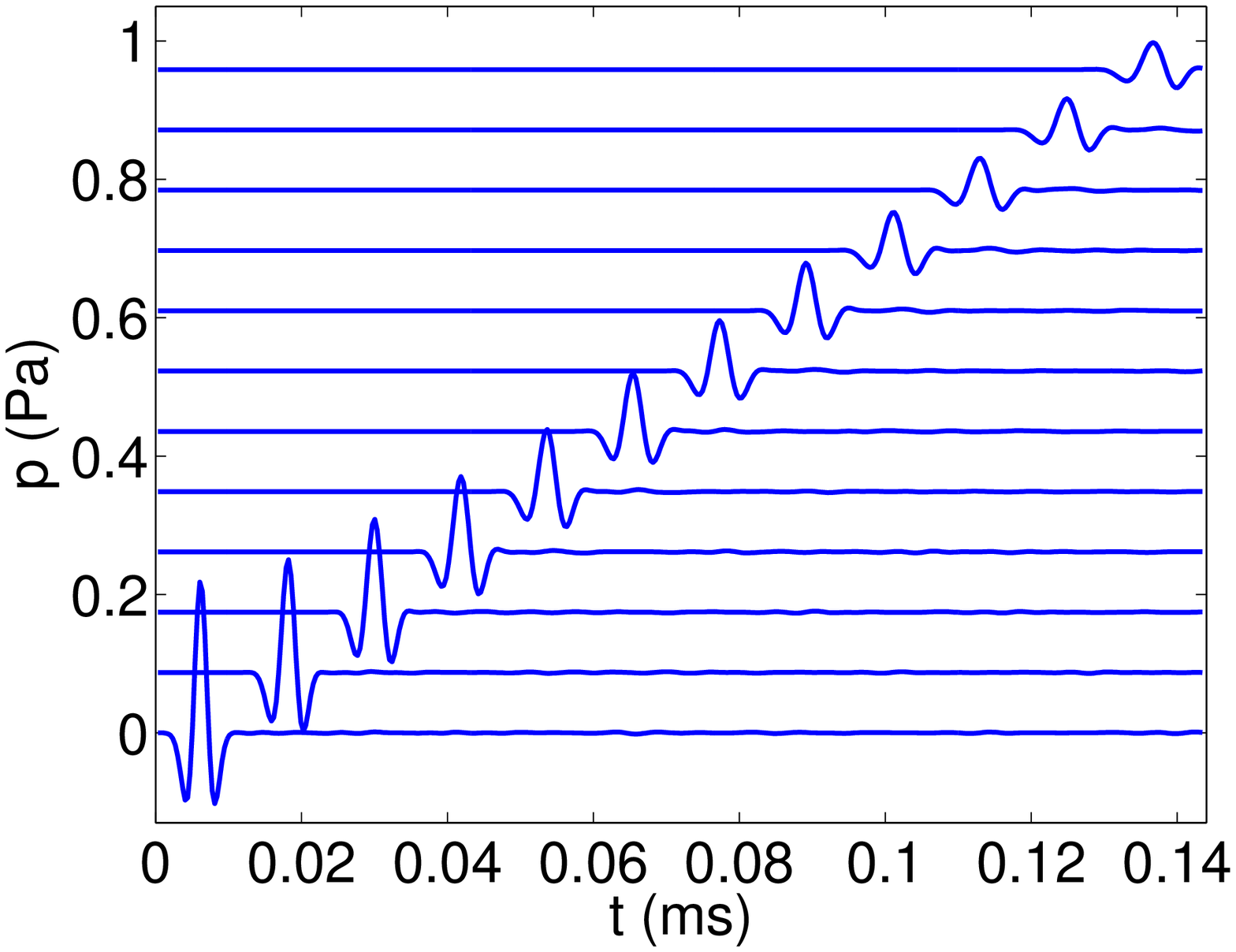}
\end{tabular}
\end{center}
\caption{test 4. Incident plane $qP_s$-wave in a medium with $25\%$ inclusion concentration. (a): pressure recorded along an array, (b): coherent pressure obtained afer summation.}
\label{fig:sismo}
\end{figure}

\begin{figure}[htbp]
\begin{center}
\begin{tabular}{cc}
(a) & (b)\\
\includegraphics[scale=0.51]{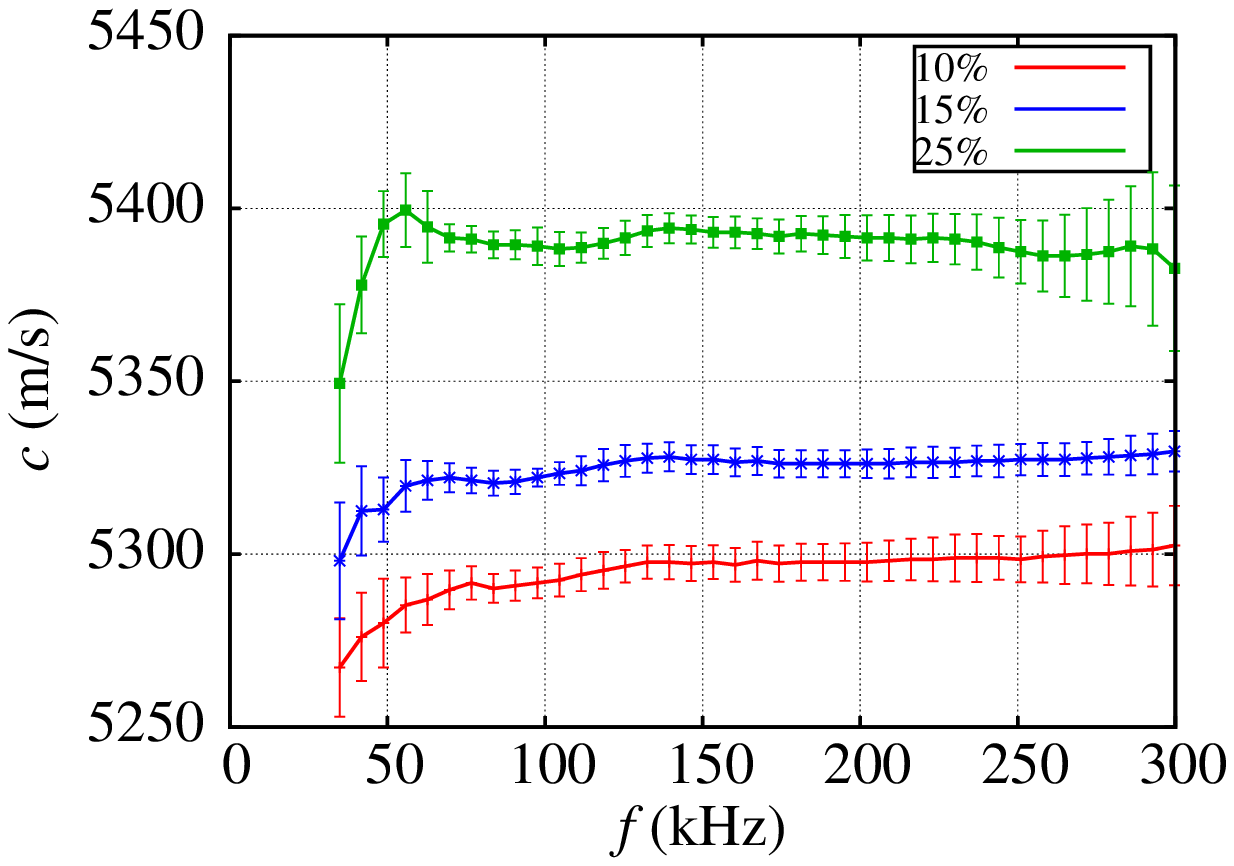} & 
\includegraphics[scale=0.51]{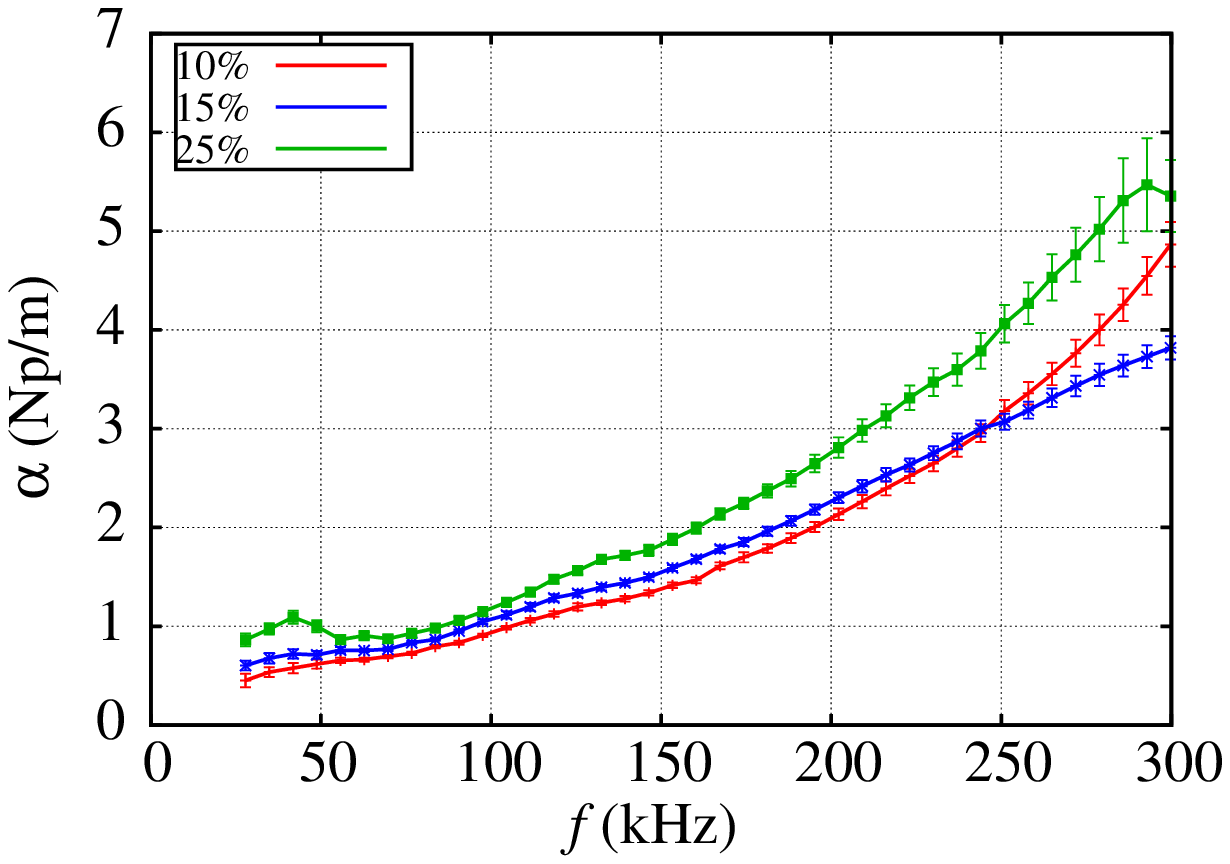}
\end{tabular}
\end{center}
\caption{test 4. Effective phase velocity (a) and effective attenuation (b) at various inclusion concentrations. The vertical lines represents the error bars. The horizontal lines in (a) give the average phase velocity weighted by the concentration.}
\label{fig:effective_properties}
\end{figure}

At each time step, the components of $\bfU_{ij}^n$ are stored inside the subdomain containing the inclusions. For this purpose, a uniform network consisting of $N_l = 800$ lines and $N_c = 25$ columns of receivers is put in the subdomain. The position of the receivers is given by $(x_i,z_j)$, where $i=0,\cdots,N_c - 1$ and $j=0,\cdots,N_l - 1$. The field $\bfU_{ij}^n$ recorded on each array (each line of receivers), represented on figure \ref{fig:sismo}-a, corresponds to a field propagating  along one horizontal line of receivers. A main wave train is clearly visible, followed by a coda. Summing the time histories of these $N_c$ arrays gives a coherent field propagating in the $x$ direction:
\begin{equation}
\overline{\bfU}_i^n = \frac{1}{N_l}\,\sum\limits_{j=0}^{N_l - 1}\bfU_{ij}^n.
\label{eq:coherent_field}
\end{equation}
On the coherent seismogram thus obtained, represented on figure \ref{fig:sismo}-b, the coda has disappeared, and the main wave train behaves like a plane wave propagating in a homogeneous (but dispersive and attenuating) medium. The coherent phase velocity $c(\omega)$, represented in figure \ref{fig:effective_properties}-a, is computed by applying a $\mathfrak{p}-\omega$ transform to the space-time data on the coherent field (\ref{eq:coherent_field}), where $\mathfrak{p} = 1/c$ is the slowness of the waves \cite{MECHAN81,MOKHTAR88}. The horizontal lines represent a simple average of the phase velocities weighted by the concentration. The coherent attenuation $\alpha(\omega)$ is estimated from the decrease in the amplitude spectrum of the coherent field during the propagation of the waves, see \ref{fig:effective_properties}-b. An error estimate is also deduced, represented in figure \ref{fig:effective_properties} by vertical lines.

%------------------------------------------------------------------------------------------
%------------------------------------------------------------------------------------------

\section{Conclusion}\label{sec:conclu}

An explicit finite-difference method has been developed here to simulate transient poroelastic waves in the full range of validity of the Biot-JKD model, which involves order $1/2$ fractional derivatives. A diffusive representation transforms the fractional derivatives, non-local in time, into a continuum of local problems, approximated by quadrature formulae. The Biot-JKD model is then replaced by an approximate Biot-DA model, much more tractable numerically. The coefficients of the diffusive approximation are determined by a nonlinear constrained optimization procedure, leading to a small number of memory variables. The hyperbolic Biot-DA system of partial differential equations is discretized using various tools of scientific computing: Strang splitting, fourth-order ADER scheme, immersed interface method. It enables to treat efficiently and accurately the propagation of transient waves in transversely isotropic porous media.

Some future lines of research are suggested:
\begin{itemize}
\item \emph{Multiple scattering.} Many theoretical methods of multiple scattering have been developed to determine the effective wavenumber of media with random scatterers; see for instance the Independent Scattering Approximation and the Waterman-Truell method \cite{WATERMAN61}. The main drawback of these methods is that their validity is restricted to small concentrations of scatterers, typically less than 10 \%. On the contrary, numerical methods do not suffer from such a limitation if suitable efforts are done. In particular, the errors due to the discretization (numerical dispersion, numerical dissipation, spurious diffractions on interfaces, ...) must be much smaller than the physical quantities of interest. In \cite{CHEKROUN12}, numerical simulations were used in the elastic case to estimate the accuracy of standard theoretical models, and also to show the improvement induced by recent models of multiple scattering \cite{CONOIR10}. As shown in test 4 of $\S$ \ref{sec:exp}, the numerical tools presented here make possible a similar study poroelastic random media and comparisons with theoretical models  \cite{TOURNAT04,LUPPE08}. 

However, realistic configurations would involve approximately $1500$ scatterers, and sizing of the experiments leads to $N_x \times N_z = 10000^2$, and $10000$ time iterations are required. Consequently, the numerical method has to be parallelized, for instance by Message Passing Interface (MPI).
\item \emph{Thermic boundary-layer.} In cases where the saturating fluid is a gas, the effects of thermal expansion of both pore fluid and the matrix have to be taken into account. In the HF regime, the thermal exchanges between fluid and solid phase occur in a small layer close to the surface of the pores. In this case, the dynamic thermal permeability is introduced \cite{LAFARGE97}, leading in the time-domain to an additional shifted fractional derivative of order $1/2$. The numerical method developed in this paper can be applied without difficulty by introducing additional memory variables.
\item \emph{Fractional derivatives in space.} The Biot theory is very efficient to predict the macroscopic behavior of long-wavelength sound propagation in porous medium with relatively simple microgeometries. However, it remains far to describe correctly the coarse-grained dynamics of the medium when the microgeometry of the porous medium become more complex, for instance fractal. For rigid-framed porous media permeated by a viscothermal fluid, a generalized macroscopic nonlocal theory of sound propagation has been developed to take into account not only temporal dispersion, but also spatial dispersion \cite{NEMATI12}. In this case, the coefficients depends on the frequency and on the wavenumber. In the space-time domain, it introduces not only time-fractional derivatives, but also space-fractional derivatives. Numerical modeling of space-fractional differential equations has been addressed by several authors \cite{LIU04,TADJERAN06}, by using a Gr\"unwald-Letnikov approximation. The diffusive approximation of such derivatives constitutes an interesting challenge.
\end{itemize}

%------------------------------------------------------------------------------------------
%-----------------------------------------------------------------------------------------

\section*{Acknowledgments}

The authors wish to thank Dr Mathieu Chekroun (LAUM, France) for his insights about multiple scattering and for computing the coherent phase velocity and attenuation with the $\mathfrak{p}-\omega$ transform in test 4.

%------------------------------------------------------------------------------------------
%------------------------------------------------------------------------------------------

\begin{appendix}

\section{Proof of proposition \ref{prop:nrjJKD}} \label{annexe:proof_nrj}

The equation (\ref{eq:biot_dynamique_a}) is multiplied by $\bfVs^T$ and integrated
\begin{equation}
\int_{\mathbb{R}^2}\left( \rho\,\bfVs^T\,\frac{\partial\,\bfVs}{\partial\,t} + \rho_f\,\bfVs^T\,\frac{\partial\,\bfW}{\partial\,t} - \bfVs^T\,(\nabla.\underline{\bfSigma})\right)\,dx\,dz = 0.
\label{eq:nrj_proof1}
\end{equation}
The first term in (\ref{eq:nrj_proof1}) is written
\begin{equation}
\int_{\mathbb{R}^2}\rho\,\bfVs^T\,\frac{\partial\,\bfVs}{\partial\,t}\,dx\,dz = \frac{d}{dt}\,\frac{1}{2}\,\int_{\mathbb{R}^2}\rho\,\bfVs^T\,\bfVs\,dx\,dz.
\label{eq:nrj_proof2}
\end{equation}
Integrating by part and using (\ref{eq:biot_comportement_bis}), we obtain
\begin{equation}
\begin{array}{l}
\displaystyle -\int_{\mathbb{R}^2}\bfVs^T\,(\nabla.\underline{\bfSigma})\,dx\,dz \displaystyle = \int_{\mathbb{R}^2}\bfSigma^T\,\frac{\partial\,\bfEps}{\partial\,t}\,dx\,dz = \displaystyle \int_{\mathbb{R}^2}\bfSigma^T\,\left(\bfC^{-1}\,\frac{\partial\,\bfSigma}{\partial\,t} - \bfC^{-1}\,\bfBeta\,\frac{\partial\,p}{\partial\,t}\right)\,dx\,dz,\\
[12pt]
%\hspace{0.6cm} \displaystyle = \int_{\mathbb{R}^2}\bfSigma^T\,\left(\bfC^{-1}\,\frac{\partial\,\bfSigma}{\partial\,t} - \bfC^{-1}\,\bfBeta\,\frac{\partial\,p}{\partial\,t}\right)\,dx\,dz,\\
%[12pt]
\hspace{0.6cm} \displaystyle = \frac{d}{dt}\,\frac{1}{2}\, \int_{\mathbb{R}^2}\bfSigma^T\,\bfC^{-1}\,\bfSigma\,dx\,dz + \int_{\mathbb{R}^2}\bfSigma^T\,\bfC^{-1}\,\bfBeta\,\frac{\partial\,p}{\partial\,t}\,dx\,dz,\\
[15pt]
\hspace{0.6cm} \displaystyle = \frac{d}{dt}\,\frac{1}{2}\,\int_{\mathbb{R}^2}\left(\bfSigma^T\,\bfC^{-1}\,\bfSigma + 2\,\bfSigma^T\,\bfC^{-1}\,\bfBeta\,p\right)\,dx\,dz - \int_{\mathbb{R}^2}\left( \frac{\partial\,\bfSigma}{\partial\,t}\right) ^T\,\bfC^{-1}\,\bfBeta\,p\,dx\,dz.
\end{array}
\label{eq:nrj_proof3}
\end{equation}
Equation (\ref{eq:biot_dynamique_b}) is multiplied by $\bfW^T$ and integrated
\begin{equation}
\begin{array}{l}
\displaystyle \int_{\mathbb{R}^2} \left\lbrace \rho_f\,\bfW^T\,\frac{\partial\,\bfVs}{\partial\,t} + \bfW^T\,\mathrm{diag}\left(\rho_{wi}\right)\,\frac{\partial\,\bfW}{\partial\,t} + \bfW^T\,\nabla p\right.\\
[13pt]
\left.+ \bfW^T\,\mathrm{diag}\left(\frac{\eta}{\kappa_i}\,\frac{1}{\Omega_i}\,(D+\Omega_i)^{1/2}\right)\,\bfW \right\rbrace \,dx\,dz = 0.
\end{array}
\label{eq:nrj_proof4}
\end{equation}
The second term in (\ref{eq:nrj_proof4}) can be written
\begin{equation}
\int_{\mathbb{R}^2}\bfW^T\,\mathrm{diag}\left(\rho_{wi}\right)\,\frac{\partial\,\bfW}{\partial\,t}\,dx\,dz = \frac{d}{dt}\,\frac{1}{2}\,\int_{\mathbb{R}^2}\bfW^T\,\mathrm{diag}\left(\rho_{wi}\right)\,\bfW\,dx\,dz.
\label{eq:nrj_proof5}
\end{equation}
Integrating by part the third term of (\ref{eq:nrj_proof4}), we obtain
\begin{equation}
\begin{array}{l}
\displaystyle \int_{\mathbb{R}^2}\bfW^T\,\nabla p\,dx\,dz \displaystyle = -\int_{\mathbb{R}^2}p\,\nabla . \bfW\,dx\,dz,\\
[15pt]
\hspace{0.1cm} \displaystyle = \int_{\mathbb{R}^2}p\,\frac{\partial\,\xi}{\partial\,t}\,dx\,dz = \int_{\mathbb{R}^2}p\,\left( \frac{1}{m}\,\frac{\partial\,p}{\partial\,t} + \bfBeta^T\,\frac{\partial\,\bfEps}{\partial\,t} \right)\,dx\,dz,\\
[15pt]
%\hspace{0.1cm} \displaystyle = \int_{\mathbb{R}^2}p\,\left( \frac{1}{m}\,\frac{\partial\,p}{\partial\,t} + \bfBeta^T\,\frac{\partial\,\bfEps}{\partial\,t} \right)\,dx\,dz,\\
%[20pt]
\hspace{0.1cm} = \displaystyle \frac{d}{dt}\,\frac{1}{2}\,\int_{\mathbb{R}^2}\frac{1}{m}\,p^2\,dx\,dz + \int_{\mathbb{R}^2}p\,\bfBeta^T\,\left( \bfC^{-1}\,\frac{\partial\,\bfSigma}{\partial\,t} + \bfC^{-1}\,\bfBeta\,\frac{\partial\,p}{\partial\,t}\right)\,dx\,dz,\\
[20pt]
\hspace{0.1cm} = \displaystyle \frac{d}{dt}\,\frac{1}{2}\,\int_{\mathbb{R}^2}\frac{1}{m}\,p^2\,dx\,dz + \int_{\mathbb{R}^2}\bfBeta^T\,\bfC^{-1}\,\frac{\partial\,\bfSigma}{\partial\,t}\,p\,dx\,dz + \int_{\mathbb{R}^2}\bfBeta^T\,\bfC^{-1}\,\bfBeta\,p\,\frac{\partial\,p}{\partial\,t}\,dx\,dz,\\
[20pt]
\hspace{0.1cm} = \displaystyle \frac{d}{dt}\,\frac{1}{2}\,\int_{\mathbb{R}^2}\frac{1}{m}\,p^2\,dx\,dz + \int_{\mathbb{R}^2}\bfBeta^T\,\bfC^{-1}\,\frac{\partial\,\bfSigma}{\partial\,t}\,p\,dx\,dz + \frac{d}{dt}\,\frac{1}{2}\,\int_{\mathbb{R}^2}\bfBeta^T\,\bfC^{-1}\,\bfBeta\,p^2\,dx\,dz.
\end{array}
\label{eq:nrj_proof6}
\end{equation}
We add (\ref{eq:nrj_proof1}) and the three first terms of (\ref{eq:nrj_proof4}). Using the symmetry of $\bfC$, there remains
\begin{equation}
\int_{\mathbb{R}^2}\rho_f\,\left( \bfVs^T\,\frac{\partial\,\bfW}{\partial\,t} + \bfW^T\,\frac{\partial\,\bfVs}{\partial\,t}\right) \,dx\,dz = \frac{d}{dt}\,\frac{1}{2}\int_{\mathbb{R}^2}2\,\rho_f\,\bfVs^T\,\bfW.
\label{eq:nrj_proof7}
\end{equation}
Equations (\ref{eq:derivee_frac}) and (\ref{eq:nrj_proof1})-(\ref{eq:nrj_proof7}) yield
\begin{equation}
\frac{d}{dt}\,(E_1 + E_2) = -\int_{\mathbb{R}^2}\int_0^{\infty}\frac{\eta}{\pi\,\sqrt{\theta}}\,\bfW^T\,\mathrm{diag}\left(\frac{1}{\kappa_i\,\sqrt{\Omega_i}}\right)\,\bfPsi\,d\theta\,dx\,dz.
\label{eq:nrj_proof8}
\end{equation}
To calculate the right-hand side of (\ref{eq:nrj_proof8}), equation (\ref{eq:EDO_psi_a}) is multiplied by $\bfW^T$ or $\bfPsi^T$
\begin{equation}
\left\lbrace 
\begin{array}{l}
\displaystyle \bfW^T\,\frac{\partial\,\bfPsi}{\partial\,t} - \bfW^T\,\frac{\partial\,\bfW}{\partial\,t} + \bfW^T\,\mathrm{diag}\left(\theta + \Omega_i\right)\,\bfPsi - \bfW^T\,\mathrm{diag}\left(\Omega_i\right)\,\bfW = \bfZero,\\
[10pt]
\displaystyle \bfPsi^T\,\frac{\partial\,\bfPsi}{\partial\,t} - \bfPsi^T\,\frac{\partial\,\bfW}{\partial\,t} + \bfPsi^T\,\mathrm{diag}\left(\theta + \Omega_i\right)\,\bfPsi - \bfPsi^T\,\mathrm{diag}\left(\Omega_i\right)\,\bfW = \bfZero.
\end{array}
\right. 
\label{eq:nrj_proof10}
\end{equation}
Equation (\ref{eq:nrj_proof10}) can be written as
\begin{equation}
\begin{array}{ll}
\bfPsi^T\,\mathrm{diag}\left(\theta + 2\,\Omega_i\right)\,\bfW = & \displaystyle \frac{\partial}{\partial\,t}\,\frac{1}{2}\,(\bfW-\bfPsi)^T\,(\bfW-\bfPsi)\\
[10pt]
& + \bfPsi^T\,\mathrm{diag}\left(\theta + \Omega_i\right)\,\bfPsi + \bfW^T\,\mathrm{diag}\left(\Omega_i\right)\,\bfW.
\end{array}
\label{eq:nrj_proof11}
\end{equation}
Substituting (\ref{eq:nrj_proof11}) in (\ref{eq:nrj_proof8}) leads to the relation (\ref{eq:dEdtJKD})
\begin{equation}
\begin{array}{ll}
\displaystyle \frac{d}{dt}\,(E_1 + E_2 + E_3) = & \displaystyle -\int_{\mathbb{R}^2}\int_0^{\infty}\frac{\eta}{\pi\,\sqrt{\theta}}\, \displaystyle \left\lbrace \bfPsi^T\,\mathrm{diag}\left(\frac{\theta+\Omega_i}{\kappa_i\,\sqrt{\Omega_i}\,(\theta+2\,\Omega_i)}\right)\,\bfPsi\right.\\
[15pt]
& \displaystyle \left. + \bfW^T\,\mathrm{diag}\left(\frac{\Omega_i}{\kappa_i\,\sqrt{\Omega_i}\,(\theta+2\,\Omega_i)}\right)\,\bfW \right\rbrace\,d\theta\,dx\,dz.
\end{array}
\label{eq:nrj_proof12}
\end{equation}
It remains to prove that $E$ (\ref{eq:E1E2E3_JKD}) is a positive definite quadratic form. Concerning $E_1$, we write
\begin{equation}
\rho\,\bfVs^T\,\bfVs + \bfW^T\,\mathrm{diag}\left(\rho_{wi}\right)\,\bfW + 2\,\rho_f\,\bfVs^T\,\bfW = \mbox{\boldmath$X_1$}^T\,\,\mbox{\boldmath$H_1$}\,\mbox{\boldmath$X_1$} + \mbox{\boldmath$X_3$}^T\,\,\mbox{\boldmath$H_3$}\,\mbox{\boldmath$X_3$},
\end{equation}
where
\begin{equation}
\mbox{\boldmath$X_i$} = (v_{si}\;w_i)^T,\quad\mbox{\boldmath$H_i$} = \left(\begin{array}{cc}
\rho & \rho_f\\
[10pt]
\rho_f & \rho_{wi}
\end{array}\right),\quad i=1,3.
\end{equation}
Taking ${\cal S}_i$ and ${\cal P}_i$ to denote the sum and the product of the eigenvalues of matrix $ \mbox{\boldmath$H_i$}$, we obtain
\begin{equation}
\left\lbrace 
\begin{array}{l}
{\cal P}_i = \det\, \mbox{\boldmath$H_i$} = \rho\,\rho_{wi} - \rho_f^2 = \chi_i > 0,\\
[10pt]
{\cal S}_i = \mbox{tr}\, \mbox{\boldmath$H_i$} = \rho + \rho_w > 0.
\end{array}
\right. 
\end{equation}
The eigenvalues of $\mbox{\boldmath$H_i$}$ are therefore positive. This proves that $E_1$ is a positive definite quadratic form. The terms $E_2$, $E_3$ and $-\frac{dE}{dt}$ are obviously positive definite quadratic form because the involved matrices are definite positive.$\qquad\square$

%------------------------------------------------------------------------------------------

\section{Matrices of propagation and dissipation} \label{annexe:matABS}

The matrices in (\ref{eq:2D_ani_syst_hyp_tens_ad}) are
\begin{equation}
\bfA = \left( 
\begin{array}{ccc}
\bfZero_{4,4} & \bfA_1 & \bfZero_{4,2N}\\ 
[10pt]
\bfA_2 & \bfZero_{4,4} & \bfZero_{4,2N}\\
[10pt]
\bfZero_{2N,4} & \bfA_3 & \bfZero_{2N,2N}
\end{array}
\right) ,\quad
\bfA_3 =
\left( 
\begin{array}{cccc}
\displaystyle \frac{\rho_f}{\chi_1} & 0 & 0 & \displaystyle \frac{\rho}{\chi_1}\\
[10pt]
0 & \displaystyle \frac{\rho_f}{\chi_3} & 0 & 0\\
[10pt]
\vdots & \vdots & \vdots & \vdots \\
[10pt]
\displaystyle \frac{\rho_f}{\chi_1} & 0 & 0 & \displaystyle \frac{\rho}{\chi_1}\\
[10pt]
0 & \displaystyle \frac{\rho_f}{\chi_3} & 0 & 0
\end{array}
\right),
\label{eq:matrixA_ani_ad}
\end{equation}
$$
\bfA_1 = \left( 
\begin{array}{cccc}
\displaystyle -\frac{\rho_{w1}}{\chi_1} & 0 & 0 & \displaystyle -\frac{\rho_f}{\chi_1}\\
[10pt]
0 & \displaystyle -\frac{\rho_{w3}}{\chi_3} & 0 & 0\\
[10pt]
\displaystyle \frac{\rho_f}{\chi_1} & 0 & 0 & \displaystyle \frac{\rho}{\chi_1}\\
[10pt]
0 & \displaystyle \frac{\rho_f}{\chi_3} & 0 & 0
\end{array}
\right) ,\quad
\bfA_2 = \left( 
\begin{array}{cccc}
-c_{11}^u & 0 & -\beta_1\,m & 0\\
[10pt]
0 & -c_{55}^u & 0 & 0\\
[10pt]
-c_{13}^u & 0 & -\beta_3\,m & 0\\
[10pt]
\beta_1\,m & 0 & m & 0
\end{array}
\right),
$$
\begin{equation}
\bfB = \left( 
\begin{array}{ccc}
\bfZero_{4,4} & \bfB_1 & \bfZero_{4,2N}\\ 
[10pt]
\bfB_2 & \bfZero_{4,4} & \bfZero_{4,2N}\\
[10pt]
\bfZero_{2N,4} & \bfB_3 & \bfZero_{2N,2N}
\end{array}
\right) ,\quad
\bfB_3 =
\left( 
\begin{array}{cccc}
0 & \displaystyle \frac{\rho_f}{\chi_1} & 0 & 0\\
[10pt]
0 & 0 & \displaystyle \frac{\rho_f}{\chi_3} & \displaystyle \frac{\rho}{\chi_3}\\
[10pt]
\vdots & \vdots & \vdots & \vdots \\
[10pt]
0 & \displaystyle \frac{\rho_f}{\chi_1} & 0 & 0\\
[10pt]
0 & 0 & \displaystyle \frac{\rho_f}{\chi_3} & \displaystyle \frac{\rho}{\chi_3}
\end{array}
\right),
\label{eq:matrixB_ani_ad}
\end{equation}
$$
\bfB_1 = \left( 
\begin{array}{cccc}
0 & \displaystyle -\frac{\rho_{w1}}{\chi_1} & 0 & 0\\
[10pt]
0 & 0 & \displaystyle -\frac{\rho_{w3}}{\chi_3} & \displaystyle -\frac{\rho_f}{\chi_3}\\
[10pt]
0 & \displaystyle \frac{\rho_f}{\chi_1} & 0 & 0\\
[10pt]
0 & 0 & \displaystyle \frac{\rho_f}{\chi_3} & \displaystyle \frac{\rho}{\chi_3}
\end{array}
\right),\quad
\bfB_2 = \left( 
\begin{array}{cccc}
0 & -c_{13}^u & 0 & -\beta_1\,m\\
[10pt]
-c_{55}^u & 0 & 0 & 0\\
[10pt]
0 & -c_{33}^u & 0 & -\beta_3\,m\\
[10pt]
0 & \beta_3\,m & 0 & m
\end{array}
\right),
$$
and $\bfS$ is the diffusive matrix
\begin{equation}
\bfS = \left( \begin{array}{ccc}
{\bf 0}_{4,4} & {\bf 0}_{4,4} & {\bf S}_1\\
[10pt]
{\bf 0}_{4,4} & {\bf 0}_{4,4} & {\bf 0}_{4,2N}\\
[10pt]
{\bf S}_3 & {\bf 0}_{2N,4} & {\bf S}_2
\end{array}
\right),\quad
\bfS_3 =
\left( 
\begin{array}{cccc}
0 & 0 & -\Omega_1 & 0\\
[10pt]
0 & 0 & 0 & -\Omega_3 \\
[10pt]
\vdots & \vdots & \vdots & \vdots \\
[10pt]
0 & 0 & -\Omega_1 & 0\\
[10pt]
0 & 0 & 0 & -\Omega_3
\end{array}
\right) ,
\label{eq:matrixS_ani_ad}
\end{equation}
\vspace{10pt}
$$
\bfS_1 =
\left( 
\begin{array}{ccccc}
\displaystyle -\frac{\rho_f}{\rho}\,\gamma_1\,a_1^1 & 0 & \cdots & \displaystyle -\frac{\rho_f}{\rho}\,\gamma_1\,a_N^1 & 0 \\
[10pt]
0 & \displaystyle -\frac{\rho_f}{\rho}\,\gamma_3\,a_1^3 & \cdots & 0 & \displaystyle -\frac{\rho_f}{\rho}\,\gamma_3\,a_N^3 \\
[10pt]
\displaystyle \gamma_1\,a_1^1 & 0 & \cdots & \displaystyle \gamma_1\,a_N^1 & 0 \\
[10pt]
0 & \displaystyle \gamma_3\,a_1^3 & \cdots & 0 & \displaystyle \gamma_3\,a_N^3
\end{array}
\right),
$$
\vspace{10pt}
$$
\bfS_2 =\left( 
\begin{array}{ccccc}
\gamma_1\,a_1^1 + (\theta_1^1+\Omega_1) & 0 & \cdots & \gamma_1\,a_N^1 & 0 \\
[10pt]
0 & \gamma_3\,a_1^3 + (\theta_1^3+\Omega_3) & \cdots & 0 & \gamma_3\,a_N^3 \\
[5pt]
\vdots & \vdots & \vdots & \vdots & \vdots \\
[10pt]
\gamma_1\,a_1^1 & 0 & \cdots & \gamma_1\,a_N^1 + (\theta_N^1+\Omega_1) & 0 \\
[10pt]
0 & \gamma_3\,a_1^3 & \cdots & 0 & \gamma_3\,a_N^3 + (\theta_N^3+\Omega_3)
\end{array}
\right).
$$

%------------------------------------------------------------------------------------------

\section{Proof of proposition \ref{prop:diffusive_part_vpS}} \label{annexe:proof_vpS}

We denote $\bfP_{{\cal B}}$ the change-of-basis matrix satisfying
\begin{equation}
\bfU = \bfP_{{\cal B}}\,\left( \bfU_1\,,\,\bfU_3\,,\,\bfSigma\,,\,p \right)^T,
\label{eq:diffu_change_of_basis}
\end{equation}
with
\begin{equation}
\bfU_i = (v_{si}\,,\,w_i\,,\,\psi_1^i\,,\,\cdots\,,\,\psi_N^i)^T,\quad i=1,3.
\label{eq:vecteur_etat_reduced}
\end{equation}
The matrix $\bfP_{{\cal B}}$ is thus invertible, and the matrices $\bfS$ (\ref{annexe:matABS}) and $\bfS_{{\cal B}} = \bfP_{{\cal B}}^{-1}\,\bfS\bfP_{{\cal B}}$ are similar. The matrix $\bfS_{{\cal B}}$ writes
\begin{equation}
\bfS_{{\cal B}} = \left( 
\begin{array}{cccc}
\tilde{\bfS_1} & \bfZero_{N+2,N+2} & \bfZero_{N+2,3} & \bfZero_{N+2,1}\\
[10pt]
\bfZero_{N+2,N+2} & \tilde{\bfS_3} & \bfZero_{N+2,3} & \bfZero_{N+2,1}\\
[10pt]
\bfZero_{3,N+2} & \bfZero_{3,N+2} & \bfZero_{3,3} & \bfZero_{3,1}\\
[10pt]
\bfZero_{1,N+2} & \bfZero_{1,N+2} & \bfZero_{1,3} & 0
\end{array}
\right)
\end{equation}
with ($i=1,3$)
\begin{equation}
\tilde{\bfS_i} = \left( \begin{array}{cc|cccc}
0 & 0 & \displaystyle -\frac{\rho_f}{\rho}\,\gamma_i\,a_1^i & \displaystyle -\frac{\rho_f}{\rho}\,\gamma_i\,a_2^i & \cdots & \displaystyle -\frac{\rho_f}{\rho}\,\gamma_i\,a_N^i\\
[10pt]
\rule[-2mm]{0mm}{2mm} 0 & 0 & \gamma_i\,a_1^i & \gamma_i\,a_2^i & \cdots & \gamma_i\,a_N^i\\ \hline
\rule[0mm]{0mm}{4mm} 0 & -\Omega_i & \gamma_i\,a_1^i + (\theta_1^i + \Omega_i) & \gamma_1\,a_2^i & \cdots & \gamma_i\,a_N^i\\
[10pt]
0 & -\Omega_i & \gamma_i\,a_1^i & \gamma_i\,a_2^i + (\theta_2^i + \Omega_i) & \cdots & \gamma_i\,a_N^i\\
[10pt]
\vdots & \vdots & \vdots & \vdots & \vdots & \vdots\\
[10pt]
0 & -\Omega_i & \gamma_i\,a_1^i & \gamma_i\,a_2^i & \cdots & \gamma_i\,a_N^i + (\theta_N^i + \Omega_i)
\end{array}
\right).
\label{eq:matrix_Sx_reduced}
\end{equation}
\medskip
The characteristic polynomial of $\bfS$ is
\begin{equation}
P_{\bfS}(s) = s^4\,P_{\tilde{\bfS_1}}(s)\,P_{\tilde{\bfS_3}}(s),
\label{eq:poly_cara_S}
\end{equation}
where $P_{\tilde{\bfS_i}}(s)$ denotes the characteristic polynomial of the matrix $\tilde{\bfS_i}$, i.e. $\tilde{\bfS_i}(s) = \det(\tilde{\bfS_i} - s\,\mbox{\boldmath$I$}_{N+2})$ with $\mbox{\boldmath$I$}_{N+2}$ the $(N+2)$-identity matrix. This $(N+2)$-determinant is expanded along the first column. The line $I$ and the column $J$ of the $(N+1)$-determinant thus obtained are denoted $L_I$ and $C_J$, respectively ($0\leqslant I,J \leqslant N$). The following algebraic manipulations are then performed successively:
\begin{equation}
\begin{array}{ll}
(i) & L_{\ell} \leftarrow L_{\ell} - L_0,\quad \ell = 1,\cdots,N,\\
[5pt]
(ii) & C_0 \leftarrow C_0\,\prod\limits_{\ell=1}^N(\theta_{\ell}^i+\Omega_i-s),\\
(iii) & C_0 \leftarrow C_0 - (s-\Omega_1)\,\mathop{\prod\limits_{k=1}^N}\limits_{k\neq \ell} (\theta_k^i+\Omega_i-s)\,C_{\ell},\quad \ell = 1,\cdots,N.
\end{array}
\end{equation}
One deduces
\begin{equation}
P_{\tilde{\bfS_i}}(s) = -s\,{\cal Q}_i(s) = s^2\,\prod\limits_{\ell=1}^N(\theta_{\ell}^i+\Omega_i-s) + \gamma_i\,s\,(s-\Omega_i)\,\sum\limits_{\ell=1}^N a_{\ell}^i\,\mathop{\prod\limits_{k=1}^N}\limits_{k\neq \ell} (\theta_k^i+\Omega_i-s).
\label{eq:poly_carac_Sx_4}
\end{equation}
From equation (\ref{eq:poly_carac_Sx_4}), one has $P_{\tilde{\bfS_i}}(0) \neq 0$ while ${\cal Q}_i(0) \neq 0$, therefore $0$ is an eigenvalue of the matrix $\tilde{\bfS_i}$ with multiplicity $1$. In what follows, the positivity of the coefficients $\theta_{\ell}^i$, $a_{\ell}^i$ of the diffusive approximation is used. In the limit $s \rightarrow 0^+$, then asymptotically
\begin{equation}
P_{\tilde{\bfS_i}}(s) \mathop{\sim}\limits_{s\rightarrow 0^+} -\gamma_i\,\Omega_i\,s\,\sum\limits_{\ell=1}^N a_{\ell}^i\,\mathop{\prod\limits_{k=1}^N}\limits_{k\neq \ell} (\theta_k^i+\Omega_i) \Rightarrow \mbox{sgn}\left(P_{\tilde{\bfS_i}}(0^+)\right) = -1.
\label{eq:vp_Sx_zero}
\end{equation}
Moreover, using (\ref{eq:poids_croissant}), then at the quadrature abscissae one has for all $\ell=1,\cdots,N$
\begin{equation}
P_{\tilde{\bfS_i}}(\theta_{\ell}^i + \Omega_i) = \gamma_i\,\theta_{\ell}^i\,(\theta_{\ell}^i + \Omega_i)\,a_{\ell}^i\,\mathop{\prod\limits_{k=1}^N}\limits_{k\neq \ell} (\theta_k^i-\theta_{\ell}^i) \Rightarrow \mbox{sgn}\left(P_{\tilde{\bfS_i}}(\theta_{\ell}^i + \Omega_i)\right) = (-1)^{\ell+1}.
\label{eq:vp_Sx_theta}
\end{equation}
Finally, the following limit holds
\begin{equation}
P_{\tilde{\bfS_i}}(s) \mathop{\sim}\limits_{s\rightarrow +\infty} (-1)^N\,s^{N+2} \Rightarrow \mbox{sgn}\left(P_{\tilde{\bfS_i}}(+\infty)\right) = (-1)^N.
\label{eq:vp_Sx_infini}
\end{equation}
We introduce the following intervals
\begin{equation}
I_N^i = ]\theta_N^i + \Omega_i,+\infty[,\; I_{\ell}^i = ]\theta_{\ell}^i,\theta_{\ell+1}^i + \Omega_i],\;\mbox{for}\,\ell = 1,\cdots,N-1,\; I_0^i = ]0,\theta_1^i + \Omega_i].
\label{eq:interval_eigen_S}
\end{equation}
The real-valued continuous function $P_{\tilde{\bfS_i}}$ changes of sign on each interval $I_{\ell}^i$. Consequently, according to the intermediate value theorem, $P_{\tilde{\bfS_i}}$ has at least one zero in each interval. Since $P_{\tilde{\bfS_i}}$ has at the most $N+1$ distinct zeros in $]0,+\infty[$, we deduce that $\exists\, ! \, s_{\ell}^i \in I_{\ell}^i / P_{\tilde{\bfS_i}}(s_{\ell}^i)=0,\quad \ell=1,\cdots,N+1$. Using equation (\ref{eq:poly_cara_S}), the characteristic polynomial of $\bfS$ (\ref{eq:poly_carac_Sx_4}) is therefore
\begin{equation}
P_{\bfS}(s) = s^6\,\prod\limits_{\ell=1}^{N+1}(s-s_{\ell}^1)\,(s-s_{\ell}^3),
\label{eq:poly_carac_Sx_5}
\end{equation}
which concludes the proof.$\qquad\square$

%------------------------------------------------------------------------------------------

\section{Semi-analytical solution of Test 2} \label{SecAppExact}

In test 2, we consider the interaction of a plane wave with a plane interface at normal incidence. This 1D case can be solved semi-analytically, by using Fourier analysis and poroelastic constitutive equations. Note that no shear wave is involved here. The general overview of the algorithm is as follows:
\begin{itemize}
\item writing the potentials of fluid and elastic motions in terms of the potentials of fast and slow compressional waves. To do so, diagonalize the vectorial Helmholtz decomposition of Biot equations \cite{BOURBIE}. The coefficient 
\begin{equation}
Y(\omega)=\frac{\textstyle \left((1-\phi)\,\rho_s+\rho_f\,\beta\,({\cal T}-1)\right)\,\omega^2-\left(\lambda_f+2\,\mu-m\,\beta^2\right)\,k^2+i\,\omega\,\phi\,\beta\,\frac{\textstyle \eta}{\textstyle \kappa}}{\textstyle \ds \rho_f\,({\cal T}\,\beta-\phi)\,\omega^2-i\,\omega\,\phi\,\beta\,\frac{\textstyle \eta}{\textstyle \kappa}F(\omega)}
\label{CoeffY}
\end{equation}
is introduced by the change of basis, where $k$ is the wavenumber and $F(\omega)$ is the JKD frequency correction;
\item deduce each field (velocity, stress, pressure) from the expressions of potentials and from the poroelastic constitutive equations;
\item for a given Fourier mode, the reflected and transmitted waves can be written
\begin{equation}
{\bf U}(x,\,\omega)=
\left(
\begin{array}{c}
\mp 1\\
0\\
\pm \phi(1-Y)\\
0\\
\ds
\frac{k}{\omega}\left(\lambda_f+2\mu+\beta\,m\,\phi(Y-1)\right)\\
0\\
\ds
\frac{k}{\omega}\left(\lambda_f+\beta\,m\,\phi(Y-1)\right)\\
[6pt]
\ds
-\frac{k}{\omega}m\,\left(\beta+\phi(Y-1)\right)
\end{array}
\right)\,e^{i(\omega t-kx)}\,{\hat g}(\omega),
\label{ModeFourier}
\end{equation}
where ${\hat g}(\omega)$ is the Fourier transform of the source (\ref{eq:ricker}). The sign $\pm$ depends whether the wave is a right-going incident or transmitted wave (+) or a left-going reflected wave (-);
\item the four reflected and transmitted slow and fast waves (\ref{ModeFourier}) must be multiplied by a reflection or transmitted coefficient. The four coefficients are computed by applying the jump conditions and by solving the resulting linear system;
\item an inverse Fourier transform yields an approximate time-domain solution.
\end{itemize}

\end{appendix}

%------------------------------------------------------------------------------------------
%------------------------------------------------------------------------------------------

\section*{References}

\bibliographystyle{elsarticle-num}
\bibliography{allbiblio}

\end{document}